\documentclass[conf]{new-aiaa} %for conference papers
\usepackage[utf8]{inputenc}
\usepackage{textcomp}
\usepackage[dvips]{graphicx}
\usepackage{float}
\usepackage{amsmath}
\usepackage[version=4]{mhchem}
\usepackage{siunitx}
\usepackage{longtable,tabularx}
\usepackage{comment}
\usepackage{indentfirst}
\usepackage{here}
\usepackage{multicol}
\usepackage{multirow}
\usepackage[subrefformat=parens]{subcaption}
\usepackage{color}

\usepackage{fancyhdr}

\fancypagestyle{plain}{
  \fancyhf{} % リセット
  \fancyhead[C]{\footnotesize
  This is the author's accepted manuscript of an article published in the
  AIAA Journal of Guidance, Control, and Dynamics.\\
  The version of record is available at \textit{https://arc.aiaa.org/doi/10.2514/1.G005873}.}
}

\newcommand{\bm}[1]{{\mbox{\boldmath $#1$}}}
\title{Kinematics Control of Electromagnetic Formation Flight\\Using Angular-Momentum Conservation Constraint}
\author{Yuta Takahashi \footnote{Ph.D. Candidate, Department of Mechanical Engineering, Tokyo Institute of Technology, 2-12-1 $\#$I3-17 Ookayama Meguro-ku, Tokyo 152-8552,
Japan; takahashi.y.cl@m.titech.ac.jp, AIAA Student Member.} and Hiraku Sakamoto\footnote{Associate Professor, Department of Mechanical Engineering, Tokyo Institute of Technology, 2-12-1 $\#$I3-17 Ookayama Meguro-ku, Tokyo 152-8552, Japan, AIAA Senior Member.}}
\affil{Tokyo Institute of Technology, Ookayama, Meguro, Tokyo, 152-8550, Japan}
\author{Shin-ichiro Sakai\footnote{Professor, Department of Spacecraft Engineering, Institute of Space and Astronautical Science (ISAS), Japan Aerospace Exploration Agency
(JAXA), Yoshinodai 3-1-1 Chuo-ku, Sagamihara 252-5210 Japan.}}
\affil{Japan Aerospace Exploration Agency (JAXA), 3-1-1 Yoshinodai, Chuo-ku, Sagamihara, Kanagawa, 252-5210, Japan}

\begin{document}
\maketitle
\thispagestyle{plain}  % ★ ここを追加：表紙だけ plain スタイルにする
%\begin{comment
\begin{abstract}
Electromagnetic formation flight (EMFF) uses the electromagnetic force to control the relative positions of multiple satellites without using conventional fuel-based propulsion. To compensate for the electromagnetic torque generated alongside the electromagnetic force, in most previous studies, all satellites were assumed to have reaction wheels (RWs) besides electromagnetic coils. However, the RW-loaded angular momentum becomes non-uniformly distributed among the satellites, because the electromagnetic torque usually differs between satellites. Without a proper control scheme, this deviation increases over time, and the RWs become saturated quickly, preventing the attitudes of the satellites from being controlled. In this study, a new controller is proposed that enables the electromagnetic force and torque to be controlled simultaneously. The EMFF kinematics derived from the conservation of angular momentum are used for the controller design. This controller can control $n$ satellites without saturating the RWs, and only one set of RWs is required among all satellites. The combination of the proposed controller with a simple unloading control exclusive to the chief satellite results in the elimination of the accumulation of angular momentum in the entire system. The effectiveness of the proposed controller is demonstrated through numerical simulations of the formation maintenance and formation reconfiguration of a five-satellite system.
%This paper presents a new kinematics control law for EMFF that enables the relative positions, absolute attitudes, and RW angular momentum to be controlled without saturation and equipping all satellites with RWs. This control lawcan theoretically eliminate the accumulation of angular momentum for the entire system by using a simple unloading control law performed only by the chief satellite.
 %Then, there is a problem of the RWs saturating relatively quickly because electromagnetic torque goes out of control and cause the non-uniform of the angular momentum among the attitude actuator of each satellite. As a result, the attitudes of each satellite cannot be controlled.
%Finally, to achieve a larger-scale Satellite Formation Flying 
%and Spacecraft Swarm using the EMFF system, 
%a decentralized control strategy is presented.
\end{abstract}
\section*{Nomenclature}
%\noindent(Nomenclature entries should have the units identified)
Abbreviations
{\renewcommand\arraystretch{1.0}
    \noindent\begin{longtable*}{@{}l @{\quad=\quad} l@{}}
        AC  & alternating current\\
        DC  & direct current\\
        EMFF & electromagnetic formation flight\\
        MTQs & 3-axis magnetic torques\\
        RWs & 3-axis reaction wheels\\
        Redundant EMFF & EMFF with more than one RWs in the system
\end{longtable*}}
List of Symbols
{\renewcommand\arraystretch{1.0}
\noindent\begin{longtable*}{@{}l @{\quad=\quad} l@{}}
%$A$  & area enclosed by the coil, m$^2$ \\
$\tilde{a}$ & cross-prodct matrix related to an arbitrary vector $\bm a$, where $\begin{bmatrix}a_x\\a_y\\a_z\end{bmatrix}\bm{\tilde{}}$\ =
$\begin{bmatrix}
  0&-a_z&a_y\\
  a_z&0&-a_x\\
  -a_y&a_x&0
\end{bmatrix}$\\
$\{\bm{a}\}$ & basis vector of an arbitrary frame (A); $\{\bm{a}\}=\{\bm{a}_x^{\mathrm{T}}, \bm{a}_y^{\mathrm{T}},\bm{a}_z^{\mathrm{T}}\}^{\mathrm{T}}$ \\
$^xa$ & component of an arbitrary vector $\bm a$ in an arbitrary frame ($X$)\\
$C^{Y/X}$ & coordinate transformation matrix from an arbitrary frame ($X$) to an arbitrary frame ($Y$)\\
%$\{\mathbb{B}_i\}$& base vectors of body-fixed frame of ith satellite, m\\
%$c$  & current strength, A\\
$\frac{^X\mathrm{d}}{\mathrm{d}t}\bm{a}$& time derivative of an arbitrary vector $\bm{a}$ in an arbitrary frame ($X$)\\
$\dot{a}$& time derivative of an arbitrary component $a$\\
$E_x$ & $x\times x$ unit matrixes\\
%$\{\mathbb{I}\}$& base vectors of inertial frame, m\\
$\bm f_{EMj}$ & electromagnetic force imparted by the system to the $j$-th satellite, N\\
$\bm h_j$ & reaction wheel angular momentum of the $j$-th satellite, N$\cdot$m$\cdot$s\\
$^b h$ & reaction wheel angular momentum of the system, $^b h=
    \left[
   \ ^{b_1}h_1^{\mathrm{T}},
   \cdots,
   \ ^{b_n}h_{n}^{\mathrm{T}}
    \right]^{\mathrm{T}}\in \mathbb{R}^{3n}$\\
$\bm h_{jd}$ & target reaction wheel angular momentum of the $j$-th satellite, N$\cdot$m$\cdot$s\\
$\bm I_j$ & inertial dyadic of the $j$-th satellite, kg$\cdot$m$^2$\\
$J_j$ & inertia of the $j$-th satellite in the body-fixed frame, kg$\cdot$m$^2$\\
$\bm L$ & angular momentum of the system, N$\cdot$m$\cdot$s\\
$\bm L_d$ & target angular momentum of the system, N$\cdot$m$\cdot$s\\
$m$ & number of reaction wheels in the system\\
$m_j$ & mass of the $j$-th satellite, kg\\
%$N_t$ & number of coil turns\\
$n$ & number of satellites\\
%$\bm n$ & unit vector perpendicular to the plane of the coil, m\\
$\dot{q}$ & time derivative of the generalized coordinates, $\dot{q}=\left[
   \ ^i\dot{r}^{\mathrm{T}},
   \dot{\sigma}^{\mathrm{T}},
   \ ^{b}h^{\mathrm{T}}
    \right]^{\mathrm{T}}\in \mathbb{R}^{6n+3m-3}$\\
%$\bm{R}$ & position vector from the center of the Earth to the reference position, m\\
%$R_o$ & distance from the center of mass of Earth to the center of mass of the formation, m\\
$\bm r_j$ & position vector of the $j$-th satellite with respect to the center of mass of the formation, m\\
$\bm r_{jk}$ & position vector of the $j$-th satellite viewed from the $k$-th satellite; $\bm r_{jk}=\bm r_{j}- \bm r_{k}$, m\\ 
$^i{r}$& position vectors of the system with respect to the center of mass of the formation, $
    ^i{r}=
    \left[
   \ ^i{r}_2^{\mathrm{T}},
   \cdots,
   \ ^i{r}_{n}^{\mathrm{T}}
    \right]^{\mathrm{T}}\in \mathbb{R}^{3n-3}$\\
$\bm\tau_{EMj}$ & electromagnetic torque imparted by the system to the $j$-th satellite, N$\cdot$m\\
$v$ & controlled states of the EMFF states $\zeta$, $v=
    \begin{bmatrix}
   \ ^i\dot{r}^{\mathrm{T}},
   \ ^{b}\omega^{\mathrm{T}},
   \ ^{b_1}\xi_1^{\mathrm{T}},
   \cdots,
   \ ^{b_{m-1}}\xi_{m-1}^{\mathrm{T}}
    \end{bmatrix}^{\mathrm{T}}\in \mathbb{R}^{6n+3m-6}$\\
%$T$ & one cycle to change the amplitude of alternating current, s\\
$\zeta$ & EMFF states, $\zeta=
    \left[
   ^i\dot{r}^{\mathrm{T}},\ 
   ^b\omega^{\mathrm{T}},\ 
   ^b\xi^{\mathrm{T}}
    \right]^{\mathrm{T}}\in \mathbb{R}^{(6n+3m-3)\times 1}$\\
%$\mu_e$ & dipole moment of Earth's geomagnetic field, 8.1 $\times$ 10$^{22}$ A$\cdot$m$^2$\\
$\mu_g$ & Earth's gravitational parameter, 3.986 $\times$ 10$^{14}$ m$^3$/s$^2$\\
$\mu_j$ & dipole moment of the $j$-th satellite, A$\cdot$m$^2$\\
$\mu_0$ & permeability constant, 4$\pi\times 10^{-7}$T$\cdot$ m/A\\
$\bm\xi_j$ & variables that are related to $\bm h_{jd}$ and $\bm L_d$, $\bm\xi_j=\left(\bm h_j -\frac{1}{m}\bm L\right)$, N$\cdot$m$\cdot$s\\
$^b\xi$ & variables of the system, $^b\xi=
    \left[
   \ ^{b_1}\xi_1^{\mathrm{T}},
   \cdots,
   \ ^{b_m}\xi_{m}^{\mathrm{T}}
    \right]^{\mathrm{T}}\in \mathbb{R}^{3m}$\\
$\sigma_j$ & modified Rodrigues parameters (MRPs) of the $j$-th satellite\\ 
$\sigma$ & MRPs of the system, $
    \sigma=
    \left[
    {\sigma}_1^{\mathrm{T}},
   \cdots,
   {\sigma}_{n}^{\mathrm{T}}
    \right]^{\mathrm{T}}\in \mathbb{R}^{3n}
$\\
%$^j\omega$ & angular velocity of frame J's componets, s$^{-1}$\\ 
%$\omega_o$ & orbital frequency $\omega_o=\sqrt{\mu_g /\|\bm R\|^3}$, s$^{-1}$\\
$\omega_{fj}$ & AC frequency of the $j$-th dipole moment, s$^{-1}$\\
$\bm \omega_j$ & angular velocity vector of the $j$-th satellite with respect to an inertial frame, s$^{-1}$\\
$^b\omega$ & angular velocity vector of the system, $^b\omega=
    \left[
   \ ^{b_1}\omega_1^{\mathrm{T}},
   \cdots,
   \ ^{b_n}\omega_{n}^{\mathrm{T}}
    \right]^{\mathrm{T}}\in \mathbb{R}^{3n}$\\
\end{longtable*}}
%\begin{multicols}{2}
\section{Introduction}
\label{sec1}
%燃料を使わずに長期間複数衛星のフォーメンションの安定性や制御性の維持を行うことは
%より柔軟で機能的な将来の宇宙ミッションを実現するために望ましい技術である．
%Without using fuel, maintaining stability and controllability of the multiple satellites' formation 
%for a long period is a desirable technology for realizing a more flexible and functional future space mission. %to realize innovative future space missions
\lettrine{M}{aintaining} the stability and control 
of satellite formations for long periods 
without consuming fuel creates the potential for more flexible and functional future space missions.
Satellite formation flying involves controlling the relative positions 
and absolute attitudes of multiple satellites. % \cite{Skinner,Aung,Lawson,Arya,Quadrelli,Kristiansen,Kapila,Foust,Xu,Morgan2,Sakai,Kwon,Porter,Youngquist,Nurge,Sunny,Fan,Huang2,Ahsun,Elias,Schweighart,Ramirez-Riberos,Ayyad,Kaneda,Abbasi,Abbasi2,Abbasi3,Zhang,Quadrelli1,Quadrelli2,Sakaguchi,Fabacher}.
%\cite{Kapila,Scharf,Scharf2,Alfriend,Kristiansen}. 
This technology provides many advantages for prospective space missions, 
including increased flexibility, scalability, accuracy, feasibility, fault tolerance, and cost reduction.
Furthermore, it enables new space developments, which cannot be realized using monolithic satellites \cite{Skinner}. Such examples include space interferometers \cite{Quadrelli1,Lawson}, space-based solar power systems \cite{Arya}, and
distributed antenna arrays \cite{Quadrelli}. However, owing to the many variables involved in maintaining satellite formation in orbit, continuous control is required to maintain the desired formation. Therefore, the mission period is often limited by fuel consumption. Increasing the initial fuel reserve increases the fuel mass and, therefore, the rocket launch cost. Moreover, the plume impingement from the thruster can cause disturbances and particle contamination \cite{Lawson}, which adversely affect the formation stability and performance of satellite optical systems. Therefore, the development of a system that can operate independently of fuel is one of the top targets in satellite flight research.

Electromagnetic formation flight (EMFF) is considered one of the most mature 
and promising propellant-free satellite formation flying methods
\cite{Sakai,Kwon,Porter,Youngquist,Nurge,Sunny,Fan,Fabacher,Huang2,Ahsun,Elias,Schweighart,Ramirez-Riberos,Ayyad,Kaneda,Abbasi3,Zhang,Sakaguchi}.
It operates on the principle that multiple satellites can be controlled by the electromagnetic forces and torques between the satellites that result from the interactions between coil-generated electromagnetic fields. %
%EMFF offers good controllability that can modulate the force in any direction by controlling the magnitude of the electromagnetic coil current. 
Previous research has proposed both the direct current (DC) \cite{Fabacher,Kwon,Fan,Huang2,Ahsun,Elias,Schweighart,Ramirez-Riberos,Sakaguchi}
 and alternating current (AC) methods \cite{Abbasi3,Porter,Sakai,Ayyad,Zhang,Youngquist,Nurge,Sunny,Kaneda} for the electromagnetic coil current. The DC method has the advantage of low power consumption, whereas the AC method provides higher functionality than the DC method by adjusting both the phase \cite{Kaneda,Sakai,Ayyad} and frequency \cite{Ayyad,Youngquist,Nurge,Abbasi3,Sunny}. For example, the coupling between the AC method control and Earth's magnetic field can be neglected compared to the control of EMFF \cite{Ahsun,Ayyad,Kaneda,Sakai,Zhang}, and the strong coupling of the EMFF dynamics can be separated \cite{Ahsun,Ayyad,Zhang,Kaneda,Sakai}. Further, many variables can be used to optimize the angular momentum and thermal buildup \cite{Ayyad}, and the relative position and relative attitude between satellites can be determined by measuring the magnetic fields of the satellites \cite{Nurge}. Multiple experiments have been conducted to test the performance of the DC method by ground testbeds using high temperature superconducting (HTS) wire \cite{Kwon}, while the validity of the AC method has been examined using experimental equipment on the ground and in space using testbeds \cite{Sakai,Youngquist,Nurge,Sunny,Porter,Kaneda}.

However, despite multiple demonstrations of EMFF technology, several persisting problems have prevented its practical application. One major problem is the uncontrollability of the electromagnetic torque generated by the electromagnetic force. %As explained %using the polynomial representation of the DC method in Sec.~\ref{sec2-3}, 
EMFF using the DC method cannot control the electromagnetic force and torque simultaneously owing to the lack of variables of the dipole moment \cite{Schweighart}. Moreover, both the electromagnetic force and torque can only take values that do not change the linear and angular momentum of the system. Owing to these constraints, most previous studies on the AC method have used only the electromagnetic force to control the relative position and regarded the electromagnetic torque as a disturbance. In the account of the electromagnetic torque, most previous studies assumed that all satellites have an additional attitude actuator, typically 3-axis reaction wheels (RWs). %, in addition to the electromagnetic coil. 
Although the total electromagnetic effects do not change the angular momentum of the entire system, the value of the electromagnetic torque acting on each satellite is different. Therefore, the angular momentum values loaded in the RWs (RW angular momentum) are distributed non-uniformly among the satellites. Without a proper control scheme, this deviation grows with time. After that, the RWs saturate relatively quickly; as a result, the attitudes of each satellite cannot be controlled \cite{Kwon}. Unloading via a fuel-based thruster \cite{Fabacher} usually consumes a significant amount of fuel. Moreover, unloading via Earth's magnetic field is not easy because simultaneous unloading by all satellites generates disturbances caused by interaction between the magnetic fields of the satellites. For the same reason, removing the RW angular momentum accumulated by external torque from the space environment is also difficult. 

To solve these non-uniform distribution and accumulation problems, the electromagnetic torque minimizing \cite{Schweighart,Fan,Ramirez-Riberos,Ayyad,Huang2}, 
polarity switch of all dipoles \cite{Ahsun}, %relative equilibrium shape invariance
torque-free formation \cite{Huang2}, desaturation using Earth's magnetic field \cite{Fan,Zhang,Ayyad,Schweighart,Fabacher}, and sinusoidal excitation to remove angular momentum buildup due to Earth's magnetic field \cite{Ahsun,Ayyad,Kaneda,Sakai,Zhang} have been studied. Nevertheless, these problems have only been mitigated, potentially saturating the attitude actuators during long-term operation. In addition, some of the corresponding solutions are unsuitable for certain scientific programs because they limit the formation shapes and are based on the existence of the Earth's magnetic field. In fact, the state of RWs depends on input history because the electromagnetic torque absorbed by RWs could not be controlled. This suggests that controlling the electromagnetic torque prevents the non-uniform distribution of the RW angular momentum. %Therefore, the simultaneous control law of the electromagnetic force and torque is desired to solve the problem fundamentally.} 
%but the problem has not been solved in general cases.

Therefore, this study proposes a new control law for EMFF systems that simultaneously controls the electromagnetic force and torque based on the AC method to address the non-uniform distribution and accumulation problems fundamentally. The contributions of the proposed control law are as follows. First, the proposed control law theoretically eliminates the non-uniform distribution of the RW angular momentum for each satellite. Second, the relative positions, absolute attitudes, and RW angular momentum of $n$ satellites are controlled assuming the system has only one set of RWs. Third, combining this control law with a simple unloading control in the chief satellite theoretically eliminates the accumulation of angular momentum throughout the entire system without previously using complicated algorithms \cite{Zhang,Ayyad}. The effectiveness of the proposed control law is demonstrated through three simulations. It should be noted that complete knowledge of the relative positions and attitude of all satellites along with their linear and angular velocities is assumed in this study to investigate the effectiveness of the proposed control law.

The remainder of the paper is organized as follows. In Sec.~\ref{sec2}, the relative translational and absolute attitude dynamics for the controller design are formulated. After the EMFF system description of the DC method and two constraints of EMFF, the AC modulation technique used in the latter half of this paper is introduced. Sec.~\ref{sec3} presents the conditions for achieving simultaneous control and EMFF kinematics which is derived from the angular momentum conservation constraint. Specifically, the polynomial representation of the AC method shows its extensibility for generating an arbitrary electromagnetic force and torque. ``Redundant EMFF'' is defined, and then, the control objectives to guarantee the existence of smooth state feedback while avoiding the non-uniform distribution problem are introduced. Subsequently, the kinematics of EMFF is derived from the conservation of angular momentum. Finally, the averaged dynamics of AC-based EMFF are computed.
%First, using the polynomial representation of the AC method, the extensibility of the AC method for outputting arbitrary electromagnetic forces and torques is shown. Then, the RW condition for realizing smooth feedback and the control target for removing the non-uniform distribution of the RW angular momentum are summarized from the viewpoint of nonholonomic mechanical systems. Finally, the kinematics of the EMFF, which has not been considered in previous research, is derived from the conservation of angular momentum of the entire system. 
Sec.~\ref{sec3-4} explains a new kinematic control law based on the EMFF kinematics for enabling the simultaneous control of the electromagnetic force and torque. In Sec.~\ref{sec4}, the validity of the proposed control law is confirmed via the formation maintenance and formation reconfiguration for five satellites. Our conclusions are presented in Sec.~\ref{sec5}.%of the formation maintenance and formation reconfiguration  %The remaining problems are then clarified. 
%In Sec.~\ref{sec5}, the problems for a larger-scale 
%Satellite Formation Flying and Spacecraft Swarm using the EMFF system 
%are outlined, and the decentralized control strategy 
%using the proposed method is presented.
%In Sec. II, the swarm reconﬁguration problem is formulated as a constrained, nonlinear optimal control problem and converted to nonlinear optimization. In Sec. III, the swarm reconﬁguration problem is broken into an assignment problem and a trajectory optimization problem. Then, the assignment is solved using VSDAA. In Sec. IV, the trajectory optimization problem is converted to convex optimization and the SCP algorithm is described. Additionally, the SCP algorithm is shown to converge to a trajectory that satisﬁes the Karush-Kuhn-Tucker (KKT) conditions34,35 of the nonconvex problem. In Sec. V, MPC is used to integrate VSDAA and SCP, and to implement a ﬁnite horizon so that the resulting algorithm, SATO, can be run onboard each agent in real-time with a disconnected communication network. In Sec. VI, SATO is run for both a 2-D, double integrator dynamics scenario and a 3-D, relative orbit dynamics scenario. The results of the two scenarios are analyzed and discussed
%The main contribution of this paper is
\section{Basic Formulations for Electromagnetic Formation Flight Control}
\label{sec2}
This section outlines the relative translational and absolute attitude dynamics for the controller design of satellite formation flying. After the system description and introduction of two constraints of EMFF, the AC modulation technique used in the latter half of this paper is introduced.
%First, the definition of the coordinate frame and the relative translational dynamics of satellite formation flying is outlined. Then, the attitude representation and dynamics of monolithic rigid satellites are outlined. 
%Moreover, the dynamics of the EMFF system and the control techniques proposed in previous studies are introduced.
In the last section, the AC modulation technique used in the latter half of this paper is introduced. 
\subsection{Relative Translational Dynamics}
\label{sec2-1}
\begin{figure}[tb]
\centering
\includegraphics[width=.5\textwidth]{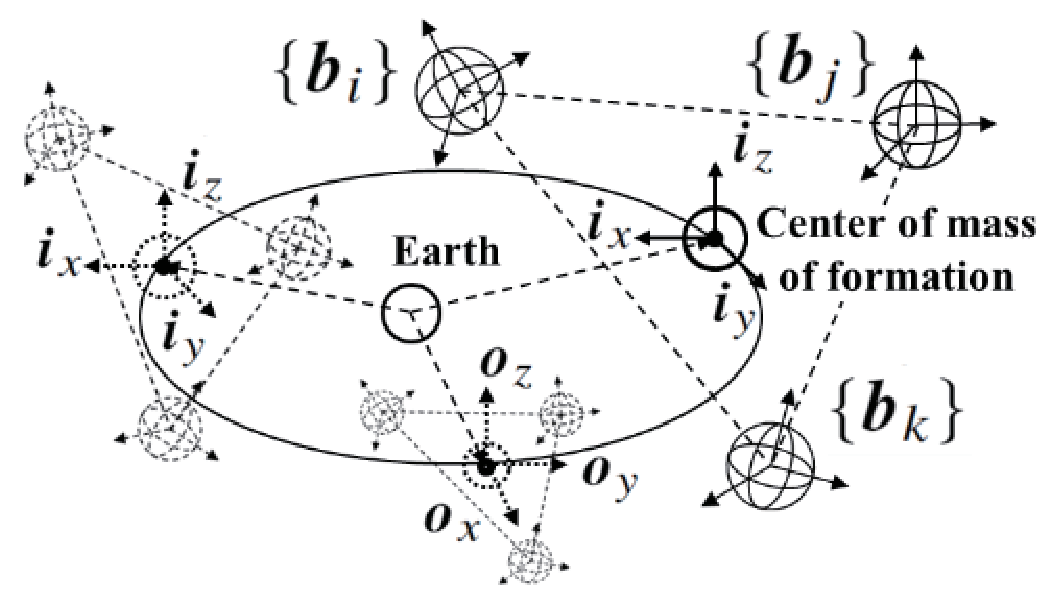}
%{graph}
\caption{Coordinate systems mainly used in this paper.}
\label{sec2:fig1}
\end{figure}

This subsection outlines the coordinate systems (see Fig.~\ref{sec2:fig1}) and relative translational dynamics of satellite formation flying. In particular, ``the invariant orientation frame (I)'' which is mainly used in this paper, is determined. It should be noted that the Earth-centered inertial (ECI) frame \cite{Ahsun} is defined as an inertial frame, for simplicity.
%\end{multicols}
\begin{comment}
\begin{figure}[tb]
\centering
%\includegraphics[width=9truecm]
\includegraphics[width=.5\textwidth]{geometry-of-the-system.eps}
%{graph}
\caption{Coordinate systems mainly used in this paper.}
\label{sec2:fig1}
\end{figure}
%\begin{multicols}{2}
\end{comment}

%First, an Earth-centered inertial (ECI) frame is introduced. The x-axis is directed toward the vernal equinox, the z-axis is directed along Earth's rotation axis toward the celestial north pole, and the y-axis completes a right-handed orthogonal frame. The basis vector is denoted as $\{\bm{ECI}\}$.
The first coordinate system is the body-fixed frame of the $j$-th satellite (B$_j$), which is located at the center of mass of the $j$-th satellite; its basis vector is defined as $\{\bm{b}_j\}$.
The second and third coordinate systems are the orbital reference frame (O) and the invariant orientation frame (I). Both frames are located at the center of mass of the formation, and their basis vectors are denoted as $\{\bm{o}\}$ and $\{\bm{i}\}$, respectively. The difference between the two frames lies in the definition of the $x$-axes. The $x$-axis $\bm o_x$ in (O) points in the orbit radial direction, whereas the $x$-axis $\bm i_x$ in (I) points to an invariant orientation within the orbital plane, such as the radial direction of the orbit at the initial time. In both (O) and (I) frames, the $z$-axes correspond to the orbit normal vector, and the $y$-axes complete a right-handed orthogonal frame. Therefore, the frame (I) is defined as a frame that does not rotate to the frame (ECI). As a result, the time derivative at frame (I) coincides with that at the inertial frame $\left(\frac{^{ECI}\mathrm{d}}{\mathrm{d}t}=\frac{^I\mathrm{d}}{\mathrm{d}t}\right)$. This property is useful for formulating the angular momentum of the system.
%%%%%%%%%%%%%%%%%%%%%%%%%%%%%%%%%%%%%%%%%%%%%%%%%%%%%%%%%%%%%%%%%%%%%%%%%%%%%
\begin{comment}
Although this reference frame could also be treated as an orbital frame 
attached to the specific satellite or the center of mass of the formation,
%the satellite will be maneuvering so 
its orbital velocity would not remain constant in a real application, and
%While for electromagnetic formations under the condition of no external force, 
the center of mass of the formation cannot be manipulated using electromagnetic actuation. 
\end{comment}
%%%%%%%%%%%%%%%%%%%%%%%%%%%%%%%%%%%%%%%%%%%%%%%%%%%%%%%%%%%%%%%%%%%%%%%%%%%%%
%Although this reference frame could also be treated as an orbital frame attached to the specific satellite, its orbital velocity would not remain constant because its satellite will be maneuvering. Furthermore, an orbital frame attached to the center of mass of the formation 
%While for electromagnetic formations under the condition of no external force, 
%the center of mass of the formation cannot be manipulated using electromagnetic actuation. Therefore, using an inertial reference frame (I) provides greater accuracy.}

Next, the relative translational dynamics for circular and elliptical orbits around the Earth is 
\begin{equation}
  \label{sec2:eq1}
    \frac{^I\mathrm{d}^2}{\mathrm{d}t^2}\bm{r}_j=\frac{1}{m_j}\bm{f}_{jc}+\frac{\mu_g}{\|\bm R\|^3}\left (3\bm{o}_x\bm{o}_x\cdot \bm{r}_j
      %\{\bm{O}\} ^{\mathrm{T}}\begin{bmatrix}^ox_{i}\\0\\0\end{bmatrix}
    -\bm{r}_j \right ).
\end{equation} assuming that $\|\bm r_j\|\ll \|\bm R\|$, where $\bm{R}$ is the position vector from the center of the Earth to the reference position. Unlike exact nonlinear models \cite{Morgan2}, this study does not include the perturbations of the center of mass of the system. Since EMFF can only output internal forces, the absolute position control in orbit requires additional actuators, such as thrusters. To simplify the derivation of the conservation of angular momentum, this study represents the relative translational dynamics, as shown in Eq.~\eqref{sec2:eq1}, as components of the invariant orientation frame (I).
\subsection{Absolute Attitude Dynamics of Rigid Spacecraft with Reaction Wheels}
In this subsection, the attitude representation and dynamics of monolithic rigid satellites are outlined. The modified Rodrigues parameters (MRPs) vector $\sigma$=\{$\sigma_1,\sigma_2,\sigma_3$\}$^{\mathrm{T}}$ is used to represent the attitude of the satellite \cite{Schaub}. The absolute attitude of the $j$-th satellite frame (B$_j$) and its target frame (B$_{jd}$) with respect to the invariant orientation frame (I) are expressed by $\sigma_j$ and $\sigma_{jd}$ as the MRPs, respectively. Based on the definitions, the relative attitude $\delta\sigma_j$ and the relative angular velocity ${^{b_j}\delta\omega_j}$ of the $j$-th satellite with respect to the target are expressed as 
\begin{equation}
  \begin{aligned}
       \delta\sigma_j&%=\sigma_j\otimes \sigma_{jd}^{-1}
       =\frac{(1-\|\sigma_{jd}\|^2)\sigma_{j}-(1-\|\sigma_{j}\|^2)\sigma_{jd}+2\sigma_{j}\times \sigma_{jd}}{1+\|\sigma_{j}\|^2\|\sigma_{jd}\|^2+2\sigma_{j}^{\mathrm{T}}\sigma_{jd}},\\
       {^{b_j}\delta\omega_j}&%={^{b_j}\omega_j}-{^{b_{jd}}\omega_{jd}}
       ={^{b_j}\omega_j}-C^{B_{j}/B_{jd}}\ {^{b_{jd}}\omega_{jd}}.
\end{aligned}
\end{equation}
Then, the MRPs kinematic differential equation for the $j$-th satellite is expressed as
\begin{equation}
  \label{sec2:eq3}
  \begin{aligned}
     \dot{\sigma}_j=Z(\sigma_j){^{b_j}\omega_j},\qquad Z(\sigma_j)=\frac{1}{4}\left[(1-\sigma_j^{\mathrm{T}}\sigma_j)E_3+2\tilde{\sigma}_j+2\sigma_j\sigma_j^{\mathrm{T}}\right].\\
\end{aligned}
\end{equation}
Note that the relative attitude variable pair ($\delta\dot{\sigma}_j$,${^{b_j}\delta\omega_j}$) also satisfies this MRP kinematics \cite{Wen}.
Although the MRPs vector takes a singular value when the principal rotation angle $\Phi$ is $\pm$360 degrees, this geometric singularity can be avoided by the shadow set of MRPs $\sigma^s =-\sigma/(\sigma^{\mathrm{T}}\sigma)$. %which defined in Eq.~\eqref{sec2:eq1-2-2}.
%\begin{equation}
%  \label{sec2:eq1-2-2}
%  \begin{aligned}
%     \sigma^s =\frac{-\sigma}{\sigma^2}.
%\end{aligned}
%\end{equation}
As $\sigma^s$ takes a singular value at $\Phi = 0$, switching $\sigma$ with $\sigma^s$ eliminates the singularity. Therefore, matrix $Z$ in Eq.~\eqref{sec2:eq3} is definite in any attitude and globally regular matrix. Similarly, using the following regular matrix $[Z]\in \mathbb{R}^{3n\times 3n}$, the MRPs kinematic differential equation of the system is expressed as
\begin{equation}
  \label{sec2:eq4}
   \dot\sigma=
   \begin{bmatrix}
    \dot\sigma_1\\
    \vdots\\
    \dot\sigma_n\\
  \end{bmatrix}
   =[Z]\begin{bmatrix}
    {^{b_1}\omega_1}\\
    \vdots\\
    {^{b_n}\omega_n}\\
  \end{bmatrix},\qquad [Z]=
    \begin{bmatrix}
    Z(\sigma_1)&&0\\
    &\ddots&\\
    0&&Z(\sigma_n)\\
  \end{bmatrix}.
\end{equation}
Subsequently, the attitude dynamics of a rigid spacecraft with RWs \cite{Schaub} is expressed as
\begin{equation}
 \label{sec2:eq5}
  \begin{aligned}
    \bm{I}_j\cdot \frac{^{B_j}\mathrm{d}}{\mathrm{d}t}\bm{\omega}_j+\bm{\omega}_j\times (\bm{I}_j\cdot\bm{\omega}_j+\bm{h}_j)&=\bm{\tau}_{jc}-\bm{\tau}_{j(RW)}+\bm{\tau}_{jd},\\
     \frac{^{B_j}\mathrm{d}}{\mathrm{d}t}\bm{h}_j&=\bm{\tau}_{j(RW)},
  \end{aligned}
  \end{equation}
where $\bm{\tau}_{jd}$ is the disturbance torque of $j$-th satellite.
In this study, the gravity gradient torque $\bm{\tau_g}$ and Earth's magnetic field $\bm{B_e}$ are considered as the sources of disturbances to the attitude of the satellite. The gravity gradient torque is one of the most important external effects in orbit. %Although gravity gradient effects seem complex to compute, 
They can be approximated accurately as follows \cite{Schaub}:
\begin{equation}
   \label{sec2:eq6}
\bm{\tau_g}=3\frac{\mu_g}{R^5}(\bm{R} \times \bm{I} \cdot \bm{R}).
\end{equation}
%where $\mu_g$=3.986$\times$10$^{14}$$m^{3}/s^{2}$ is the gravitational parameter of the Earth, $\bm{R}$ is the position vector from the center of the Earth to the reference position, and $\bm{I}$ is the inertial dyadic of the satellite.
The dipolar approximation model of Earth's magnetic field that can provide a sufficient estimate of geomagnetism is % for most places on Earth. 
\begin{equation}
  \label{sec2:eq6-2}
  \begin{aligned}
     \bm{B}_e (\bm\mu_e,\bm R)=\frac{\mu_0}{4\pi} \left(
        \frac{3\bm R(\bm{\mu}_e \cdot \bm R)}{\|\bm R\|^5}-\frac{\bm{\mu}_e}{\|\bm R\|^3}
        \right).
      \end{aligned}
\end{equation}
This approximates the magnetic properties of Earth as those of a magnetic dipole with a moment of $\mu_e \approx$ 8.1$\times$ 10$^{22}$ A$\cdot$ m$^2$ and tilted by approximately 11$^\circ$ from the geographic poles \cite{Poedts}:

%where $\mu_0=4\pi\times 10^{-7}$ T$\cdot$ m/A is the permeability constant.
\subsection{Direct Current Method for Electromagnetic Formation Flight System}
\label{sec2-3}
This subsection describes the generation of the electromagnetic force and torque between electromagnetic coils. 
In an EMFF system, each satellite is equipped with 3-axis electromagnetic coils and 
controlled by adjusting the value of the ``dipole moment,'' which is proportional to the current $c$. For a circular coil, the relationship between the dipole moment $\bm\mu$ and the current $c$ is given by:
\begin{equation}
  \label{sec2:eq7}
  \begin{aligned}
     \bm\mu&=N_tcA\bm n,
\end{aligned}
\end{equation}
where $N_t$ is the number of coil turns; $c$ is the current strength; $A$ is the area enclosed by the coil; $\bm n$ is the unit vector perpendicular to the plane of the coil. The dipole moment of the $j$-th satellite $\bm\mu_j$ is defined by the sum of three different orthogonal circular coils in the satellite. It should be noted that the magnetic dipole is assumed to be located precisely at the center of the mass of the satellites. Therefore, disturbing torques caused by the lever-arm between the coil and their centers of mass are not taken into account, unlike in previous study \cite{Fabacher}. %The fixed coils on each satellite can be approximated as a single steerable coil, 
Following the DC approach for EMFF, the magnetic field $\bm B (\bm\mu_k, \bm{r}_{jk})$, 
electromagnetic force $\bm f (\bm\mu_k,\bm\mu_j, \bm{r}_{jk})=\bm\mu_j \cdot \nabla \bm B (\bm\mu_k,\bm r_{jk})$, and 
electromagnetic torque $\bm\tau (\bm\mu_k,\bm\mu_j, \bm r_{jk})= \bm\mu_j \times \bm B (\bm\mu_k,\bm r_{jk})$ 
exerted on the $j$-th dipole
by the $k$-th dipole are expressed as \cite{Schweighart}:
\begin{equation}
  \label{sec2:eq8}
  \begin{aligned}
     B (\bm\mu_k,\bm r_{jk})=\frac{\mu_0}{4\pi} \left(
        \frac{3\bm r_{jk}(\bm\mu_k \cdot \bm r_{jk})}{\|\bm r_{jk}\|^5}-\frac{\bm\mu_k}{\|\bm r_{jk}\|^3}
        \right),
      \end{aligned}
\end{equation}
\begin{equation}
  \label{sec2:eq9}
  \begin{aligned}
    \bm{f}(\bm\mu_k,\bm\mu_j, \bm r_{jk})
    =&\frac{3\mu_0}{4\pi}
    \left(
            \frac{\bm\mu_k \cdot \bm\mu_j}{\|\bm{r}_{jk}\|^5} \bm r_{jk}+
            \frac{\bm\mu_k \cdot \bm r_{jk}}{\|\bm r_{jk}\|^5}\bm\mu_{j}+
            \frac{\bm\mu_j \cdot \bm r_{jk}}{\|\bm r_{jk}\|^5}\bm\mu_{i}
            -5\frac{(\bm\mu_k \cdot \bm r_{jk})(\bm\mu_j\cdot \bm r_{jk})}{\|\bm r_{jk}\|^7}\bm r_{jk}    
           \right),
   \end{aligned}
\end{equation}
and
\begin{equation}
  \label{sec2:eq10}
  \begin{aligned}
     \bm\tau (\bm\mu_k, \bm\mu_j, \bm r_{jk}) 
          &=\frac{\mu_0}{4\pi} \bm\mu_j \times \left(
        \frac{3\bm r_{jk}(\bm\mu_k \cdot \bm r_{jk})}{\|\bm r_{jk}\|^5}-\frac{\bm\mu_k}{\|\bm r_{jk}\|^3}
        \right),
\end{aligned}
\end{equation}
where $\bm r_{jk}=\bm r_{j}- \bm r_{k}$ is the position vector of the $j$-th satellite viewed from the $k$-th satellite. The magnetic field $B (\bm\mu_k,\bm r_{jk})$ in Eq.~\eqref{sec2:eq8} is given by a far-field approximation \cite{Schweighart}, which is accurate only when the distance between the satellites exceeds two times the radius of the coil \cite{Sakaguchi}. Numerical calculations in this study are conducted to the extent that this approximation condition is satisfied. %This provides accurate results only when the distance between the satellites is sufficiently large relative to the radius of the coils.
Otherwise, the ratios of exact to far-field forces \cite{Sakaguchi} and the exact model \cite{Schweighart}, which is computationally expensive, can be used. It should be noted that the electromagnetic force and torque satisfy the equality and inequality in Eq.~\eqref{sec2:eq12}: %because the EMFF system only generates internal force.
\begin{equation}
  \label{sec2:eq12}
     \bm{f}(\bm\mu_k,\bm\mu_j, \bm r_{jk})+\bm{f}(\bm\mu_j,\bm\mu_k, \bm r_{ji})=\bm 0,\qquad \bm\tau (\bm\mu_k, \bm\mu_j,\bm r_{jk})+\bm\tau(\bm\mu_j, \bm\mu_k,\bm r_{ji})\neq \bm 0.
\end{equation}
% using a far-field approximation of the magnetic field \cite{Schweighart}
Finally, the electromagnetic force $\bm f_{EMj}$ and torque $\bm\tau_{EMj}$ imparted by the system to the $j$-th satellite can be derived by the summation of the electromagnetic force and torque on the $j$-th satellite.
\begin{equation}
  \label{sec2:eq10-2}
     \bm {f}_{EMj}=\sum_{i=1}^n \bm f (\bm\mu_k, \bm\mu_j, \bm r_{jk}),\qquad \bm{\tau}_{EMj}=\sum_{i=1}^n \bm\tau (\bm\mu_k, \bm\mu_j, \bm r_{jk}). 
\end{equation}

At each time step, EMFF derives dipole moments $\bm\mu$, known as ``dipole solutions,'' to realize the target outputs calculated by the controller. Then, the derived values of $\bm\mu$ are assigned to all the electromagnetic coils available in the system. This derivation process is called ``dipole inversion.'' A calculation example of this process is as follows
\begin{equation}
  \label{sec2:eq11-2}
  \left \{
  \begin{aligned}
    &Minimize : J ={^b\tau_{EM}^{\mathrm{T}}}{^b\tau_{EM}}\\
       &Subject\ to :\  ^if_{cj} =\  ^if_{EMj}
  \end{aligned}
  \right.
\end{equation}
This equation describes the nonlinear nonconvex optimization problem with evaluation function $J={^b\tau_{EM}^{\mathrm{T}}}{^b\tau_{EM}}$ to minimize the generated $\tau_{EM}$ \cite{Schweighart,Fan,Ramirez-Riberos,Ayyad,Huang2}.
Note that the derived dipole moment (not control values) can take discontinuous values because the solution can move from one local optimum to another. Although this dipole inversion is processed at each time step, initializing the nonlinear solver with the solution of the previous time step significantly reduces the convergence time, facilitating real-time implementation \cite{Ayyad}. In addition to general nonlinear programming algorithms, the calculation using the homotopy (continuation) method \cite{Morgan} is proposed for DC-based EMFF, to find all possible solutions for deriving optimal $\tau_{EM}$ \cite{Schweighart} or to find the evolution of the one viable guidance solution in time \cite{Fabacher}. 
\subsection{Dipole Modulation of Alternating Current Method}
\label{sec2-4}
%In the previous chapter, the electromagnetic force, and torque corresponding to the dipole moment 
%and the distance between satellites were shown. 
This subsection outlines 
%the dipole moment assignment when controlling an electromagnetic system and, 
the modulation technique of the AC method \cite{Ayyad}. 
%制御器によって目標出力$f_c, \tau_c$が決定されると，この出力を実現するため，システムで利用可能なすべての電磁コイルに磁気モーメントが割り当てられる．この割り当ては「dipole inversion」と呼ばれる．
%交流電流駆動のダイポールAのダイポールモーメント$\mu_a(t)$は式Eq()で表される．
The AC-driven dipole moments of the sine wave $\bm{\mu}_a(t)$ and $\bm{\mu}_b(t)$ are expressed in Eq.~\eqref{sec2:eq13}:
\begin{equation}
  \label{sec2:eq13}
  \begin{aligned}
        \bm{\mu}_a(t)=\bm{\mu}_{(sin)a}\sin(\omega_{fa}t+\theta),\qquad \bm\mu_b(t)=\bm\mu_{(sin)b}\sin(\omega_{fb}t), \\
        %\bm f (\bm\mu_a(t),\bm\mu_b(t), \bm r_{ba})=\sin(\omega_{fa}t+\theta)\sin(\omega_{fb}t)\bm f(\bm{\mu}_{(\sin ) a}, \bm\mu_{(\sin ) b}, \bm r_{a b}),
  \end{aligned} 
\end{equation}
where $\bm\mu_{(sin)a}$ and $\bm\mu_{(sin)b}$ are the amplitudes of the sine wave for the $a$-th and $b$-th dipoles, respectively, and $\theta$ is the phase difference between two dipole moments. Using the periodic dipole moment, the time-varying electromagnetic force $\bm f(\bm\mu_a(t),\bm\mu_b(t), \bm r_{ba})$ and torque $\bm\tau(\bm\mu_a(t),\bm\mu_b(t), \bm r_{ba})$ are derived.
\begin{equation}
  \label{sec2:eq13-2}
  \left \{
  \begin{aligned}
    &\bm f(\bm\mu_a(t),\bm\mu_b(t), \bm r_{ba})=(\kappa_1+\kappa_2)\bm f(\bm\mu_{(sin)a},\bm\mu_{(sin)b},\bm r_{ba})\\
    &\bm\tau(\bm\mu_a(t),\bm\mu_b(t), \bm r_{ba})=(\kappa_1+\kappa_2)\bm\tau(\bm\mu_{(sin)a},\bm\mu_{(sin)b},\bm r_{ba})\\
%&\kappa_1+\kappa_2=\bm f(\bm\mu_a(t),\bm\mu_b(t), \bm r_{ba})/\bm f(\bm\mu_{(sin)a},\bm\mu_{(sin)b},\bm r_{ba})\\&=\bm\tau(\bm\mu_a(t),\bm\mu_b(t), \bm r_{ba})/\bm\tau(\bm\mu_{(sin)a},\bm\mu_{(sin)b},\bm r_{ba})\\
&\kappa_1,\kappa_2=\pm\frac{1}{2}\cos((\omega_{fa}\mp\omega_{fb})t+\theta)\\
     % &\kappa_1=\frac{1}{2}\cos((\omega_{fa}-\omega_{fb})t+\theta)\\
     % &\kappa_2=-\frac{1}{2}\cos((\omega_{fa}+\omega_{fb})t+\theta)\\
      \end{aligned}
      \right.
\end{equation}

Because the discussion of the stability of the time-varying systems is complex, they are usually approximated based on first-order averaging \cite{Sanders} as described in previous AC EMFF studies \cite{Porter,Sakai,Ayyad,Zhang,Youngquist,Nurge,Sunny,Kaneda}. Note that the following discussion solely refers to the electromagnetic force, but similar averaging can be applied to the electromagnetic torque. The average electromagnetic force $\bm f_{avg}$ 
%and torque $\bm\tau_{EM(avg)}$
%exerted by the $a$-th dipole on the $b$-th dipole 
during the period $T=2N_{\pm}\pi/(\omega_{fa}\pm \omega_{fb}) \ (N_+,N_-\in\mathbb{Z})$ is expressed by Eq.~\eqref{sec2:eq14}. 
%=&\frac{1}{2}(\cos\theta-\cos(2\omega_{fa}t+\theta))\\
%&T=2N_+\pi/(\omega_{fa}+\omega_{fb}) =2N_-\pi/(\omega_{fa}-\omega_{fb}) \ (N_+,N_-\in\mathbb{Z})
\begin{equation}
  \label{sec2:eq14}
  %\left \{
  \begin{aligned}
      \bm f_{avg}(\bm\mu_a(t),\bm\mu_b(t), \bm r_{ba})=&\left (\frac{1}{T}\int_0^T \left (\kappa_1 +\kappa_2 \right ) dt\right )\bm f(\bm\mu_{(sin)a},\bm\mu_{(sin)b},\bm r_{ba}),\\
     %\bm \tau_{avg}(\bm\mu_a(t),\bm\mu_b(t), \bm r_{ba})=&\left(\frac{1}{T}\int_0^T \left(\kappa_1 +\kappa_2 \right) dt\right) \tau(\bm\mu_{(sin)a},\bm\mu_{(sin)b},\bm r_{ba}),\\
      %T=2N_{\pm}\pi/(\omega_{fa}\pm \omega_{fb}) &=2N_-\pi/(\omega_{fa}-\omega_{fb}) \ (N_+,N_-\in\mathbb{Z}).\\
      \end{aligned}
      %\right .
\end{equation}
This approximation holds when either the AC frequency or the dynamic frequency is sufficiently large for each. %when the AC frequency or the dynamic frequency is sufficiently high.  when is sufficiently higher than the AC frequency. 
Equation~\eqref{sec2:eq14} implies that, on average, the AC-driven magnetic fields do not interact with each other for different AC frequencies $\omega_f$ or AC phases $\theta$ that are offset by $90^\circ$. Following the same principle, the coupling between the AC control and Earth's magnetic field, which is a constant value for short intervals, can be neglected compared to the control of EMFF \cite{Ahsun,Zhang}. Therefore, if the two dipoles are driven at the same value of frequency $\omega_f$ with no phase difference, the averaged electromagnetic force during $T=N\pi/\omega_f\ (N\in\mathbb{Z})$ can be expressed as
\begin{equation}
  \label{sec2:eq14-2}
  \begin{aligned}
      \bm f_{avg}(\bm\mu_a(t),\bm\mu_b(t), \bm r_{ba})=&\underbrace{\frac{1}{2}\bm f(\bm\mu_{(sin)a},\bm\mu_{(sin)b},\bm r_{ba})}_{\text {average term}} \underbrace{-\frac{1}{2}\cos(2\omega_{f}t)\bm f(\bm\mu_{(sin)a},\bm\mu_{(sin)b},\bm r_{ba})}_{\text {oscillatory term $\approx$ 0}}\\
      \approx &\frac{1}{2}f(\bm\mu_{(sin)a},\bm\mu_{(sin)b},\bm r_{ba}).
      \end{aligned}
\end{equation}
Herein, the oscillation term in Eq.~\eqref{sec2:eq14-2} is set to zero by averaging during T; however, it appears as a control error. This error is shown to be tolerable by setting the value of the AC frequency $\omega_f$ based on the frequency response analysis \cite{Sakai,Kaneda}.
\section{Simultaneous Control Conditions and Electromagnetic Formation Flight Kinematics}
\label{sec3}
%This section derives a new control law enabling the electromagnetic force and torque to be controlled simultaneously. 
%\subsection{Similarities between Electromagnetic Formation Flight and Redundant Manipulators}
To achieve simultaneous control of the electromagnetic force and torque, this section details the required conditions, and the EMFF kinematics are derived from the angular momentum conservation constraint. In particular, this study considers the EMFF as a monolithic system, such as a redundant manipulator, where all satellites are connected via the conservation of angular momentum. Then, the EMFF with more than one RWs in the system is called the ``Redundant EMFF.''

%Therefore, the system is quite similar to the redundant manipulator where all parts are connected
%Specifically, the following is discussed in this section. First, the polynomial representation of the AC method shows its extensibility for generating an arbitrary electromagnetic force and torque. Then, after reviewing the nonholonomic system and defining ``Redundant EMFF,'' the RW condition for realizing smooth state feedback is introduced. Subsequently, the control target of ``Redundant EMFF'' for removing the non-uniform distribution of the RW angular momentum is presented. Finally, the kinematics of EMFF is derived from the null space of the conservation of angular momentum. This kinematics is used in Sec.~\ref{sec3-4} to derive a new control law that can control the electromagnetic force and torque concurrently.

\subsection{Electromagnetic Formation Flight Constraints}
\label{sec2-3-2}
To understand the EMFF structure, EMFF constraints are described using the holonomic and nonholonomic constraints. In particular, properties of nonholonomic constraints for smooth stabilization are presented.

EMFF is a system with the conservation of linear and angular momentum. The values of $\bm{f}_{EMj}$ and $\bm{\tau}_{EMj}$ in Eq.~\eqref{sec2:eq10-2} cannot change the linear momentum, $\bm P$, and angular momentum, $\bm L$, of the system. Hence, using Eqs.~\eqref{sec2:eq9} and \eqref{sec2:eq10}, the relationships shown in
\begin{equation}
  \label{sec2:eq10-3}
     \sum_{j=1}^n \bm f_{EMj}=\frac{^I\mathrm{d}}{\mathrm{d}t}\bm P=\bm 0,\qquad \sum_{j=1}^n\left(\bm r_j \times \bm f_{EMj} + \bm\tau_{EMj}\right)=\frac{^I\mathrm{d}}{\mathrm{d}t}\bm L=\bm 0. 
\end{equation}
are automatically satisfied.
%where $\bm r_j$ is the position vector from the center of mass of the system to the $j$-th satellite. 
Therefore, in the absence of external forces, EMFF holds the linear and angular momentum conservation constraints
\begin{equation}
 \label{sec2:eq11}
    \sum_{i=1}^n m_j \frac{^I\mathrm{d}}{\mathrm{d}t}\bm{r}_j = {\rm const},\qquad  
    \sum_{i=1}^n \bm I_j \cdot \bm\omega_j +m_j \bm r_j\times \frac{^I\mathrm{d}}{\mathrm{d}t}{\bm{r}}_j = {\rm const}.
\end{equation}
Note that the angular momentum of the system includes the angular momentum caused by the motion of each satellite around the center of mass of the system.

In general, the constraints are classified into two types, and these constraints have properties that are relevant to control system design. The first constraints can be expressed as:
\begin{equation}
   \label{sec3:eq2-2}
C(\bm q)=\bm 0,
\end{equation}
which, for the generalized coordinate vector, $\bm q$, are commonly called ``holonomic constraints.'' These can be removed from the dynamics by eliminating specific values of $\bm q$. However, certain states in holonomic systems cannot be reached irrespective of the control method used. In the case of EMFF, the conservation of linear momentum corresponds to the holonomic constraint, and the center of mass of the system cannot always be changed by EMFF control.

All constraints that cannot expressed as holonomic constraints, such as 
\begin{equation}
   \label{sec3:eq2-3}
C(\bm q,\dot{\bm q})=\bm 0,
\end{equation}
and are called ``nonholonomic constraints,'' which includes the conservation of angular momentum \cite{Nakamura}. To remove nonholonomic constraint terms in the dynamics, kinematics are derived from nonholonomic constraints \cite{Fierro,Bloch}. Unlike holonomic constraints, arbitrary states can possibly be reached using specific control inputs depending on the system.

However, nonholonomic systems cannot be asymptotically stabilized to a single equilibrium point via smooth state feedback \cite{Brockett}. In the case of EMFF, all states cannot be asymptotically stabilized at once via smooth state feedback. Control systems with $x$-nonholonomic constraints can be smoothly stabilized only for smooth manifolds of an $x$-dimensional smooth manifold \cite{Bloch}.%This scheme has also been employed in typical space systems, as will be discussed later in Sec.~\ref{sec3-2-2}.}
\subsection{Control Objectives for ``Redundant Electromagnetic Formation Flight'' to Avoid Saturation of RWs}
\label{sec3-2-2}
This subsection clarifies the cause of the non-uniform distribution problem and the appropriate control objectives to avoid this problem.
%to avoid the non-uniform distribution problem while guaranteeing the existence of smooth state feedback.
Previous studies on EMFF \cite{Sakai,Kwon,Porter,Fan,Fabacher,Huang2,Ahsun,Elias,Schweighart,Ramirez-Riberos,Ayyad,Kaneda,Zhang,Sakaguchi,Youngquist,Nurge,Sunny,Abbasi3} guaranteed the existence of smooth state feedback by not considering the non-uniform distribution problem. %from the viewpoint of nonholonomic mechanical systems. 
Finally, EMFF, which uses more than one set of RWs, is defined as ``Redundant EMFF.''
% and proved from the matrix properties in \textbf{Theorem 3} of Sec.~\ref{sec3-4}.} 
%Control schemes of previous studies on EMFF and typical space systems are included as the discussion on appropriate control objectives.
%As described in Sec.~\ref{sec2-3-2}, the relative positions and absolute attitudes of EMFF cannot be asymptotically stabilized via smooth state feedback.
%Therefore, non-smooth stabilization or different control objectives are addressed. In previous studies, it has been reported that specific controls such as smooth time-varying state feedback \cite{Coron} can asymptotically stabilize the control-linear nonholonomic systems, i.e. driftless systems, although these require complex control laws specific to each system. 
%As another control scheme, an example of different control objectives is stabilization to a suitably defined manifold, including an equilibrium manifold \cite{Nakamura,Bloch}. 

Control objectives of typical space systems, which have the conservation of angular momentum as well as EMFF, have been stabilized to an appropriate equilibrium manifold to avoid the non-existence of smooth state feedback. For example, the monolithic satellites of a satellite formation flying system is equipped with attitude actuators, such as RWs, to control $\bm\omega_j$ via smooth state feedback. It should be noted that the RW angular momentum $\bm h_j$ is considered as uncontrolled values and eventually absorbs the excess angular momentum $\bm h_{jd}$ 
%\begin{equation}
%   \label{sec3-2-2:eq1}
% \lim_{t \to \infty} \bm\omega = \bm\omega_d \Rightarrow \lim_{t \to \infty} \bm h =  \bm h_d.
%\bm{N}_e=\{(\bm\omega,\bm h) \mid  \bm\omega=\bm\omega_{d}\} \Rightarrow \lim_{t \to \infty} \bm h =  \bm h_d.
%\end{equation}
%Similarly, for a satellite formation flying system without EMFF, 
%Therefore, each RW can absorb the angular momentum of the system $h_{jd}$ via the same principle as the monolithic satellite, as expressed by Eq.~\eqref{sec3-2-2:eq2}:
\begin{equation}
   \label{sec3-2-2:eq2}
 \lim_{t \to \infty} \bm\omega_j = \bm\omega_{id}\ (j=1,\ldots,n) \Rightarrow \lim_{t \to \infty} \bm h_j =  \bm h_{id}\ (j=1,\ldots,n),
 %\bm{N}_{ej}=\{(\bm\omega_j,\bm h_j) \mid  \bm\omega_j=\bm\omega_{jd}\} \Rightarrow \lim_{t \to \infty} \bm h_j =  \bm h_{jd}.
\end{equation}
because the conservation of angular momentum holds ``in each satellite.''
In other words, typical space systems that have 3-nonholonomic constraints are smoothly stabilized for a 3-dimensional equilibrium manifold 
\begin{equation}
   \label{sec3-2-2:eq2-2}
 \bm{N}_{e}=\{(\bm\omega_j,\bm h_j) \mid  \bm\omega_j=\bm\omega_{jd}\}
\end{equation}
to guarantee the existence of smooth state feedback.

Previous studies of EMFF did not consider the conservation of angular momenta in control design by all $\bm h_j$ as uncontrolled values such as typical space systems \cite{Sakai,Kwon,Porter,Fan,Fabacher,Huang2,Ahsun,Elias,Schweighart,Ramirez-Riberos,Ayyad,Kaneda,Zhang,Sakaguchi} or treating the control objectives as relative attitudes of the satellite instead of absolute attitudes \cite{Youngquist,Nurge,Sunny,Abbasi3}. Unlike typical space systems, EMFF holds the conservation of angular momentum in ``the entire system.'' Therefore, if all $\bm h_j$ are regarded as uncontrolled values, the total value of the RW angular momentum $\sum_{i=0}^n \bm h_j$ is guaranteed to converge the excess angular momentum $L_d$
\begin{equation}
   \label{sec3-2-2:eq3}
\lim_{t \to \infty} \bm\omega_j = \bm\omega_{id}\ (j=1,\ldots,n) \Rightarrow \lim_{t \to \infty} \sum_{i=0}^n \bm h_j =  \bm L_d,
 %\bm{N}_{e}=\{(\bm\omega_j,\bm h_j) (j=1,\ldots,n)\mid  \bm\omega_j=\bm\omega_{jd}\} \Rightarrow \lim_{t \to \infty} \sum_{i=0}^n \bm h_j =  \bm L_d. 
\end{equation} and each value of $\bm h_j$ depends on input history of uncontrolled electromagnetic torque. This is the source of the non-uniform distribution problem in previous EMFF studies. In this case, the control objectives are a $3n$-dimensional equilibrium manifold
\begin{equation}
   \label{sec3-2-2:eq3-2}
%\lim_{t \to \infty} \bm\omega_j = \bm\omega_{id}\ (j=1,\ldots,n) \Rightarrow \lim_{t \to \infty} \sum_{i=0}^n \bm h_j =  \bm L_d. 
 \bm{N}_{e}=\{(\bm\omega_j,\bm h_j) (j=1,\ldots,n)\mid  \bm\omega_j=\bm\omega_{jd}\}. 
\end{equation}

Based on above discussion, considering the RW angular momentum $\bm h_j$ as controlled states except for one RWs is essential to prevent the non-uniform distribution problem. For example, if $m$ satellites (from a total of $n$) have RWs ($1 \leq m\leq n$), $\bm h_1\sim\bm h_{(m-1)}$ are controlled, and only $\bm h_m$ is not controlled. In this case, $\bm h_m$ eventually absorbs the excess angular momentum $\bm h_{md}$ by controlling $\bm\omega_j\ (j=1,\ldots,n)$ and $\bm h_j\ (j=1,\ldots,(m-1))$ using smooth state feedback
\begin{equation}
   \label{sec3-2-2:eq4}
 \lim_{t \to \infty} \bm\omega_j = \bm\omega_{id}\ (j=1,\ldots,n),\qquad \lim_{t \to \infty} \bm h_j = h_{id}\ (j=1,\ldots,(m-1)) \Rightarrow \lim_{t \to \infty} \bm h_m = \bm h_{md},
  %\bm{N}_{e}=\{(\bm\omega_j,\bm h_j) (j=1,\ldots,n)\mid \bm\omega_j = \bm\omega_{id}\ (j=1,\ldots,n), \bm h_j = \bm h_{id}\ (j=1,\ldots,(m-1))\} \Rightarrow \lim_{t \to \infty} \bm h_m = \bm h_{md}.  
\end{equation}
and the appropriate control objectives to prevent the non-uniform distribution problem is implied to be a $3$-dimensional equilibrium manifold
\begin{equation}
   \label{sec3-2-2:eq4-X2}
 %\lim_{t \to \infty} \bm\omega_j = \bm\omega_{id}\ (j=1,\ldots,n),\ \bm h_j = h_{id}\ (j=1,\ldots,(m-1)) \Rightarrow \lim_{t \to \infty} \bm h_m = \bm h_{md}. 
  \bm{N}_{e}=\{(\bm\omega_j,\bm h_j) (j=1,\ldots,n)\mid \bm\omega_j = \bm\omega_{id}\ (j=1,\ldots,n),\qquad \bm h_j = \bm h_{id}\ (j=1,\ldots,(m-1))\}.  
\end{equation}
%On the other hand, these studies have not shown a control scheme to fundamentally prevent the non-uniform distribution of the RW angular momentum.
%The objective of asymptotic stabilization might still be achievable by nonsmooth feedback. However, results in the field of nonsmooth stabilization are quite recent, and systematic procedures to build nonsmooth stabilizing control laws exist only in special cases (e.g. for two-dimensional systems [14]).
%Similar to the monolithic satellite, to converge the state of the EMFF system using smooth state feedback, the number of RWs must equal the number of nonholonomic constraints. EMFF possesses nonholonomic constraints with only three degrees of freedom throughout the system. Therefore, by considering the $n$ satellites of EMFF as one system connected by the conservation of angular momentum, the $n$ satellites of the EMFF can theoretically be controlled only by a single set of 3-axis reaction wheels. 
%\begin{equation}
%\bm{M}_e=\{({q}_{s},{v},\bm\xi_m) \mid  {q}_{s}={q}_{sd},{v}={v}_d,\bm\xi_m=\bm\xi_{md})\}
%\end{equation}

This subsection showed that EMFF can guarantee the existence of smooth state feedback while avoiding the non-uniform distribution problem if EMFF uses at least one set of RWs (the proof is presented in Sec.~\ref{sec3-4}). %Although it is complicated to investigate the exact conditions \cite{Brockett} for realizing smooth state-feedback of the EMFF system (i.e. drift system). 
In contrast, most previous studies \cite{Sakai,Kwon,Porter,Fan,Fabacher,Huang2,Ahsun,Elias,Schweighart,Ramirez-Riberos,Ayyad,Kaneda,Zhang,Sakaguchi} assumed that all satellites use RWs in addition to electromagnetic coils. Then, the EMFF with more than one RWs in the system is called the ``Redundant EMFF.''
\subsection{Extensibility of Alternating Current Method for Outputting Arbitrary Electromagnetic Force and Torque}
\label{sec3-1}
By using the polynomial representation of the AC method, this subsection demonstrates how the AC-based EMFF can generate an arbitrary electromagnetic force and torque for multiple satellites, unlike the DC-based EMFF. The dipole modulation of the AC method using sine and cosine wave currents is derived. 
%\subsection{Non-Existence of Dipole Moments for Simultaneous Control in Direct Current Method}
%In this subsection, As shown in Sec.~\ref{sec2-3}, when there are more than 3 satellites in the EMFF of DC method, simultaneous control of electromagnetic force and torque can not be realized because of the lack of variables. 

To realize simultaneous control of the electromagnetic force and torque, an arbitrary electromagnetic force and torque calculated by the controller are generated. In this case, the ``dipole solution'', i.e., the amplitude value of the AC, can be derived from a nonlinear nonconvex optimization problem using an arbitrary cost function $J$ based on Eq.~\eqref{sec3:eq0}: 
\begin{equation}
  \label{sec3:eq0}
    \left \{
  \begin{aligned}
    &Minimize : J\\ 
  &Subject\ to :{\bm f_{jc}}={\bm f_{(avg)j}}({\bm\mu}),\quad {\bm\tau_{jc}}={\bm\tau_{(avg)j}}({\bm\mu}). 
  \end{aligned}
  \right .
\end{equation}
%As shown in Sec.~\ref{sec2-3}, owing to the lack of variables, there is no ``dipole solution'' to solve the nonlinear EMFF model for the DC method. By contrast, the modulation technique for the AC method described in Sec.~\ref{sec2-4} provides an additional degree of freedom for the control of EMFF. %By adopting modulation technique in Sec.~\ref{sec2-4}, 
Suppose that each satellite presumably generates the sine and cosine waves with angular frequencies $\omega_f$ and no phase difference. In this case, $\bm\mu_j(t)$, $\bm f_{EM(avg)j}$, and $\bm\tau_ {EM(avg)j}$, which are obatined by the system to the $j$-th dipole, are
\begin{equation}
  \label{sec2:eq15}
  \left \{
  \begin{aligned}
        \bm\mu_j(t)&=\bm\mu_{(sin)j}\sin(\omega_{f}t)+\bm\mu_{(cos)j}\cos(\omega_{f}t)\\
        \bm f_{EM(avg)j}&=\frac{1}{2}\sum_{j=1}^{n} \left(\bm{f}(\bm\mu_{(sin)k},\bm\mu_{(sin)j}, \bm r_{jk})+\bm{f}(\bm\mu_{(cos)k},\bm\mu_{(cos)j}, \bm r_{jk})\right)\\
        \bm{\tau}_{EM(avg)j}&=\frac{1}{2}\sum_{j=1}^{n} \left(\bm{\tau}(\bm\mu_{(sin)k},\bm\mu_{(sin)j}, \bm r_{jk})+\bm{\tau}(\bm\mu_{(cos)k},\bm\mu_{(cos)j}, \bm r_{jk})\right)
  \end{aligned} 
  \right.
\end{equation}

%The existence of the dipole moment $\bm\mu$ that gives rise to an only arbitrary $\bm f_{EM}$ in the DC-based EMFF is described.
%Most previous studies use only $\bm f_{EM}$ for relative position control and treat the generated $\tau_{EM}$ as a disturbance torque. 
%This is explained by the relationship between the degree of freedom of the control force and the number of dipole moment variables.

The EMFF outputs can be simplified to a multivariate bilinear system of the polynomial equations. With $\bm\mu_{j}={\{\bm{i}\}^{\mathrm{T}}}{[{^i\mu_{j1}}, {^i\mu_{j2}}, {^i\mu_{j3}}]^{\mathrm{T}}}$ as the variable, the DC-based EMFF outputs of $\bm f_{EMj}={\{\bm{i}\}^{\mathrm{T}}}{[{^if_{EMj1}}, {^if_{EMj2}}, {^if_{EMj3}}]^{\mathrm{T}}}$ and $\bm\tau_{EMj}={\{\bm{i}\}^{\mathrm{T}}}{[{^i\tau_{EMj1}}, {^i\tau_{EMj2}}, {^i\tau_{EMj3}}]^{\mathrm{T}}}$ (see Eq.~\eqref{sec2:eq10-2}) can be described \cite{Schweighart}:
\begin{equation}
  \label{sec2:eq12-2}
     %\begin{aligned}
         %{^if_{EMj1}}\sim {^if_{EMj2}}\sim{^if_{EMj3}}
         {^if_{EMjw}}\sim {^i\tau_{EMjw}}(w=1,2,3)
         =\sum_{k=1}^3\sum_{\substack{l=1\\(l\neq j)}}^{n}\sum_{m=1}^{3}
         {^i\mu_{jk}}C{^i\mu_{lm}},
\end{equation}
where the constant $C$ with different values depends on the state of the system, and the symbol $\sim$ indicates that these constants are represented by the same type of polynomial. After these notations, the AC-based EMFF outputs of $\bm f_{EM(avg)j}$ and $\bm\tau_{EM(avg)j}$ of Eq.~\eqref{sec2:eq15} can also be simplified as 
\begin{equation}
  \label{sec3:eq1}
  \begin{aligned}
%&{^if_{EM(avg)jx}}\sim {^if_{EM(avg)jy}}\sim {^if_{EM(avg)jz}}\\
%\sim&{^i\tau_{EM(avg)jx}}\sim {^i\tau_{EM(avg)jy}}\sim {^i\tau_{EM(avg)jz}}\\
{^if_{EM(avg)jw}}\sim{^i\tau_{EM(avg)jw}}(w=1,2,3)
         =\sum_{k=1}^3\sum_{\substack{l=1\\(l\neq j)}}^{n}\sum_{m=1}^{3}
         ({^i\mu_{(sin)jk}}C{^i\mu_{(sin)lm}}+{^i\mu_{(cos)jk}}C{^i\mu_{(cos)lm})}.
  \end{aligned}
\end{equation}

The relationship between the number of dipole moment variables and the degree of freedom of the control values is useful to investigate the existence of dipole moments in AC-based EMFF for simultaneous control. As shown in Eqs.~\eqref{sec2:eq12-2} and \eqref{sec3:eq1}, the DC-based and AC-based EMFF outputs have $3n$ and $6n$ variables, respectively. By contrast, these EMFF outputs have $6n-6$ degrees of freedom according to EMFF constraints in Eq.~\eqref{sec2:eq11}. %Therefore, the dipole inversion of DC-based EMFF in Eq.~\eqref{sec2:eq11-2} shows the nonlinear programming for $3(n-1)$ equations with $3n$ variables, where that of AC-based EMFF in Eq.~\eqref{sec3:eq0} describes the nonlinear programming of $6(n-1)$ equations for $6n$ variables. due to the lack of variables
Therefore, the dipole inversion of DC-based EMFF in Eq.~\eqref{sec2:eq11-2} have ``dipole solution'' to generate arbitrary electromagnetic forces for $n$ satellites. In this case, simultaneous control can only be realized for a maximum of two satellites. As a result, most previous studies \cite{Sakai,Kwon,Porter,Fan,Fabacher,Huang2,Ahsun,Elias,Schweighart,Ramirez-Riberos,Ayyad,Kaneda,Zhang,Sakaguchi}, which included AC-based EMFF, used only $\bm f_{EM}$ for relative position control and considered the generated $\tau_{EM}$ as a disturbance torque.
%As shown in Sec.~\ref{sec2-3}, owing to the lack of variables, there is no ``dipole solution'' to solve the nonlinear EMFF model for the DC method. By contrast, the modulation technique for the AC method described in Sec.~\ref{sec2-4} provides an additional degree of freedom for the control of EMFF. %By adopting modulation technique in Sec.~\ref{sec2-4}, 
%shows the nonlinear programming for $3(n-1)$ equations with $3n$ variables, where that of AC-based EMFF in Eq.~\eqref{sec3:eq0} describes the nonlinear programming of $6(n-1)$ equations for $6n$ variables.
%which generates an arbitrary electromagnetic force. 
%However, there are only $3n$ variables for a total of $6(n-1)$ degrees of freedom for the independent electromagnetic force and torque. 
Contrarily, simultaneous control in AC-based EMFF can be realized even for formations comprising $n$ satellites because there are more variables than equations. Therefore, AC-based control demonstrates greater versatility than the DC method.
%where each $C$ is a constant whose value depends on the state of the system, with symbol $\sim$ indicating that the constants are represented by the same type of polynomial. %In the invariant orientation frame (I), $\ ^if_{EM(avg)jk}$, $\ ^i\tau_{EM(avg)jk}$, $\ ^i\mu_{(sin)jk}$, and $\ ^i\mu_{(cos)jk}$ ($k$=1,2,3) are the (x, y, z) components of $\bm f_{EM(avg)j}$, $\bm\tau_{EM(avg)j}$, $\bm\mu_{(sin)j}$, and $\bm\mu_{(cos)j}$, respectively. 
%As shown in Eq.~\eqref{sec3:eq1}, $\bm f_{EM(avg)j}$ and $\bm\tau_{EM(avg)j}$ are represented by $6n$ variables of $\bm\mu_{(sin)}$ and $\bm\mu_{(cos)}$.
%On the other hand, $\bm f_{EM(avg)}$ and $\bm\tau_{EM(avg)}$ each have $3(n-1)$ degrees of freedom according to the conservation of linear and angular momentums in Eq.~\eqref{sec2:eq11}.  Therefore, Eq.~\eqref{sec3:eq0} 

Moreover, the arbitrary evaluation function $J$ can be optimized while realizing simultaneous control. In general, the number of solutions of $k$ polynomial equations containing $k$ variables is equal 
to the degree $d$ of the entire polynomial as follows: %\cite{Morgan}:
\begin{equation}
  \label{sec3:eq2}
d=\prod_{l=1}^k d_l,
\end{equation}
where $d_l$ is the maximum degree of the $l$-th polynomial equation. As shown in Eq.~\eqref{sec3:eq1}, the maximum degree, $d_l$, of each polynomial equation of EMFF is 2.
Therefore, Eq.~\eqref{sec3:eq0} has infinite solutions or $2^{(6n-6)}$ solutions with 6 ``free dipoles'' that can be determined arbitrarily.
Nevertheless, it should be noted that these solutions contain ``solutions at infinity'' \cite{Morgan}, 
which cannot be utilized for real applications. Using this property, the following nonlinear nonconvex optimization problem with evaluation function to optimize power consumption while achieving simultaneous control is used:
\begin{equation}
  \label{sec3:eq2-1-2}
   \left \{
  \begin{aligned}
    &Minimize : J ={^i\mu_{(sin)}^{\mathrm{T}}}{^i\mu_{(sin)}}+{^i\mu_{(cos)}^{\mathrm{T}}}{^i\mu_{(cos)}}\\ 
  &Subject\ to : {^if_{jc}}={^if_{(avg)j}},\quad {^i\tau_{jc}} ={^i\tau_{(avg)j}}, 
  \end{aligned}
  \right ..
\end{equation}

%\begin{equation}
%  \label{eq:4-2}
%  \begin{aligned}
%         &f_{jx}\sim f_{jy}\sim f_{jz}\sim\tau_{jx}\sim\tau_{jy}\sim\tau_{jz}\\
%         &=\begin{bmatrix}
%  \mu_{1}&\cdots&\mu_{k}
%\end{bmatrix}
%\begin{bmatrix}
%C&&0\\&\ddots&\\0&&C
%\end{bmatrix}
%\begin{bmatrix}
%  \mu_4\\\vdots\\\mu_{3n}
%\end{bmatrix}+\cdots+
%  \end{aligned} 
%\end{equation}
\subsection{Electromagnetic Formation Flight Kinematics Based on Nonholonomic Constraints}
\label{sec3-3}
In this subsection, the kinematics of EMFF is derived using the nonholonomic constraint of the EMFF system to develop the control design (see Sec.~\ref{sec3-4}). The term kinematics refers to the tangent space of manifold satisfying the conservation of angular momentum. %to achieve simultaneous control of the electromagnetic force and torque. 
In addition, a new EMFF state is defined for ``Redundant EMFF.''
%It is shown that the EMFF using AC method can generate the arbitrary electromagnetic force and torque calculated by the controller in Sec.~\ref{sec3-1}. However, the electromagnetic force and torque must have values that satisfy the conservation of linear and angular momentum of the system because the EMFF system can only output internal force. 
%To achieve simultaneous control of the electromagnetic force and torque, this subsection derive the control direction to achieve convergence to the control target while satisfying the constraints of the system. 
%As discussed in Sec.~\ref{sec3-2}, there are inputs that reach arbitrary states 
%while satisfying the the conservation of linear and angular momentum of the system. 
%The conservation of linear momentum $\bm P$ in the system is expressed as: 
%\begin{equation}
%   \label{sec3:eq2-2}
%    \sum_{i=1}^n m_j \frac{^I\mathrm{d}}{\mathrm{d}t}{\bm r}_j = \bm P.
%\end{equation} 
Note that, in this study, the linear momentum constraint is not implemented by eliminating $\bm r_1$ from the position vector $\bm r$ as follows:
\begin{equation}
   \label{sec3:eq2-3-2}
    ^ir=
    \begin{bmatrix}
   {^i{r}_2^{\mathrm{T}}},
   \cdots,
   {^i{r}_{n}^{\mathrm{T}}}
    \end{bmatrix}^{\mathrm{T}}\in \mathbb{R}^{3n-3}.
\end{equation}
%According to the conservation of linear momentum, $\bm r_1$ can be eliminated from the position vector $\bm r$, as shown in Eq.~\eqref{sec3:eq2-3-2}, i.e.: 

The kinematics of EMFF are derived using the conservation of angular momentum.
Assuming that $m$ satellites (from a total of $n$) are equipped with RWs $(1 \leq m \leq n)$, the angular momentum $\bm L$ are expressed as:
\begin{equation}
   \label{sec3:eq2-3}
   \begin{aligned}
   \sum_{j=1}^n \left( m_j(\bm r_{j}-\bm r_{1})\times \frac{^I\mathrm{d}}{\mathrm{d}t}{\bm r}_j + \bm I_j \cdot \bm\omega_j\right) + \sum_{j=1}^m\bm h_j = \bm L.\\
   %\sum_{j=1}^n \left( m_j(\bm r_{jd}-\bm r_{1d})\times \frac{^I\mathrm{d}}{\mathrm{d}t}{\bm r}_{jd} + \bm I_{jd} \cdot \bm\omega_{jd}\right) = \bm L_d, 
   \end{aligned}
\end{equation} 
%The orbital motion and the attitude motion can be separated by considering 
%the angular momentum around the center mass of the foramtion.
%Now, assuming $\bm L_d$ is the target angular momentum,
%the vector representation of $\bm L_d$ can be expressed as :
%\begin{equation}
%   \label{sec3:eq10-2}
%   %\begin{aligned}
%   \bm L_d  = \sum_{j=1}^n \left( m_j(\bm r_{jd}-\bm r_{1d})\times \frac{^Id}{dt}{\bm r}_{jd} + \bm I_j \cdot \bm\omega_{jd}\right)
%\end{equation}
%ce the RWs of each satellite has the difference 
%between the angular momentum $\bm L$ and the target angular momentum $\bm L_d$ at target state, 
Then, the components of $\bm L$ in the invariant orientation frame (I) can be expressed as:
\begin{equation}
   \label{sec3:eq3}
   \begin{aligned}
    %&\sum_{j=1}^n \left( m_j\left(\bm r_j-\bm r_1\right)\times \frac{^Id}{dt}{\bm r}_j + \bm I_j \cdot \bm\omega_j\right) + \sum_{j=1}^m\bm h_j = \bm L\\
        %\Leftrightarrow   
        %&\sum_{j=1}^n  \left(m_j\left(^i\tilde{r}_j-^i\tilde{r}_1\right)\ ^i\dot{r}_j + C^{I/B_j}J_j\ ^{b_j}\omega_j\right)+ \sum_{j=1}^m \left(C^{I/B_j}\left(\ ^{b_j}h_j -\ ^{b_j}h_{jd}\right) \right) = ^iL_d \\
       %\Leftrightarrow 
       \sum_{j=1}^n  \left(m_j\left(^i\tilde{r}_j-^i\tilde{r}_1\right){^i\dot{r}_j}+ C^{I/B_j}J_j{^{b_j}\omega_j}\right)+ \sum_{j=1}^m \left(C^{I/B_j}\left({^{b_j}h_j}-\frac{1}{m}{^{b_j}L}\right) \right) = 0 \\
       %  \Leftrightarrow &\{\mathbb{I}\}^{\mathrm{T}} \left(\sum_{j=1}^n  \left(m_jC^{I/O}(^i\tilde{r}_j-\ ^i\tilde{r}_1)^ir_j + C^{I/B_j}J_j\ ^{b_j}\omega_j
    % + C^{I/B_i}(\ ^{b_i}h_i -\frac{1}{n}\ ^{b_i}L(t)) \right) =0\right)\\
    \end{aligned}.
\end{equation}
Let $^b\xi=\left[
   {^{b_1}\xi_1^{\mathrm{T}}},
   \cdots,
   {^{b_m}\xi_{m}^{\mathrm{T}}}
    \right]^{\mathrm{T}}\in \mathbb{R}^{3m}$ be defined as new variables related to $\bm h_{j}$ and $\bm L$, as expressed in Eq.~\eqref{sec3:eq4}. %These variables indicate the error from $\bm h_{jd}$ when $\bm L_d$ = $\bm 0$.
\begin{equation}
   \label{sec3:eq4}
   \begin{aligned}
   ^b\xi
    &=
    \left[
   \left({^{b_1}h_1}-\frac{1}{m}{^{b_1}L}\right)^{\mathrm{T}},
   \cdots,
   \left({^{b_m}h_m}-\frac{1}{m}{^{b_m}L}\right)^{\mathrm{T}}
    \right]^{\mathrm{T}}.
\end{aligned}
\end{equation}
As mentioned in Sec.~\ref{sec3-2-2}, the RW angular momentum is considered to be a controlled state in this study to overcome the non-uniform distribution problem. Therefore, the new EMFF states $\zeta$ are defined in Eq.~\eqref{sec3:eq4-2}:
\begin{equation}
   \label{sec3:eq4-2}
\zeta=
    \left[
   ^i\dot{r}^{\mathrm{T}},\ 
   ^b\omega^{\mathrm{T}},\ 
   ^b\xi^{\mathrm{T}}
    \right]^{\mathrm{T}}\in \mathbb{R}^{6n+3m-3}.
\end{equation}
However, not all states can be controlled from the viewpoint of a nonholonomic system. As explained in Sec.~\ref{sec3-2-2}, EMFF requires uncontrolled RWs with the same degree of freedom as the angular momentum of the system. Let $\bm\xi_m$ define the uncontrolled states. In this case, the controlled states $v$ of the EMFF state $\zeta$ are defined as
\begin{equation}
   \label{sec3:eq6-2}
\begin{aligned}
   v
    &=
    \begin{bmatrix}
   {^i\dot{r}^{\mathrm{T}}},
   {^{b}\omega^{\mathrm{T}}},
   {^{b_1}\xi_1^{\mathrm{T}}},
   \cdots,
   {^{b_{m-1}}\xi_{m-1}^{\mathrm{T}}}
    \end{bmatrix}^{\mathrm{T}}\in \mathbb{R}^{6n+3m-6}\\
     %\zeta&=Sv.
\end{aligned}
\end{equation}
Now, the kinematics of EMFF satisfying the constraints of the system can be derived using the relationship between $\zeta$ and $v$. By using the new EMFF states $\zeta$, the angular momentum of the system (see Eq.~\eqref{sec3:eq3}) is expressed as:
\begin{equation}
    \label{sec3:eq5}
     \left \{
   \begin{aligned}
    A\zeta=&0\\
   A=&\left [
   m_2{^i\tilde{r}_{21}}\right . ,\cdots,m_n{^i\tilde{r}_{n1}},C^{I/B_1}J_1,\cdots,C^{I/B_{n}}J_n,\\
   &C^{I/B_{1}},\cdots,\left . C^{I/B_{m}} \right ]\in \mathbb{R}^{3\times(6n+3m-3)}
    \end{aligned}
    \right.
  \end{equation}  
%ここで行列$A$の零空間として滑らかで線形独立なベクトル場のフルランク行列$S(q)$を定義する i.e.,
Let $S\in \mathbb{R}^{(6n+3m-3)\times(6n+3m-6)}$ be defined as a smooth and linearly independent vector field 
full rank matrix corresponding to the null space of matrix $A$, i.e.:
\begin{equation}
   \label{sec3:eq6}
   \left \{
   \begin{aligned}
  S=&
 \begin{bmatrix}
      E_{(6n+3m-6)} \\
      -C^{B_{m}/I}A_s
    \end{bmatrix}\in Null\ Space(A)\\
   A_s=&\left [
   m_2{^i\tilde{r}_{21}}\right . ,\cdots,m_n{^i\tilde{r}_{n1}},C^{I/B_1}J_1,\cdots,C^{I/B_{n}}J_n,\\
   &C^{I/B_{1}},\cdots,\left . C^{I/B_{m-1}} \right ] \in \mathbb{R}^{3\times(6n+3m-6)}
    \end{aligned}
    \right.
  \end{equation}  
%where $S$ denotes the direction in which the angular momentum of the system is not changed. 
where $S$ denotes tangent space of the manifold in which the angular momentum of the system does not change. In these definitions, the relationship between $\zeta$ and $v$,
\begin{equation}
    \zeta=Sv\in \mathbb{R}^{6n+3m-3},
\end{equation}
is established. Using this relationship and the matrix $[Z]\in \mathbb{R}^{3n\times 3n}$ of MRP’s kinematic differential equation (see Eq.~\eqref{sec2:eq3}), the kinematics of the EMFF system are given by
\begin{equation}
  \label{sec3:eq7}
  \left \{
  \begin{aligned}
    \dot{q}_s
    &=
    \begin{bmatrix}
    \hat{[Z_s]}(\sigma)&0_{(6n+3m-6)\times 3}\\
  \end{bmatrix}
   \zeta=\hat{[Z_s]}(\sigma)v\\
  \hat{[Z_s]}(\sigma)&=
  \begin{bmatrix}
    E_{3n-3}&0&0\\
    0&[Z](\sigma)&0\\
    0&0&0_{(3m-3)\times (3m-3)}\\
  \end{bmatrix}
   \end{aligned}
  \right .,
\end{equation}
where ${q}_s=
    \begin{bmatrix}
   {^i{r}^{\mathrm{T}}},
   {\sigma}^{\mathrm{T}},
   0_{3m-3}^{\mathrm{T}}%,
   %{^{b}\xi^{\mathrm{T}}}
    \end{bmatrix}^{\mathrm{T}}\in \mathbb{R}^{6n+3m-6}$ is the subvector of the generalized coordinates $q$.

In summary, the newly defined states $\dot{q}_s\in \mathbb{R}^{6n+3m-6},\ \zeta\in \mathbb{R}^{6n+3m-3},$ and $\ {^{b_{m}}\xi_{m}}\in \mathbb{R}^3$ for the ``Redundant EMFF,'' and their kinematics are listed in Table~\ref{sec3:tableX-1} using the controlled states $v=
    \begin{bmatrix}
   {^i\dot{r}^{\mathrm{T}}},
   {^{b}\omega^{\mathrm{T}}},
   {^{b_1}\xi_1^{\mathrm{T}}},
   \cdots,
   {^{b_{m-1}}\xi_{m-1}^{\mathrm{T}}}
    \end{bmatrix}^{\mathrm{T}}\in \mathbb{R}^{6n+3m-6}$.
\begin{table}[bt!]
\caption{\label{sec3:tableX-1}Summary of states definition and kinematics.}
\centering
\begin{tabular}{ccc}
\hline\hline
States&Definitions&Kinematics\\\hline%& Attitude$[\sigma_1,\sigma_2,\sigma_3]$\\\hline
$\dot{q}_s$&{${\equiv \begin{bmatrix}
   {^i\dot{r}^{\mathrm{T}}},
   \dot{\sigma}^{\mathrm{T}},
   0_{3m-3}^{\mathrm{T}}%,
   %{^{b}\xi^{\mathrm{T}}}
    \end{bmatrix}^{\mathrm{T}}}$}
    &$=\hat{[Z_s]}v$\\
$\zeta$&$\equiv 
    \left[
   ^i\dot{r}^{\mathrm{T}},\ 
   ^b\omega^{\mathrm{T}},\ 
   ^b\xi^{\mathrm{T}}
    \right]^{\mathrm{T}}$
    &$=Sv$\\
${^{b_{m}}\xi_{m}}$&$\equiv \left({^{b_m}h_m} -\frac{1}{m}{^{b_m}L}\right)^{\mathrm{T}}$&$=-C^{B_m/I}A_sv$\\
\hline\hline
\end{tabular}
\end{table}
\subsection{Averaged Dynamics of Redundant Electromagnetic Formation Flight}
This subsection describes the averaged dynamics of ``Redundant EMFF'' that averaged by dipole modulation of Section~\ref{sec2-4} for AC-based controller design in Sec.~\ref{sec3-4}. Note that the stability of the averaged system into an equilibrium point indicates the stability of the original system not averaged by dipole modulation into the corresponding periodic orbit \cite{Sanders}. For these errors, the AC frequency is set appropriately to achieve the control performance determined in the operational plan \cite{Sakai,Kaneda}.

The equation of motion describing the averaged EMFF system is
\begin{equation}
  \label{sec3:eq8}
  %\left \{
  \ [M]\dot{\zeta}+
    [C]\zeta%+
    %[G]
  =
    [B]u_c+
    u_d
    -A^{\mathrm{T}}\eta
  %\ ^b\dot{h}&=\ ^{b}\tau_{RW}\\
  %\frac{^{I}d}{dt}L&=Ru_{d}
   % \right,
\end{equation}
by combining the relative translational dynamics in Eq.~\eqref{sec2:eq1} with the attitude dynamics of an RW-equipped rigid spacecraft in Eq.~\eqref{sec2:eq5}.
%Section 2で導出したSatellite Formation FlyingのTranslational and attitude equationに基づいて
%EMFFシステムの運動方程式とシステム全体の角運動量$L$はシステムの状態量$\zeta$と
%行列$R\in \mathbb{R}^{3\times (6n+3m-3)}$を用いてEq.~\eqref{sec3:eq9}で表される．
In Eq.~\eqref{sec3:eq8},
$
u_c=
\begin{bmatrix}
   {^if_c^{\mathrm{T}}},
   {^{b}\tau_c^{\mathrm{T}}},
   {^{b}\dot{h}^{\mathrm{T}}}
    \end{bmatrix}^{\mathrm{T}}
    \in \mathbb{R}^{6n+3m-3}$ is the control input, $u_{d}=
    \begin{bmatrix}
      {^if_{g}^{\mathrm{T}}},
   {^{b}\tau_{g}^{\mathrm{T}}},
   -\frac{1}{m}{^{b}\dot{L}}
\end{bmatrix}^{\mathrm{T}}\in \mathbb{R}^{6n+3m-3}$ is an external force input, and $-A^{\mathrm{T}}\eta$ is the constraint term with $\eta\in \mathbb{R}^{3}$ that represents the vector of constraint forces. $^b\dot{L}$ is the time-differentiated angular momentum vector of the system in the body-fixed frame of each satellite: 
\begin{equation}
  \label{sec3:eq11-4}
  \begin{aligned}
    {^b\dot{L}}&=\left[
   {^{b_1}\dot{L}^{\mathrm{T}}},
   \cdots,
   {^{b_m}\dot{L}^{\mathrm{T}}}
    \right]^{\mathrm{T}}\in \mathbb{R}^{3m}\\
    %&=\left[(C^{B_1/I}\ ^{i}\dot{L}-\ ^{b_1}\tilde{\omega}_1C^{B_1/I}\ ^{i}{L})^{\mathrm{T}},\cdots,(C^{B_m/I}\ ^{i}\dot{L}-\ ^{b_m}\tilde{\omega}_mC^{B_m/I}\ ^{i}{L})^{\mathrm{T}}\right]^{\mathrm{T}}\\
    &=\left[
   C^{B_1/I}({^{i}\dot{L}}-{^{i}\tilde{\omega}_1}{^{i}{L}})^{\mathrm{T}},
   \cdots,
   C^{B_m/I}({^{i}\dot{L}}-{^{i}\tilde{\omega}_m}{^{i}{L}})^{\mathrm{T}}\right]^{\mathrm{T}}\\
  \end{aligned}
  \end{equation}
where ${^i\dot{L}}$ is the component of time-differentiated $\bm L$ in the frame (I) as follows:
\begin{equation}
  \label{sec3:eq11-4-2}
  \begin{aligned}
    {^{i}\dot{L}}&=R(u_c+u_d)\in \mathbb{R}^{3},\\
%R=[(^i\tilde{r}_2-^i\tilde{r}_1),\cdots,(^i\tilde{r}_{n}-^i\tilde{r}_1),C^{I/B_1},\cdots,C^{I/B_{n}},0_1,\cdots,0_m]\\
 R&=\left [{^i\tilde{r}_{21}}\right . ,\cdots,{^i\tilde{r}_{n1}},
   \ C^{I/B_1},\cdots,C^{I/B_{n}}, \left . 0_{3 \times 3m} \right ]\in\mathbb{R}^{3\times(6n+3m-3)}. 
  \end{aligned}
  \end{equation}
  %$R(q)\in\mathbb{R}^{3\times(6n+3m-3)}$ represents the following matrix: 
%It should be noted that $u_d$ is replaced by the constraint term $-A^{\mathrm{T}}\eta$ when there is no external force $u_d$, with vector $\eta\in \mathbb{R}^{3}$ representing the vector of constraint forces.
%Matrices $[M]\in \mathbb{R}^{(6n+3m-3)\times (6n+3m-3)}$, $[C]\in \mathbb{R}^{(6n+3m-3)\times (6n+3m-3)}$, and $[B]\in \mathbb{R}^{(6n+3m-3)\times (6n+3m-3)}$ are expressed as:
The $(6n+3m-3)\times (6n+3m-3)$ matrices $[M]$, $[C]$, and $[B]$ are expressed as:
\begin{equation}
  \label{sec3:eq9}
  \begin{aligned}
  \ [M]=
  \begin{bmatrix}
    [M_p]&&0\\
    &[M_a]&\\
    0&&E_{3m}
  \end{bmatrix},\quad
  [C]=
  \begin{bmatrix}
    0_{3n-3}&&0\\
    &[C_a]&\\
    0&&0_{3m}\\
  \end{bmatrix},\\
%[G]=0_{(6n+3m-3)}\,\ 
  [B]=
  \begin{bmatrix}
      E_{3n-3}&&0\\
      &E_{3n}&-E_{3n\times3m}\\
      0&&E_{3m}
    \end{bmatrix}\in \mathbb{R}^{(6n+3m-3)\times (6n+3m-3)},
      \end{aligned}
\end{equation}
where $[M]$ and $[B]$ has full rank. As mentioned in Secs.~\ref{sec3-3}, the position of the first satellite $\bm r_1$ is removed from the generalized coordinates.
%because the relative motion of the satellites 
%with respect to the center mass of the EMFF system 
%satisfies the conservation of linear momentum.
Therefore, the mass matrices $[M_p]\in \mathbb{R}^{(3n-3)\times (3n-3)}$ of all satellites except the
first satellite can be expressed as:
  \begin{equation}
  \label{sec2:eq2}
  \ [M_p]=
  \begin{bmatrix}
    m_2E_3&&0\\
    &\ddots&\\
    0&&m_nE_3
  \end{bmatrix}\in \mathbb{R}^{(3n-3)\times (3n-3)}.
\end{equation}
Using Eq.~\eqref{sec2:eq5}, the matrices $[M_a]\in \mathbb{R}^{3n\times 3n}$ and $[C_a]\in \mathbb{R}^{3n\times 3n}$ are defined with respect to the attitude motion 
of the entire system.
  \begin{equation}
  \label{sec2:eq5-2}
  \begin{aligned}
  \ [M_a]&=
  \begin{bmatrix}
    J_1&&0\\
    &\ddots&\\
    0&&J_n
  \end{bmatrix},\quad 
 \ [C_a]&=
  \begin{bmatrix}
    -(J_1{^{b_1}\omega_1}+{^{b_1}h_1})\ \bm{\tilde{}}&&0\\
    &\ddots&\\
    0&&-(J_n{^{b_n}\omega_n}+{^{b_n}h_n})\ \bm{\tilde{}}\ \ 
  \end{bmatrix}%,\ 
  %[G_a]=0_{3n\times 1}
  \end{aligned}\in \mathbb{R}^{3n\times 3n}.
\end{equation}
%the equation of motion of the EMFF system is derived 
%using the kinematics of Eq.~\eqref{sec3:eq7}. 
Substituting $\zeta=Sv$ and $\dot{\zeta}=\dot{S}v+S\dot{v}$ 
into Eq.~\eqref{sec3:eq8} and multiplying by the matrix $S^{\mathrm{T}}$ from right yields:
\begin{equation}
\label{sec3:eq10}
\begin{aligned}
    \overline{M}(q_s)\dot{v}+\overline{C}(q_s,v)v%+\overline{G}
    &=\overline{B}(q_s)u_c+\overline{u}_d\\
\end{aligned}
\end{equation}
where $\overline{M}(q_s)=S^{\mathrm{T}}[M]S\in \mathbb{R}^{(6n+3m-6)\times (6n+3m-6)}$ is a symmetric and positive definite matrix, $\overline{C}(q_s,v)=S^{\mathrm{T}}([M]\dot{S}+[C]S)\in \mathbb{R}^{(6n+3m-6)\times (6n+3m-6)}$, $\overline{B}(q_s)=S^{\mathrm{T}}
    [B]\in \mathbb{R}^{(6n+3m-6)\times (6n+3m-3)}$, and $\overline{u}_d=S^{\mathrm{T}}u_d\in \mathbb{R}^{6n+3m-6}$.
 %$\overline{G}=S^{\mathrm{T}}[G]$ 
It should be noted that $\dot{\overline{M}}-2\overline{C}$ is skew symmetric.
%In this case, assuming the absence of external forces, the constraint force term $-A^{\mathrm{T}}\eta$ disappears, and the constraint force owing to the conservation of angular momentum is neglected.
\section{Kinematics Controller Design for Simultaneous Control}
%\label{sec3-2}
%\subsection{Kinematics Control Formulation for Simultaneous Control of Electromagnetic Force and Torque}
\label{sec3-4}
To overcome the non-uniform distribution of the RW angular momentum, a new control law enabling the simultaneous control of the electromagnetic force and torque is introduced. %This results in the long-term stability of the $n$ satellites' formation maintenance. 
First, the properties of the proposed control law and the definition of the target states are summarized. Then, the EMFF kinematics described in Sec.~\ref{sec3-3} derive a smooth feedback control law that satisfies the conservation of angular momentum. The Lyapunov stability theory \cite{Slotine} is used to show that the state of the averaged system asymptotically stabilizes at the control target. Subsequently, this section demonstrates that EMFF requires only one set of RWs as an additional attitude actuator to control EMFF for $n$ satellites. 
%\subsection{Reaction Wheel Conditions for Smooth Feedback Stabilizability}
%\label{sec3-2}

\begin{figure}[hbt!]
  \begin{minipage}{0.5\hsize}
    \centering
    \includegraphics[width=1\textwidth]{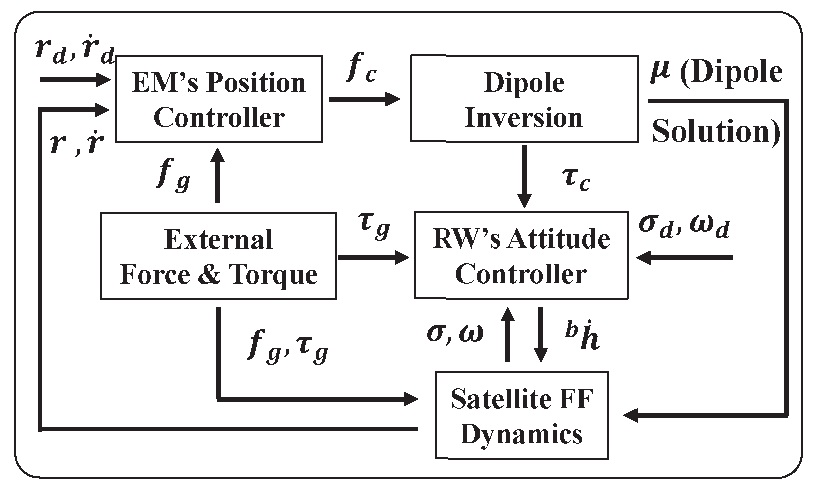}
    \subcaption{Conventional system diagram.}
    \label{sec3:fig:rte_1}
  \end{minipage}
  \begin{minipage}{0.5\hsize}
    \centering
    \includegraphics[width=1\textwidth]{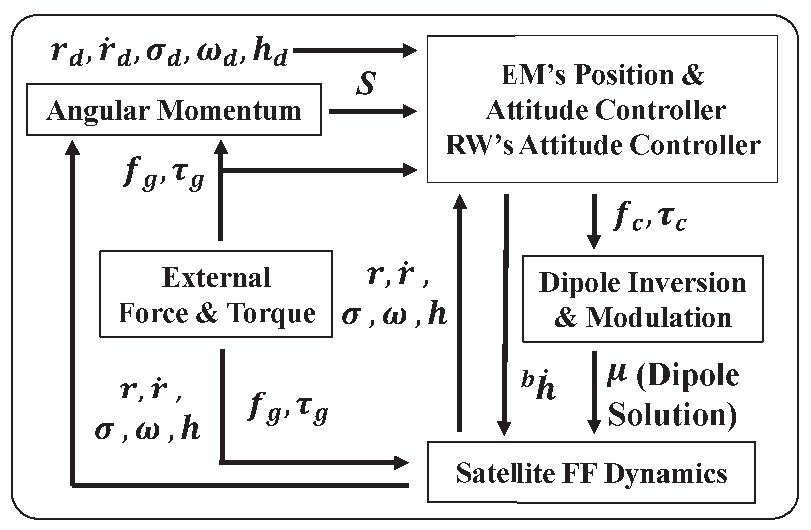}
    \subcaption{Proposed system diagram.}
    \label{sec3:fig:rte_2}
  \end{minipage}\\
  \caption{Electromagnetic formation flight system diagram.}
   \label{sec3:fig1-0}
\end{figure}

Figure ~\ref{sec3:fig1-0} shows the conventional system diagram of Fig.~\ref{sec3:fig:rte_1} \cite{Sakai,Kwon,Porter,Fan,Huang2,Ahsun,Elias,Schweighart,Ramirez-Riberos,Ayyad,Kaneda,Zhang,Sakaguchi,Fabacher} and the proposed system diagram of Fig.~\ref{sec3:fig:rte_2}.%$S$ is used for the controller design.
%In contrast to previous EMFF studies \cite{Sakai,Kwon,Porter,Youngquist,Nurge,Sunny,Fan,Huang2,Ahsun,Elias,Schweighart,Ramirez-Riberos,Ayyad,Kaneda,Abbasi,Abbasi2,Zhang}, it is shown that the proposed kinematics control law theoretically eliminates the non-uniform distribution of RW angular momentum among each satellite under conditions where external force exist. It is shown that the proposed kinematics control law can control more than three satellites without all the satellites having RWs. It is shown that the combination of the proposed kinematics control law and restricting the simple unloading control to the chief satellite mitigates the accumulation of angular momentum within the entire system. In contrast to previous reports, \cite{Zhang}, this unloading control does not require a complicated algorithm for angular momentum management of EMFF.
%In this subsection, the equation of motion of the EMFF system 
%without considering the conservation of angular momentum is derived 
%and the control law is designed to guarantee that the system converges asymptotically to the target manifold.
%In this subsection, the kinematic controller design for the concurrent control of the electromagnetic force and torque is discussed. 
The derived EMFF kinematics include a matrix $S$ that corresponds to the tangent space of the manifold in which the angular momentum of the system remains constant. As shown in Fig.~\ref{sec3:fig:rte_2}, the control law corresponding to a block of ``EM's Position \& Attitude Controller, RW's Attitude Controller'' uses the target error of the RW angular momentum of each satellite. Then, the command values of the electromagnetic force $\bm f_c$ and torque $\bm\tau_c$ that the EMFF system can output as well as the RW torques $^b\dot{\bm h}$ are calculated. Theorem 1 describes the characteristics of control laws that allow EMFF to achieve simultaneous control.

\textbf{Theorem 1}: \textit{The control law $u_c$=
$\begin{bmatrix}
   {^if_c^{\mathrm{T}}},
   {^{b}\tau_c^{\mathrm{T}}},
   {^{b}\dot{h}^{\mathrm{T}}}
    \end{bmatrix}^{\mathrm{T}}$ that is multiplied by $\left([B]^{-1}[M]S\right)$ from the left and contains an arbitrary vector $X \in \mathbb{R}^{6n+3m-6}$, such as $u_c=[B]^{-1}[M]SX$, will not change the angular momentum of the whole system, and can be output by the EMFF system.}

%the control force given by the control law in Eq.~\eqref{sec3:eq11} will not change the angular momentum of the system $\bm L$;
%Term that guarantees simultaneous control
\textbf{Proof of Theorem 1}: See Appendix A.\\

For simplicity of the following control system design, 
let $\delta{q}_s$ and $\delta\zeta$ be the error variables of $q_{s}$ and EMFF states $\zeta$ from the target value, as follows:
\begin{equation}
\label{sec3:eq10-1-2}
\left \{
   \begin{aligned}
    \delta{q}_s&=
    \begin{bmatrix}
   {^i{r}}-{^i{r}_d}\\
   \delta{\sigma}\\
   0_{3m-3}
    \end{bmatrix},\qquad \delta\zeta=\begin{bmatrix}
   ^i\delta\dot{r}\\ 
   ^b\delta\omega\\
   ^b\delta\xi
\end{bmatrix}=\begin{bmatrix}
   \delta v\\
   {^{b_m}\delta\xi_{m}}
\end{bmatrix}
    %T_1(\sigma)=\left [\hat{[Z_s]}(\sigma)\right.&, \ \left.0_{(6n-3)\times (3m-3)}\right ] \in \mathbb{R}^{(6n-3)\times(6n+3m-6)}.
   \end{aligned}
    \right..
\end{equation}
When the system has angular momentum other than RW angular momentum in the target state, such as nadir-pointed formations, the target angular momentum $\bm L_d$ is
\begin{equation}
   \label{sec3:eq2-Y3}
   \begin{aligned}
   \sum_{j=1}^n \left( m_j(\bm r_{jd}-\bm r_{1d})\times \frac{^I\mathrm{d}}{\mathrm{d}t}{\bm r}_{jd} + \bm I_{jd} \cdot \bm\omega_{jd}\right) = \bm L_d, 
   \end{aligned}
\end{equation} 
where $\bm r_{jd}$, $\bm\omega_{jd}$, and $\bm I_{jd}$ are the target relative position vectors, angular velocity vectors, and inertial dyadic of the $j$-th satellite, respectively. Based on the definition of $\bm L$ provided by Eq.~\eqref{sec3:eq2-3}, the total of target RW angular momentum $\bm h_{jd}$ and $\bm \xi_{jd}$ are
\begin{equation}
   \label{sec3:eq10-3}
  \sum_{j=1}^m \bm h_{jd}= \bm L -\bm L_d,\qquad \sum_{j=1}^m \bm \xi_{jd}= -\bm L_d.%,\ \sum_{j=1}^n \bm\eta_{j}= \bm L.
\end{equation}
 Now, to avoid the non-uniform distribution of RW angular momentum, the target values of $\bm h_j$ and $\bm \xi_j$ (see Eq.~\eqref{sec3:eq4}) are set as 
 \begin{equation}
   \label{sec3:eq4-2-1}
   \begin{aligned}
   ^{b_j}h_{jd}=\frac{1}{m}\left({^{b_j}L}-{^{b_j}L_d}\right),\qquad {^{b_j}\xi_{jd}}=-\frac{1}{m}{^{b_j}L_d},
   \end{aligned}
\end{equation}
by making $\bm h_{jd}$ equivalent. Note that $\bm h_{jd}$ can be non-zero values when the satellite has spin stiffness to ensure robustness and can be periodic when the satellite performs Earth-pointed formations without unloading control.%In this case, $^b\xi$ verify the non-uniform distribution of $\bm h_{j}$.
%when the purpose of control is to eliminate the non-uniform distribution of the RW angular momentum 
%In this case, $\bm\xi_j$, $\bm h_{jd}$ and the target state $\bm\xi_{jd}$ of $\bm\xi_j$ in (B$_j$) are as follows:

Finally, this subsection describes the derivation of a kinematics control law for ``Redundant EMFF'' that can stabilize the relative position, absolute attitude, and RW-loaded angular momentum of $n$ satellites to a constant or time-varying state without the non-uniform distribution and accumulation of RW angular momentum. This control law controls the electromagnetic force and torque simultaneously and needs only one set of 3-axis reaction wheels for all satellites. Combining the results of Sec.~\ref{sec3}, Theorem 1, and the above target states result in Theorem 2, which summarizes the main results of this study.

%\textit{
  \textbf{Theorem 2}: \textit{Given $n$ satellites with averaged EMFF kinematics and dynamics 
 \begin{equation}
\label{sec3:eq10-X}
\left\{
\begin{aligned}
  \delta\dot{q}_{s}&= \hat{[Z_s]}(\delta\sigma)\delta v\\
    \overline{M}(q_s)\dot{v}+\overline{C}(q_s,v)v%+\overline{G}
    &=\overline{B}(q_s)u_c+\overline{u}_d\\
    ^{b_m}\xi_m&=-C^{B_m/I}A_s(q_s)v
\end{aligned}
\right .
\end{equation} based on Eqs.~\eqref{sec3:eq7}, \eqref{sec3:eq10}, and \eqref{sec3:eq10-1-2}; thus, the following control law
    \begin{equation}
  \label{sec3:eq11}
  \begin{aligned}
    u_c=
\begin{bmatrix}
   {^if_c^{\mathrm{T}}},
   {^{b}\tau_c^{\mathrm{T}}},
   {^{b}\dot{h}^{\mathrm{T}}}
    \end{bmatrix}^{\mathrm{T}}
  =\overline{B}_r^{-1}
    \left(-K_1\delta q_s
    -K_2\delta v-\overline{u}_d+\overline{M}(q_s)\dot{v}_d+\overline{C}(q_s,v)v_d\right),
  \end{aligned}
\end{equation}
%Term that guarantees simultaneous control
can be output by EMFF and is applied to the system of Eq.~\eqref{sec3:eq10-X}, where $\overline{B}_r^{-1}=[B]^{-1}[M]S(S^{\mathrm{T}}[M]S)^{-1}$ is a smooth right inverse matrix of $\overline{B}$, and $K_1$ and $K_2$ are positive definite gain matrices. Then, the state of the system converges to a three-dimensional equilibrium manifold
%\begin{equation}
%  \label{sec3:eq11-2}
% \lim_{t \to \infty} v=v_d,
%\end{equation}
%\begin{equation}
%  \label{sec3:eq11-3}
%  \lim_{t \to \infty}q_s=q_{sd}.
%\end{equation}
\begin{equation}
  \label{sec3:eq11-X1}
%\bm{N}_e=\{({q}_{s},{v},{^{b_m}\xi_m}) \mid  T_1^{\mathrm{T}}K_1(q_s-q_{sd})=0, v=v_d\})\},
\bm{M}=\left\{(\delta {q}_{s},\delta {v},{^{b_m}\delta\xi_m}) \mid \delta q_{s}=0,\quad {\delta v}=0\right\}
\end{equation}
which is submanifold of manifold satisfying the conservation of angular momentum, asymptotically as $t \rightarrow \infty$. Subsequently, when at least one satellite in the system has RWs, e.g., ($1\leq m\leq n$), the system states, $q_{s}$ and $\zeta$, converge to the target values, $q_{sd}$ and $\zeta_d$, asymptotically as $t \rightarrow \infty$.
}

\textbf{Proof of Theorem 2}: See Appendix B.
\section{Numerical Calculations with Proposed Kinematics Control}
\label{sec4}
%ゲインの値を大きくすると収束性が良くなる（即応性が高くなる？）が
%・RWとDipole Inversionの値が大きくぶれた値となる．
%・評価関数を特に指定指定しない場合電流値が非常に大きい値となり発散する傾向がある
%交流電流周波数
%・位置誤差が大きい場合はダイナミクスの周波数の仮定が成立しない．
%・磁場の影響などを追加するとゲインの値によって衛星同士が衝突
\begin{figure}[bt!]
\centering
\includegraphics[width=.5\textwidth]{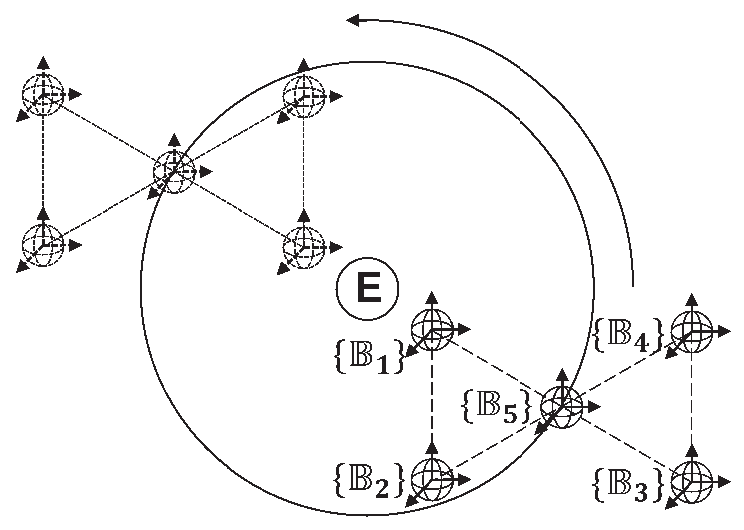}
\caption{Conceptual diagram of numerical calculation imitating situation of space interferometer project.}
\label{sec4:fig1}
\end{figure}
To identify the advantages of the proposed control law, three numerical calculations on stabilization to time-invariant formations (see Fig.~\ref{sec4:fig1}) are performed using the five-satellite system. Section~\ref{sec4-1} discusses the removability of the non-uniform distribution of the RW angular momentum for each satellite based on formation maintenance simulation. Section~\ref{sec4-2} presents the formation reconfiguration simulation, and the achievability of convergence to the target state when only some satellites are equipped with RWs is verified. Section~\ref{sec4-3} demonstrates the removability of the accumulation of angular momentum throughout the system by incorporating a simple unloading control in the chief satellite (e.g., the 5th satellite). These imitate the space interferometer project used by tethered satellite formation flying \cite{Quadrelli1,Lawson}, and the trajectory of the satellites in Fig.~\ref{sec4:fig1} is useful for data collection. Tables~\ref{sec4:table1} and \ref{sec4:tableX-1} summarize  the characteristics of each satellite in the formation and activated actuators on each simulation. Note that EMFF satellites are assumed to comprise HTS wires \cite{Kwon,Ayyad} to realize high current, and the nature of the hardware is irrelevant for the proposed control law. 
\begin{table}[bt!]
\caption{\label{sec4:table1} Characteristics of each satellite.}
\centering
\begin{tabular}{cc}
\hline\hline
Characteristics&Value\\\hline
RWs&30kg\\
Scientific equipment&70kg\\
EM coils + cooling&100kg\\
Total&200kg\\
Inertia matrix $J$&diag(107,107,134) kg m$^2$\\
\hline\hline
\end{tabular}
\end{table}
\begin{table}[bt!]
\caption{\label{sec4:tableX-1} Activated actuators on each simulation.}
\centering
\begin{tabular}{cccc}
\hline\hline
Num&\ref{sec4-1} & \ref{sec4-2}&\ref{sec4-3}\\\hline%& Attitude$[\sigma_1,\sigma_2,\sigma_3]$\\\hline
1st&RWs&RWs&RWs\\
2nd&RWs&RWs&RWs\\
3rd&RWs&RWs&RWs\\
4th&RWs&-&RWs\\
5th&RWs&-&RWs \& MTQs \\%&[0,0,0]\\
\hline\hline
\end{tabular}
\end{table}

The target system state described in this section is based on formation maintenance in an inertial frame, which means that the formation of satellites is stationary with respect to the invariant orientation frame (I). Therefore, the target relative positions $^ir_{jd}=[^ir_{jdx},^ir_{jdy},^ir_{jdz}]$ are the constant values shown in Table~\ref{sec4:tableX-2}. Moreover, the target attitude of the $j$-th satellite $\sigma_{jd}$ and the target angular momentum $\bm L_d$ in Eq.~\eqref{sec3:eq2-Y3} are set to $\bm 0$. In this case, the target RW angular momentum $^{b_j}h_{jd}$ and the target values $^{b_j}\xi_{dj}$ of the states $^{b_j}\xi$ are expressed as
\begin{equation}
   \label{sec4:eq1-3}
   ^{b_j}h_{jd}=\frac{1}{m}{^{b_j}L},\qquad {^{b_j}\xi_{dj}}=0_{(3,1)},
\end{equation}
following the definition provided in Eq.~\eqref{sec3:eq4-2-1}. These conditions establish
\begin{equation}
\label{sec3:eq11-X}
\left\{
\begin{aligned}
   \dot{v}_d&=\bm 0\\
   {^{b}\omega_d}&=\bm 0\\
\end{aligned}
\right.
\Leftrightarrow
\left\{
\begin{aligned}
   \overline{M}(q_s)\dot{v}_d&=\bm 0\\
\overline{C}(q_s,v)v_d&=\bm 0
\end{aligned}
\right. .
\end{equation}
%and these relationships simplify the kinematics control law $u_c$. 
Therefore, the control law $u_c$ and nonlinear programming of the EMFF used in this section are
\begin{equation}
  \label{sec4:eq1}
  \begin{aligned}
  &u_c
%=\begin{bmatrix}
%   \ ^if_c^{\mathrm{T}},
%   \ ^{b}\tau_c^{\mathrm{T}},
%   \ ^{b}\dot{h}^{\mathrm{T}}
%    \end{bmatrix}^{\mathrm{T}}
    %=[B]^{-1}S(S^{\mathrm{T}}S)^{-1}
    =\overline{B}_r^{-1}
    %=[B]^{-1}[M]S\overline{M}\ ^{-1}
    \left(-K_1 \delta q_s-K_2 \delta v-\overline{u}_d\right),\\
  %&K_1=
  %\begin{bmatrix}250E_{6n-3}&0\\
  %  0&0_{3m-3}
  %\end{bmatrix},\ K_2
  %=\begin{bmatrix}1250E_{6n-3}&0\\0&0.005E_{3m-3}\end{bmatrix},\\
  &K_1=
  \begin{bmatrix}
    150E_{3n-3}&0&0\\
    0&620E_{3n}&0\\
    0&0&0_{(3m-3)\times (3m-3)}
  \end{bmatrix},\qquad K_2=
  \begin{bmatrix}
    130E_{3n-3}&0&0\\
    0&530E_{3n}&0\\
    0&0&0.01E_{3m-3}
  \end{bmatrix},\\
  %=\begin{bmatrix}1250E_{6n-3}&0\\0&0.005E_{3m-3}\end{bmatrix},\\
   &\left \{
  \begin{aligned}
    &Minimize : J = {^i\mu_{(sin)}^{\mathrm{T}}}{^i\mu_{(sin)}}+{^i\mu_{(cos)}^{\mathrm{T}}}{^i\mu_{(cos)}}\\ 
  &Subject\ to :{^if_{jc}}= {^if_{(avg)j}},\quad {^i\tau_{jc}}={^i\tau_{(avg)j}}.
\end{aligned}
  \right.
   \end{aligned}
\end{equation}
\begin{table}[bt!]
\caption{\label{sec4:tableX-2} Target relative positions $^ir_d=[^ir_{dx},^ir_{dy},^ir_{dz}]$.}
\centering
\begin{tabular}{ccc}
\hline\hline
Num&\ref{sec4-1} \& \ref{sec4-2}&\ref{sec4-3}\\\hline
1st&$\left[10\cos(105^\circ),10\sin(105^\circ),-2\right]$&$\left[",","\right]$\\%&[0,0,0]\\
2nd&$\left[10\cos(165^\circ),10\sin(165^\circ),2\right]$&$\left[",",-2\right]$\\%&[0,0,0]\\
3rd&$\left[10\cos(285^\circ),10\sin(285^\circ),-2\right]$&$\left[",",2\right]$\\%&[0,0,0]\\
4th&$\left[10\cos(-15^\circ),10\sin(-15^\circ),2\right]$&$\left[",","\right]$\\%&[0,0,0]\\
5th&$\left[0,0,0\right]$&$\left[",","\right]$\\%&[0,0,0]\\
\hline\hline
\end{tabular}
\end{table}

The formation controls are assumed at an altitude of 700 km above the Earth. Eqs.~\eqref{sec2:eq1} and ~\eqref{sec2:eq5} are used to determine the relative translational dynamics and attitude dynamics of an RW-equipped rigid spacecraft, respectively. Moreover, the gravity gradient torques $\bm\tau_g$ for each satellite are considered. In these conditions, the angular momentum of the system $\bm L$ is, ideally, a periodic value with a period of approximately 6000 s. Table~\ref{sec4:tableX-3} summarizes the disturbance torque $\bm\tau_d$ in orbit and AC frequency $\omega_f$ of electromagnetic coils. In Sec.~\ref{sec4-1}, $\bm\tau_d$ due to the interaction between the EMFF control and Earth's magnetic field $\bm\mu_j\times\bm B_e$ are not considered for simplicity. Considering the frequency of these dynamics, the AC frequency $\omega_f$ is set to $8\pi$ rad/s in Sec.~\ref{sec4-1} and $16\pi$ rad/s in Secs.~\ref{sec4-2} and \ref{sec4-3}. 
\begin{table}[bt!]
\caption{\label{sec4:tableX-3} Disturbance torque $\bm\tau_d$ and AC frequency $\omega_f$.}
\centering
\begin{tabular}{ccc}
\hline\hline
Parameters&\ref{sec4-1} & \ref{sec4-2} \& \ref{sec4-3}\\\hline
$\bm\tau_d$ [N$\cdot$m]&$\bm\tau_g$&$\bm\tau_g,\ \bm\mu_j\times\bm B_e$\\%&$\bm\tau_g,\ \bm\mu_j\times\bm B_e$\\%&[0,0,0]\\
$\omega_f$ [rad/s] & $8\pi$ & $16\pi$\\% & $16\pi$\\
\hline\hline
\end{tabular}
\end{table}
\subsection{Formation Maintenance Using Five Satellites with Five Sets of RWs}
\label{sec4-1}
In this numerical calculation, the non-uniform distribution of the RW angular momentum among satellites should be eliminated. %under conditions where external force exists. 
The temporal variations of the RW angular momentum are shown for two cases: the conventional control scheme (see Fig.~\ref{sec3:fig:rte_1}) and the proposed kinematics control scheme (see Fig.~\ref{sec3:fig:rte_2}). It is assumed that the RWs of all satellites are activated, as shown in Table~\ref{sec4:tableX-1}.
%\begin{table}[bt!]
%\caption{\label{sec4:table2} Simulation conditions.}
%\centering
%\begin{tabular}{ccc}
%\hline\hline
%Num&$^ir_d=[^ir_{dx},^ir_{dy},^ir_{dz}]$&Activated\\\hline%& Attitude$[\sigma_1,\sigma_2,\sigma_3]$\\\hline
%1st&$\left[10\cos(105^\circ),10\sin(105^\circ),-2\right]$&RWs\\%&[0,0,0]\\
%2nd&$\left[10\cos(165^\circ),10\sin(165^\circ),2\right]$&RWs\\%&[0,0,0]\\
%3rd&$\left[10\cos(285^\circ),10\sin(285^\circ),-2\right]$&RWs\\%&[0,0,0]\\
%4th&$\left[10\cos(-15^\circ),10\sin(-15^\circ),2\right]$&RWs\\%&[0,0,0]\\
%5th&$\left[0,0,0\right]$&RWs\\%&[0,0,0]\\
%\hline\hline
%\end{tabular}
%\end{table}
%提案手法によって各衛星に搭載されたRWの角運動量が制御されRWの角運動量の偏りが
%起こらないことを示すため，
The conventional control law, compared with the proposed control law of Eq.~\eqref{sec4:eq1}, and the nonlinear programming of the EMFF are expressed as
\begin{equation}
  \label{sec4:eq2}
  \begin{aligned}
     {^o f_{jc}}&=-{^of_{jg}}+m_j\left(\frac{^{I}\mathrm{d}^2}{\mathrm{d}t^2}{^or_j}+\frac{^{O}\mathrm{d}^2}{\mathrm{d}t^2}(^or_{jd}-{^or_j})+\lambda_{p1}\frac{^{O}\mathrm{d}}{\mathrm{d}t}(^or_{jd}-{^or_{j}})+\lambda_{p2}U_j\right),\\
     U_j&=\frac{^O\mathrm{d}}{\mathrm{d}t}\left(^or_{jd}-{^or_i}\right)+\lambda_{p1}(^or_{jd}-{^or_i}),\\
     {^{b_j}\dot{h}_j}&=^{b_j}\tau_{jEM}-^{b_j}\omega_j \left(J_j{^{b_j}\omega_j}+{^{b_j}h}\right)-J_j{^{b_j}}\dot{\omega}_j+\lambda_{a1}\sigma_j+\lambda_{a2} {^{b_j}\omega_j},\\
\lambda_{p1}&=0.0125,\ \lambda_{p2}=0.0125,\ \lambda_{a1}=10,\ \lambda_{a2}=15,\\
  &\left \{
  \begin{aligned}
    &Minimize: J = {^b\tau_{c}^{\mathrm{T}}}{^b\tau_{c}}\\
       &Subject\ to {^if_{jc}}={^if_{(avg)j}}.
  \end{aligned}
  \right.
\end{aligned}
\end{equation}
%In addition, a model based on the Hill-Clohessy-Wiltshire (HCW) equations (see Eq.~\eqref{sec2:eq1-1}) \cite{Schaub} is used to evaluate the effectiveness of the proposed method in Sec.~\ref{sec4-1}:
%\begin{equation}
%    \label{sec2:eq1-1}
%    \Leftrightarrow\frac{^od^2}{dt^2}\ ^o{r}_j+c_j+g_j=\frac{1}{m_j}\ ^of_{ic},
%\end{equation}
%where $\bm{f}_{ic}$ is the translational control input of the $j$-th satellite. Furthermore, $c_j$ and $g_j$ are expressed by Eq.~\eqref{sec2:eq1-2} using the orbital frequency $\omega_o=\sqrt{\mu_g /R_o^3}$:
%\begin{equation}
%  \label{sec2:eq1-2}
%  c_j=
%  \begin{bmatrix}
%    0&-2\omega_o&0\\
%    2\omega_o&0&0\\
%    0&0&0
%  \end{bmatrix}
%  \begin{bmatrix}
%    ^o\dot{x}_j\\
%    ^o\dot{y}_j\\
%    ^o\dot{z}_j\\
%  \end{bmatrix},\ 
%  g_j=
%  \begin{bmatrix}
%    -3\omega_o^2&0&0\\
%    0&0&0\\
%    0&0&\omega_o^2
%  \end{bmatrix}
%  \begin{bmatrix}
%    ^o{x}_j\\
%    ^o{y}_j\\
%    ^o{z}_j\\
%  \end{bmatrix}.
%\end{equation}
The target electromagnetic force, ${^of_c}$, and the target RW control torque, $^b\dot{h}$, are derived using the terminal sliding mode controller \cite{Ahsun} and Lyapunov's direct method \cite{Slotine}, respectively. The nonlinear programming of the conventional control law is often set to minimize the magnitude of the electromagnetic torque $\bm \tau_c$ because the conventional control law cannot control $\bm \tau_c$. The AC frequency $\omega_f$ is set to $4\pi$ rad/s, which is the same as that of the proposed control law.
%First, the results of controlling $^or$ and $\sigma$ via the proposed control law are shown in Figs.~\ref{sec4:fig2} and \ref{sec4:fig3}, respectively. The thin and thick lines show the reference trajectory and the simulation result, respectively. The results of the compared control given by Eq.~\eqref{sec4:eq2} are shown in Figs.~\ref{seca:fig1} and \ref{seca:fig2}. 

\begin{figure}[bt!]
  \begin{center}
    \begin{minipage}{0.25\hsize}
        \centering
        \includegraphics[width=1\textwidth]{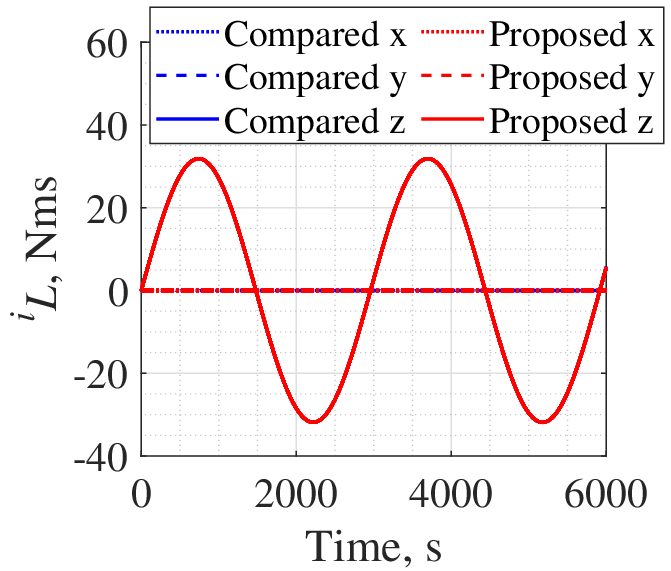}
        \subcaption{The entire system.}
        \label{sec4:fig4_a}
    \end{minipage}
    \begin{minipage}{0.25\hsize}
        \centering
        \includegraphics[width=1\textwidth]{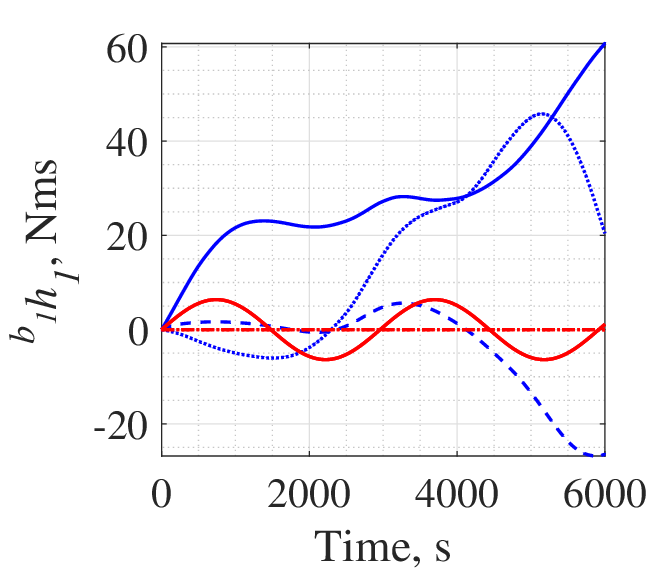}
        \subcaption{RWs of 1st satellite.}
        \label{sec4:fig4_b}
  \end{minipage}
%  \begin{comment}
  \begin{minipage}{0.25\hsize}
        \centering
        \includegraphics[width=1\textwidth]{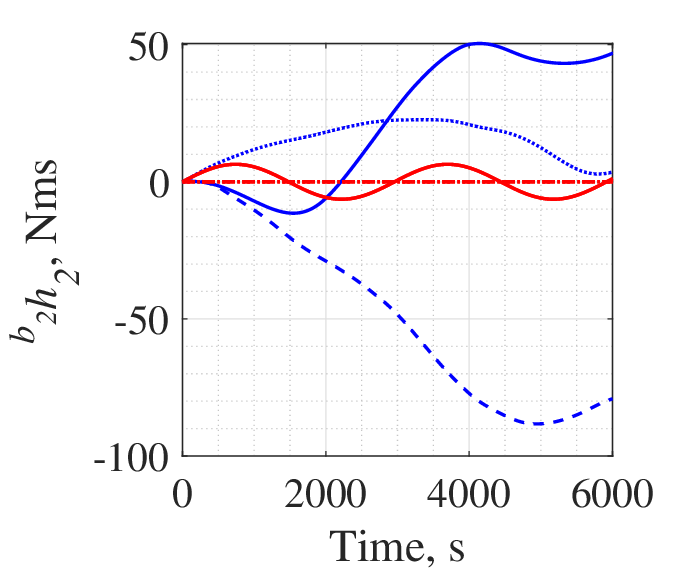}
        \subcaption{RWs of 2nd satellite.}
        \label{sec4:fig4_c}
    \end{minipage}\\
    \begin{minipage}{0.25\hsize}
        \centering
        \includegraphics[width=1\textwidth]{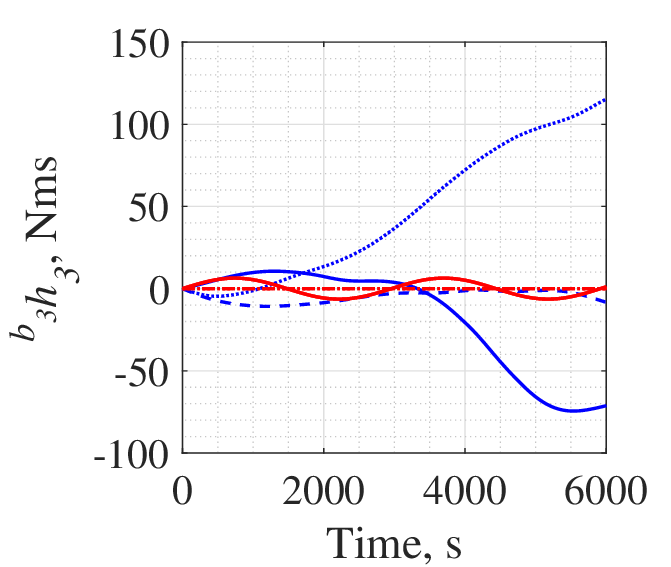}
        \subcaption{RWs of 3rd satellite.}
        \label{sec4:fig4_d}
  \end{minipage}
  \begin{minipage}{0.25\hsize}
        \centering
        \includegraphics[width=1\textwidth]{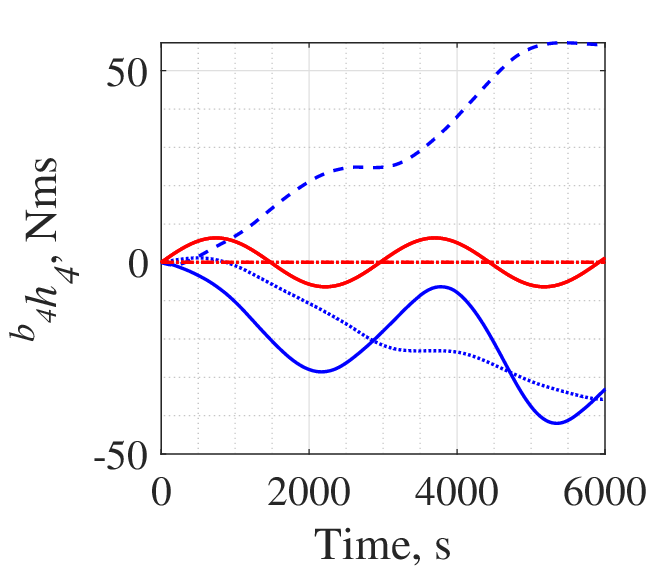}
        \subcaption{RWs of 4th satellite.}
        \label{sec4:fig4_e}
    \end{minipage}
  \begin{minipage}{0.25\hsize}
        \centering
        \includegraphics[width=1\textwidth]{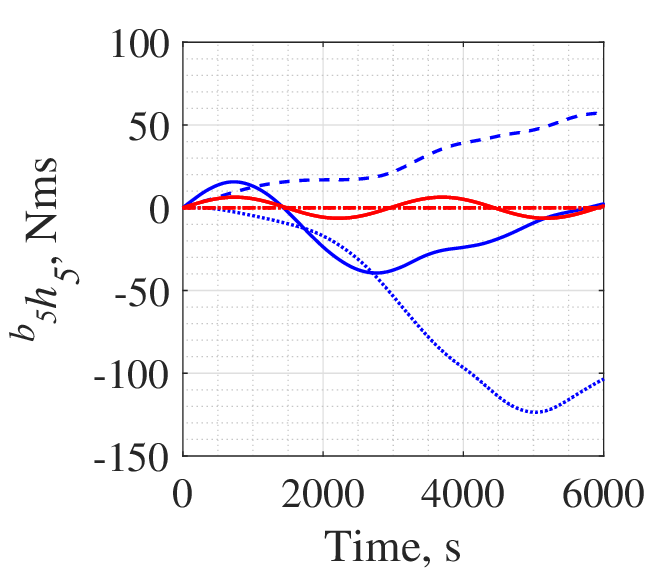}
        \subcaption{RWs of 5th satellite.}
        \label{sec4:fig4_f}
  \end{minipage}\\
%\end{comment}
  \caption{Simulated results (Section~\ref{sec4-1}) with respect to angular momentum for the: (a) entire system $^{i} L$, (b) 1st satellite $^{b_1} h_1$, (c) 2nd satellite $^{b_2} h_2$, (d) 3rd satellite $^{b_3} h_3$, (e) 4th satellite $^{b_4} h_4$, and (f) 5th satellite $^{b_5} h_5$.}
   \label{sec4:fig4}
\end{center}
\end{figure}

First, the results corresponding to the control of $^iL$ and $^{b_i}h$ are shown in Fig.~\ref{sec4:fig4}. 
%, therefore 
%it returns to the initial value 
%after going around the orbit (about 6000s). 
The blue and red lines in Figs.~\ref{sec4:fig4_b}--\ref{sec4:fig4_e} show the RW angular momentum for the compared and proposed control laws, respectively. The total values of the RW angular momentum have almost same periodic values, as shown in Fig.~\ref{sec4:fig4_a}. However, each RW angular momentum corresponding to the compared control law shows a non-uniform distribution after one completion of the orbit because the electromagnetic torques are not controlled. Conversely, for the proposed control law, the angular momentum of each RW shows uniform values; these values are obtained by dividing $\bm L$ evenly between all the satellites.
These results indicate that the proposed kinematics control law theoretically eliminates the non-uniform distribution of the RW angular momentum among satellites.
%From the control results, 
%the proposed control method functions to stabilize the EMFF system 
%to the target manifold, and its validity is shown. 

\begin{figure}[bt!]
  \begin{center}
    %\begin{minipage}{0.25\hsize}
    %   \centering
    %   \includegraphics[width=.7\textwidth]{SecA_Position_1_0.eps}
    %   \subcaption{Legend.}
%       \label{sec4:fig2_0_a}
    %\end{minipage}
    \begin{minipage}{0.25\hsize}
        \centering
        \includegraphics[width=1\textwidth]{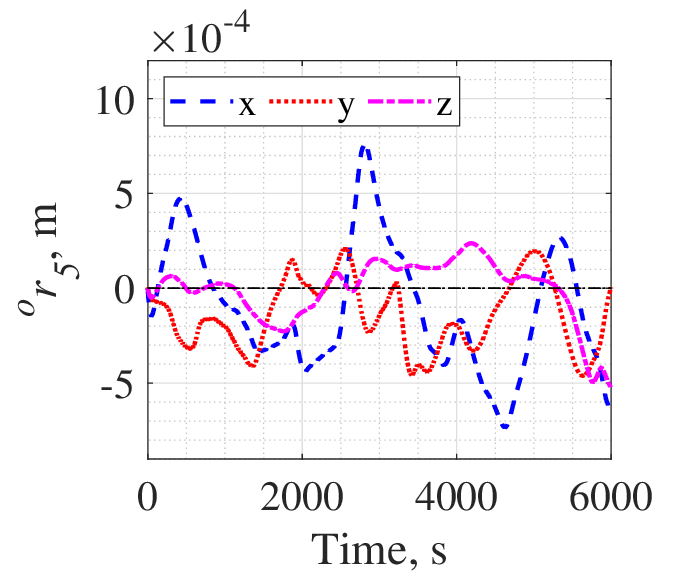}
        \subcaption{Relative position via the compared control.}
        \label{sec4:fig2_0_b}
  \end{minipage}
  \begin{minipage}{0.25\hsize}
        \centering
        \includegraphics[width=1\textwidth]{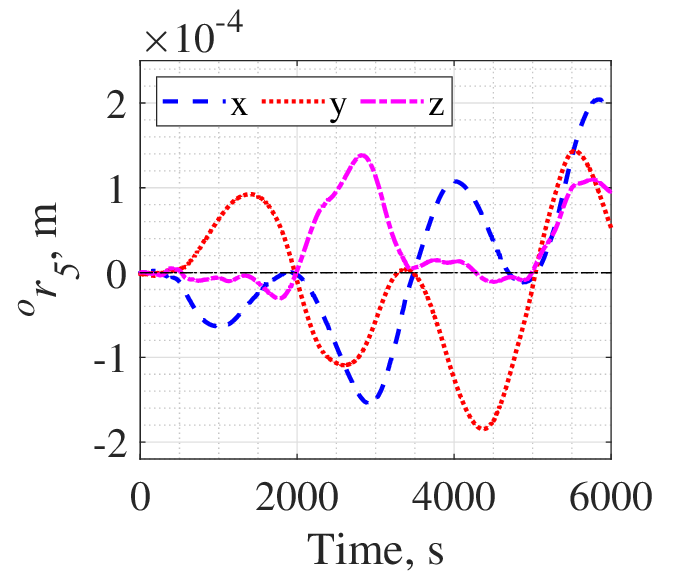}
        \subcaption{Relative position via the proposed control.}
        \label{sec4:fig2_0_c}
    \end{minipage}\\
    %\begin{minipage}{0.25\hsize}
    %   \centering
    %   \includegraphics[width=1\textwidth]{SecA_Position_4_0.eps}
    %   \subcaption{3rd satellite.}
    %   \label{sec4:fig2_0_d}
  %\end{minipage}
  \begin{minipage}{0.25\hsize}
        \centering
        \includegraphics[width=1\textwidth]{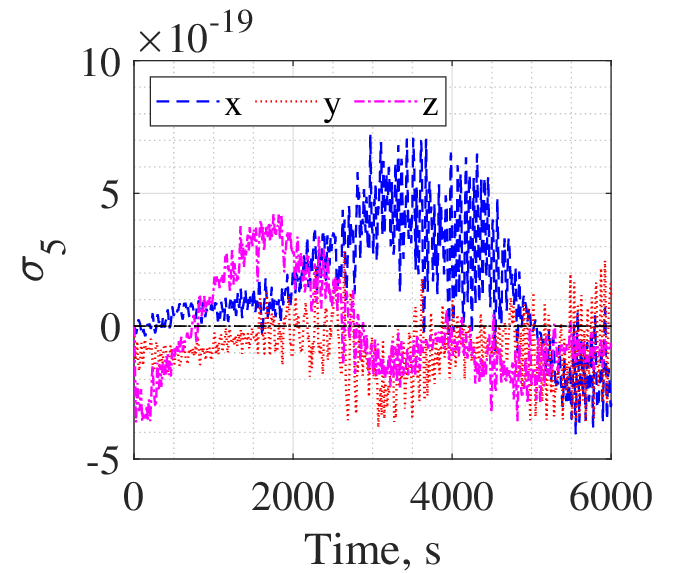}
        \subcaption{Absolute attitude via the compared control.}
        \label{sec4:fig2_0_e}
    \end{minipage}
  \begin{minipage}{0.25\hsize}
        \centering
        \includegraphics[width=1\textwidth]{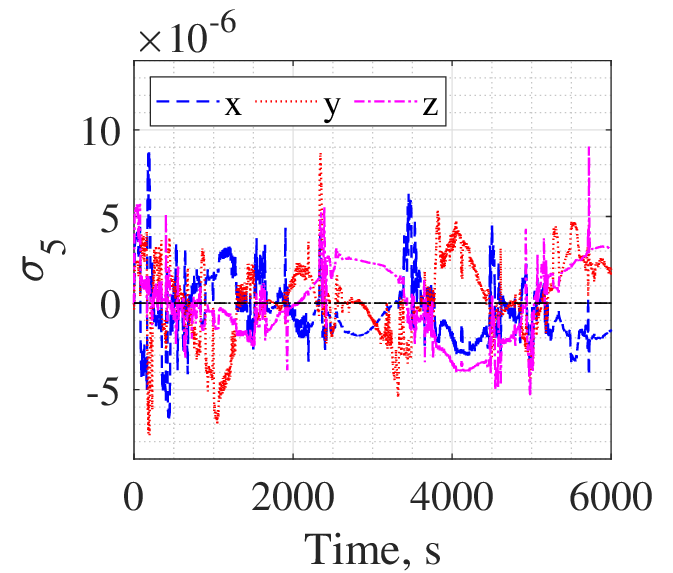}
        \subcaption{Absolute attitude via the proposed control.}
        \label{sec4:fig2_0_f}
    \end{minipage}\\
  \caption{Simulated results of the 5th satellite (Section~\ref{sec4-1}) via the compared control and the proposed control law.}
     \label{sec4:figX-1}
\end{center}
\end{figure}

Next, the formation maintenance results of the 5th satellite via the compared and proposed control law are shown in Fig.~\ref{sec4:figX-1}. Although this subsection only describes the results of the 5th satellite for brevity, it can be shown that the satellite system maintains the formation and attitude of each satellite. Figure~\ref{sec4:figX-1} shows that the results of the absolute attitude for the proposed control method has larger errors compared to the result of the compared control method. This is because the attitude control of conventional control law is performed only by RW, where the attitude control law based on the proposed control has the oscillation term in Eq.~\eqref{sec2:eq14-2} which reduces the control performance. The relative position control results include the effect of the oscillation term and therefore contain errors from the target value. 

\begin{figure}[bt!]
  \begin{center}
    \begin{minipage}{0.25\hsize}
        \centering
        \includegraphics[width=1\textwidth]{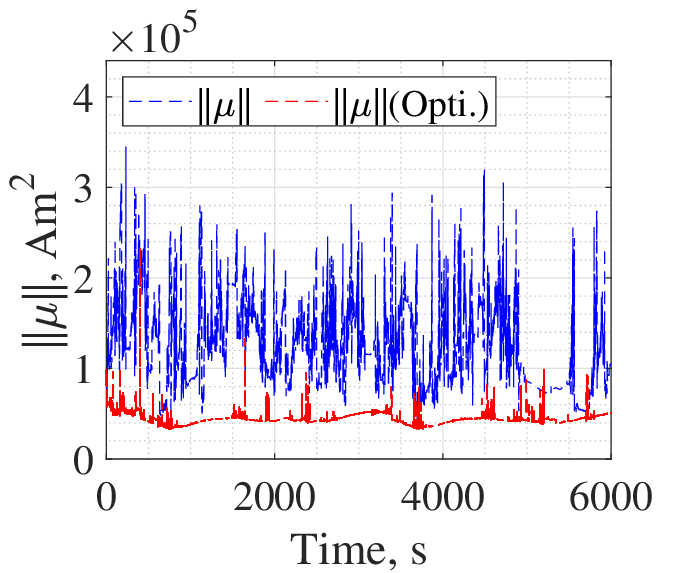}
        \subcaption{The entire system.}
        \label{sec4:fig5_a}
    \end{minipage}
    \begin{minipage}{0.25\hsize}
        \centering
        \includegraphics[width=1\textwidth]{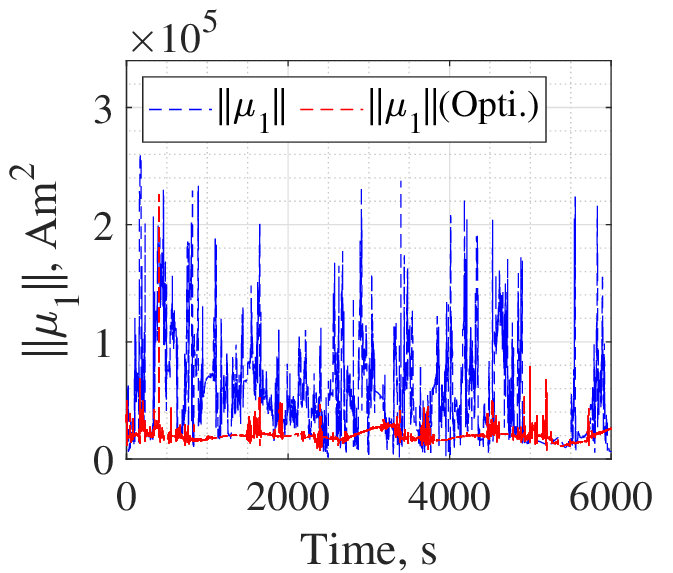}
        \subcaption{1st satellite.}
        \label{sec4:fig5_b}
  \end{minipage}
%  \begin{comment}
  \begin{minipage}{0.25\hsize}
        \centering
        \includegraphics[width=1\textwidth]{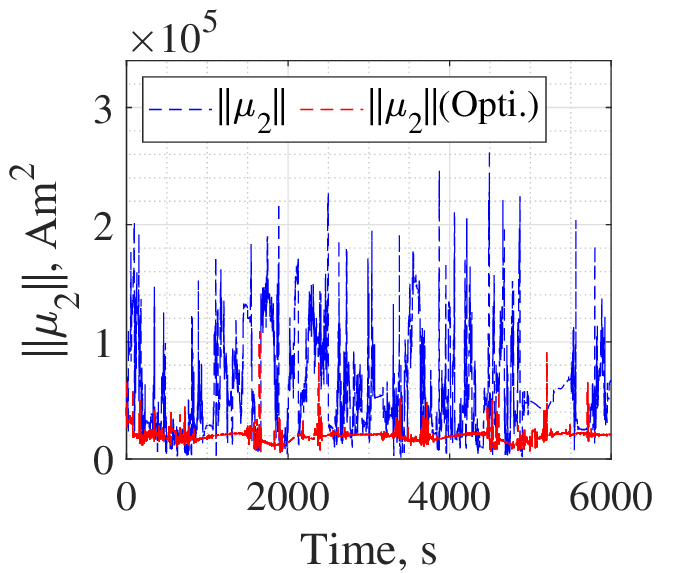}
        \subcaption{2nd satellite.}
        \label{sec4:fig5_c}
    \end{minipage}\\
    \begin{minipage}{0.25\hsize}
        \centering
        \includegraphics[width=1\textwidth]{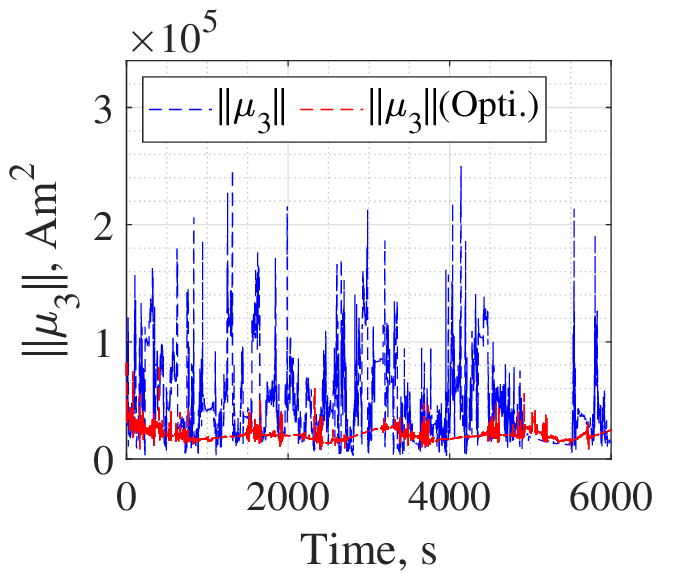}
        \subcaption{3rd satellite.}
        \label{sec4:fig5_d}
  \end{minipage}
  \begin{minipage}{0.25\hsize}
        \centering
        \includegraphics[width=1\textwidth]{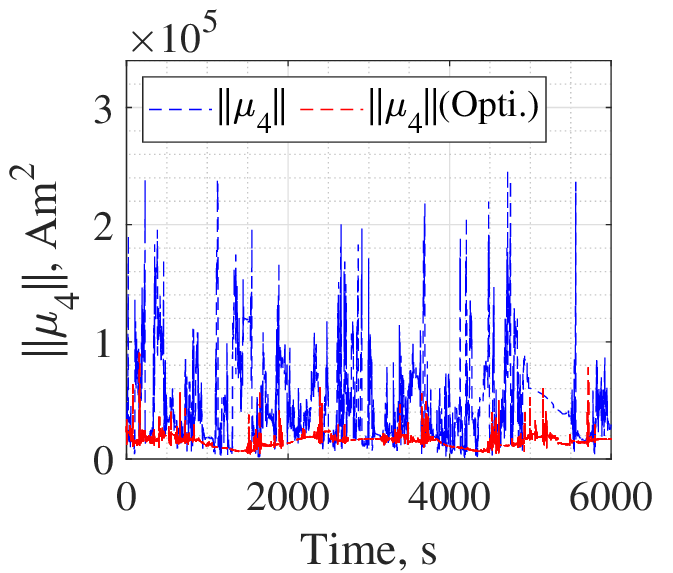}
        \subcaption{4th satellite.}
        \label{sec4:fig5_e}
    \end{minipage}
  \begin{minipage}{0.25\hsize}
        \centering
        \includegraphics[width=1\textwidth]{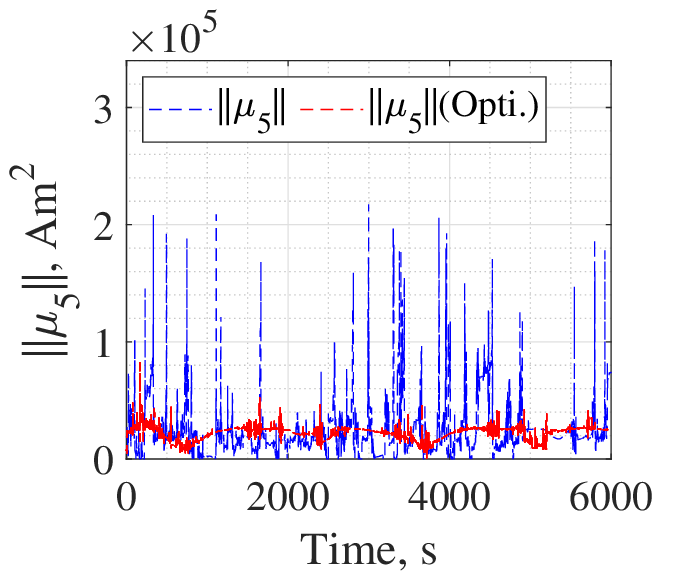}
        \subcaption{5th satellite.}
        \label{sec4:fig5_f}
  \end{minipage}\\
%\end{comment}
  \caption{Simulated results (Section~\ref{sec4-1}) with respect to the dipole moment magnitude for the: (a) entire system $\|\mu\|$, (b) 1st satellite $\|\mu_1\|$, (c) 2nd satellite $\|\mu_2\|$, (d) 3rd satellite $\|\mu_3\|$, (e) 4th satellite $\|\mu_4\|$, and (f) 5th satellite $\|\mu_5\|$.}
   \label{sec4:fig5}
\end{center}
\end{figure}

Finally, the dipole moment for generating the desired electromagnetic force and torque is shown in Fig.~\ref{sec4:fig5}. This figure compares the case without optimization ($J=0$) and the case with optimization for the objective function $J$ of Eq.~\eqref{sec4:eq1}. Numerical calculations show that there is a ``dipole solution'' that minimizes the power consumption while outputting calculated electromagnetic force and torque. In the case without optimization, initialization of dipole solution is needed because it tends to be extremely large, even though the computation time is short. Therefore, initialization is conducted when the maximum values of the case with optimization are reached. In AC-based EMFF control, the states are disturbed by vibration components. As a result, the target control value also has an oscillatory component. This component causes the value of the dipole moment to move from one local minimum to another in the process of dipole inversion.
\begin{comment}
\begin{figure}[bt!]
\centering
\includegraphics[width=.5\textwidth]{simu1-dipole-proposed-paperdate.eps}
\caption{Simulated results of the five-satellite formation maintenance to the dipole moment magnitude for the:
(a) entire system ||$\mu$||, (b) 1st satellite ||$\mu_{(\sin)1}$|| and ||$\mu_{(\cos)1}$||, (c) 2nd satellite ||$\mu_{(\sin)2}$|| and ||$\mu_{(\cos)2}$||, (d) 3rd satellite ||$\mu_{(\sin)3}$|| and ||$\mu_{(\cos)3}$||, (e) 4th satellite ||$\mu_{(\sin)4}$|| and ||$\mu_{(\cos)4}$||, and (f) 5th satellite ||$\mu_{(\sin)5}$|| and ||$\mu_{(\cos)5}$||.}
\label{sec4:fig5}
\end{figure}
\end{comment}

%\begin{multicols}{2}
\subsection{Formation Reconfiguration Using Five Satellites with Three Sets of RWs}
This subsection verify that the proposed kinematics control law can control more than three satellites without all the satellites being equipped with RWs. As shown in Table~\ref{sec4:tableX-1}, only the RWs belonging to the 1st, 2nd, and 3rd satellites are activated. For this numerical calculation, the initial conditions of $^or$ and $\sigma$ are assigned at random. 
\label{sec4-2}
%\begin{table}[bt!]
%\caption{\label{sec4:table3} Simulation conditions}
%\centering
%\begin{tabular}{ccc}
%\hline\hline
%Num&$^ir_d=[^ir_{dx},^ir_{dy},^ir_{dz}]$&Activated\\\hline%& Attitude$[\sigma_1,\sigma_2,\sigma_3]$\\\hline
%1st&$\left[10\cos(105^\circ),10\sin(105^\circ),-2\right]$&RWs\\%&[0,0,0]\\
%2nd&$\left[10\cos(165^\circ),10\sin(165^\circ),2\right]$&RWs\\%&[0,0,0]\\
%3rd&$\left[10\cos(285^\circ),10\sin(285^\circ),-2\right]$&RWs\\%&[0,0,0]\\
%4th&$\left[10\cos(-15^\circ),10\sin(-15^\circ),2\right]$&-\\%&[0,0,0]\\
%5th&$\left[0,0,0\right]$&-\\%&[0,0,0]\\
%\hline\hline
%\end{tabular}
%\end{table}

\begin{figure}[bt!]
  \begin{center}
    \begin{minipage}{0.25\hsize}
        \centering
        \includegraphics[width=.7\textwidth]{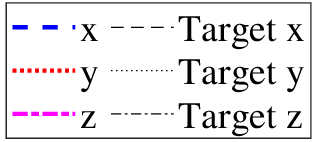}
        \subcaption{Legend.}
        \label{sec4:fig6_a}
    \end{minipage}
    \begin{minipage}{0.25\hsize}
        \centering
        \includegraphics[width=1\textwidth]{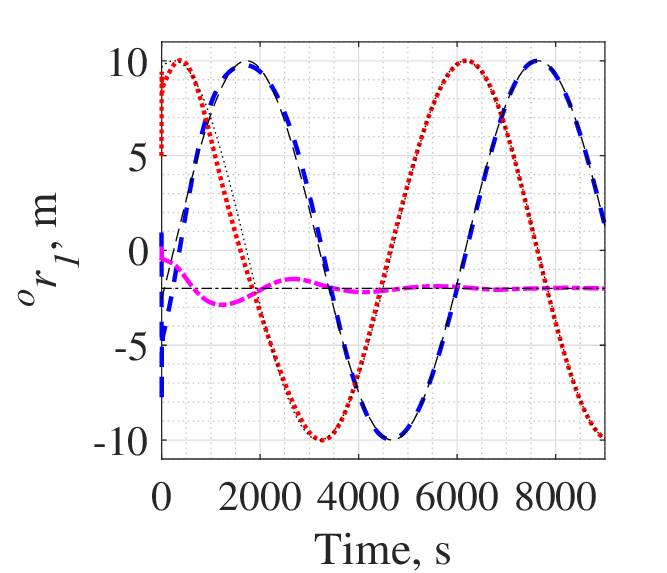}
        \subcaption{1st satellite.}
        \label{sec4:fig6_b}
  \end{minipage}
  \begin{minipage}{0.25\hsize}
        \centering
        \includegraphics[width=1\textwidth]{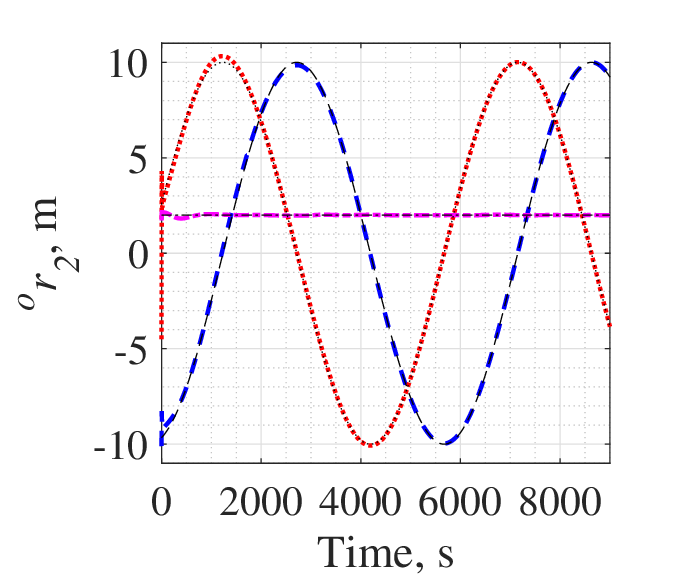}
        \subcaption{2nd satellite.}
        \label{sec4:fig6_c}
    \end{minipage}\\
    \begin{minipage}{0.25\hsize}
        \centering
        \includegraphics[width=1\textwidth]{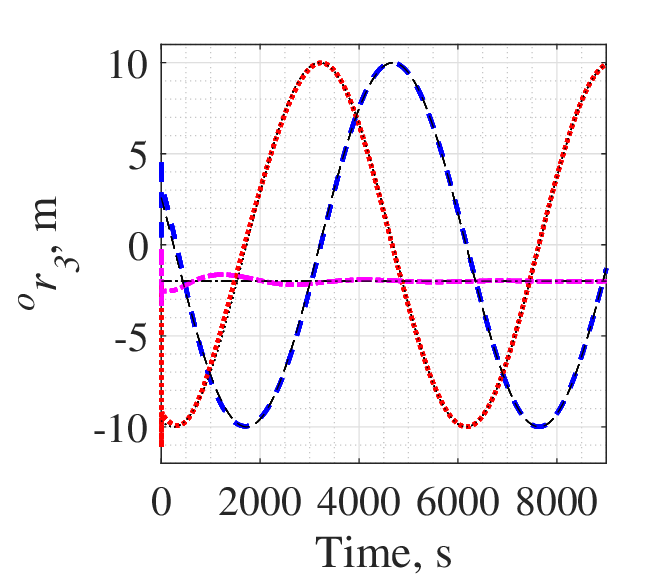}
        \subcaption{3rd satellite.}
        \label{sec4:fig6_d}
  \end{minipage}
  \begin{minipage}{0.25\hsize}
        \centering
        \includegraphics[width=1\textwidth]{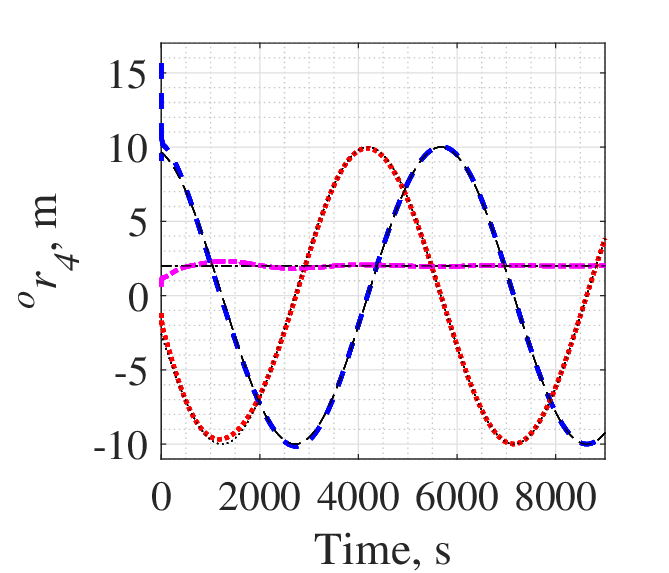}
        \subcaption{4th satellite.}
        \label{sec4:fig6_e}
    \end{minipage}
  \begin{minipage}{0.25\hsize}
        \centering
        \includegraphics[width=1\textwidth]{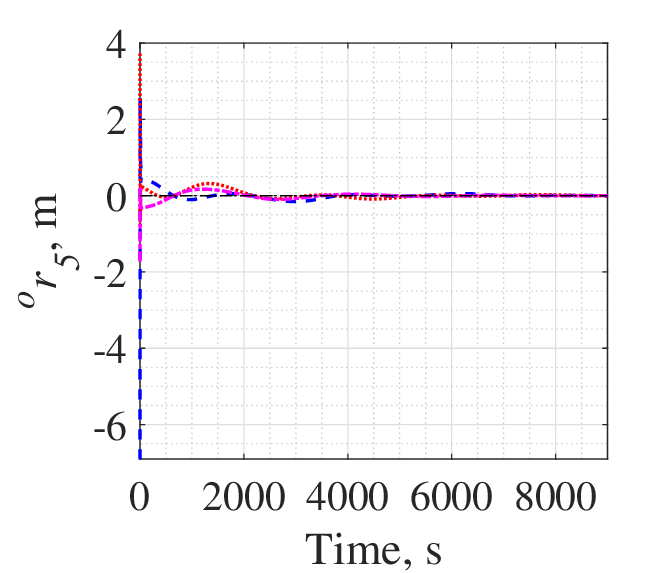}
        \subcaption{5th satellite.}
        \label{sec4:fig6_f}
    \end{minipage}\\
  \caption{Simulated results (Section~\ref{sec4-2}) with respect to the relative position for the: (a) 1st satellite $^or_1$, (b) 2nd satellite $^or_2$, (c) 3rd satellite $^or_3$, (d) 4th satellite $^or_4$, and (e) 5th satellite $^or_5$.}
   \label{sec4:fig6}
\end{center}
\end{figure}

\begin{figure}[bt!]
  \begin{center}
    \begin{minipage}{0.25\hsize}
        \centering
        \includegraphics[width=0.7\textwidth]{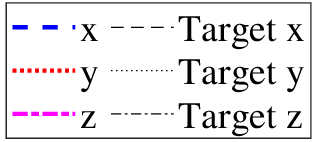}
        \subcaption{Legend.}
        \label{sec4:fig7_a}
    \end{minipage}
    \begin{minipage}{0.25\hsize}
        \centering
        \includegraphics[width=1\textwidth]{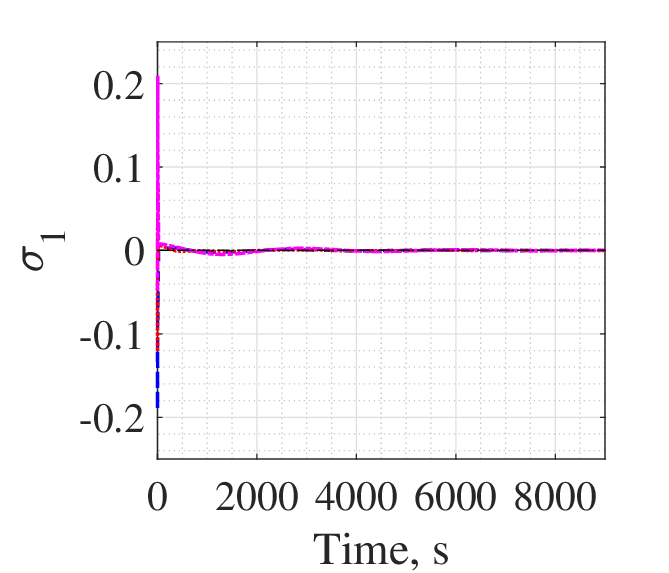}
        \subcaption{1st satellite.}
        \label{sec4:fig7_b}
  \end{minipage}
  \begin{minipage}{0.25\hsize}
        \centering
        \includegraphics[width=1\textwidth]{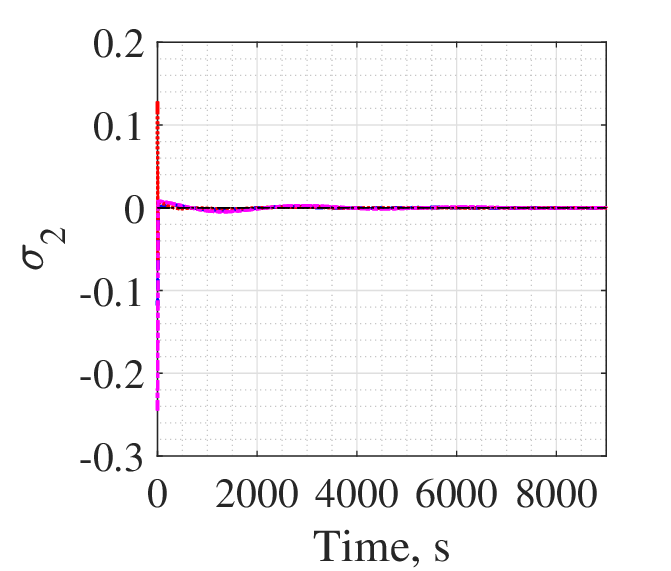}
        \subcaption{2nd satellite.}
        \label{sec4:fig7_c}
    \end{minipage}\\
    \begin{minipage}{0.25\hsize}
        \centering
        \includegraphics[width=1\textwidth]{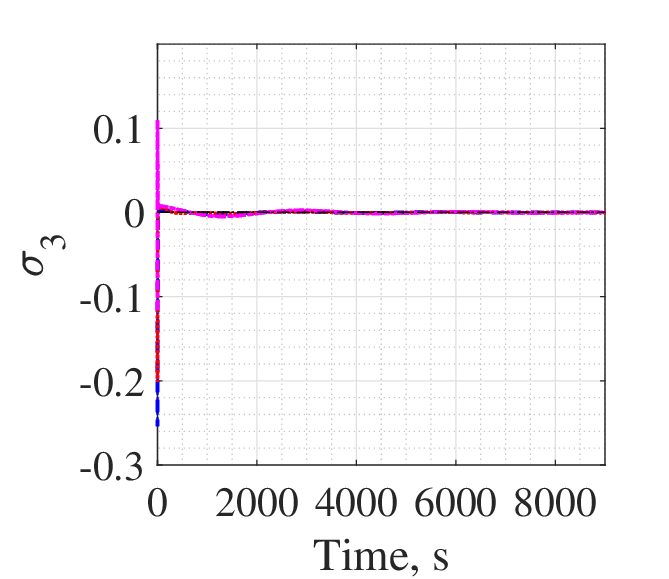}
        \subcaption{3rd satellite.}
        \label{sec4:fig7_d}
  \end{minipage}
  \begin{minipage}{0.25\hsize}
        \centering
        \includegraphics[width=1\textwidth]{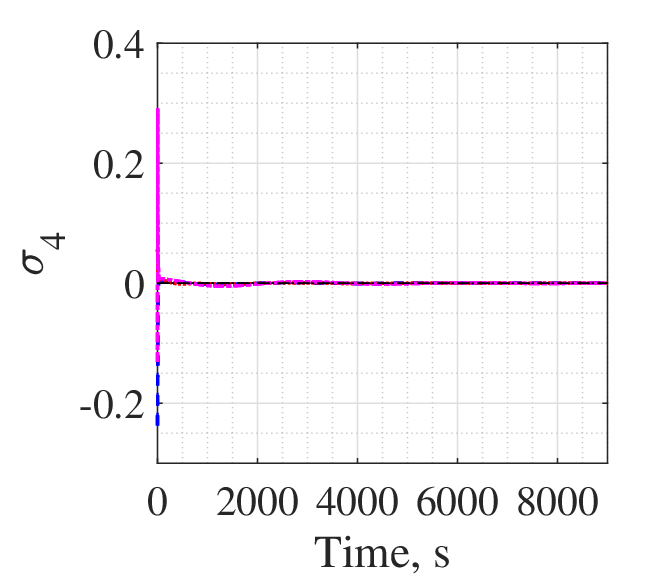}
        \subcaption{4th satellite.}
        \label{sec4:fig7_e}
    \end{minipage}
  \begin{minipage}{0.25\hsize}
        \centering
        \includegraphics[width=1\textwidth]{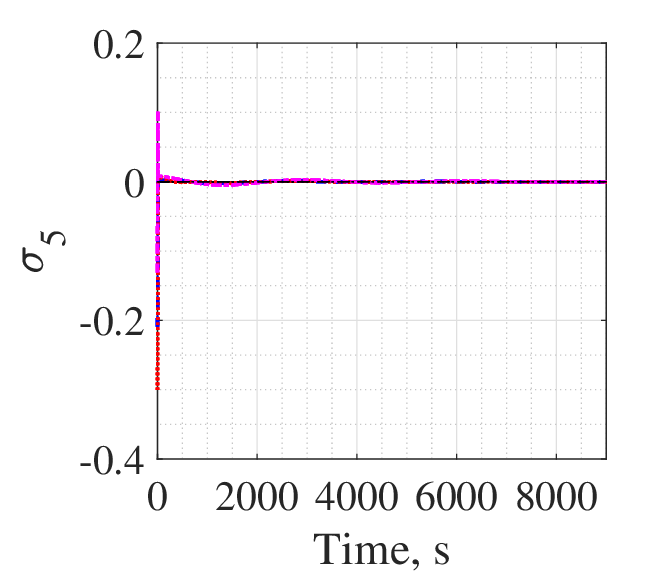}
        \subcaption{5th satellite.}
        \label{sec4:fig7_f}
    \end{minipage}\\
  \caption{Simulated results (Section~\ref{sec4-2}) with respect to the absolute attitude for the: (a) 1st satellite $\sigma_1$, (b) 2nd satellite $\sigma_2$, (c) 3rd satellite $\sigma_3$, (d) 4th satellite $\sigma_4$, and (e) 5th satellite $\sigma_5$.}
   \label{sec4:fig7}
\end{center}
\end{figure}

The results corresponding to the control of $^or$ and $\sigma$ are shown in Figs.~\ref{sec4:fig6} and \ref{sec4:fig7}, respectively, with black lines and lines of other colors showing the reference trajectory and the simulation results, respectively. These figures show that the relative position and absolute attitude of each satellite 
converge from the random initial values to the target values. 
%\begin{figure}[bt!]
%\centering
%\includegraphics[width=.5\textwidth]{simu2-attitude-proposed-paperdate.eps}
%\caption{Simulated results for five satellite formation reconfiguration to attitude: 
%(a) 1st satellite $\sigma_1$, (b) 2nd satellite $\sigma_2$, (c) 3rd satellite $\sigma_3$, (d) 4th satellite $\sigma_4$, and (e) 5th satellite $\sigma_5$.}
%\label{sec4:fig7}
%\end{figure}

\begin{figure}[bt!]
  \begin{center}
    \begin{minipage}{0.25\hsize}
        \centering
        \includegraphics[width=1\textwidth]{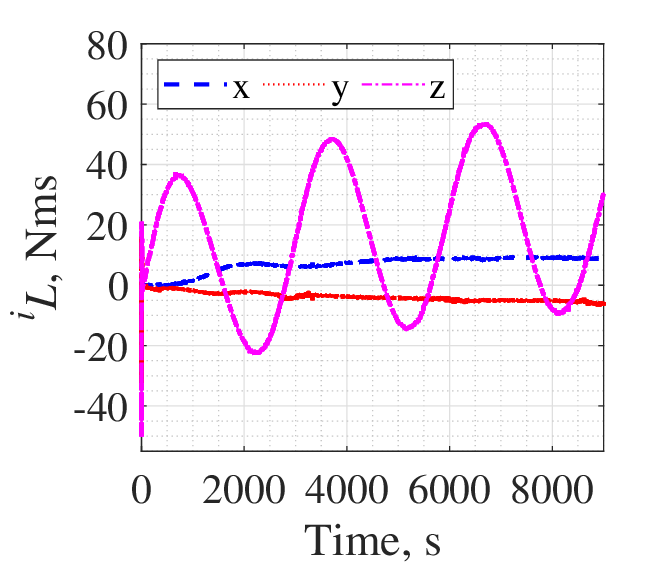}
        \subcaption{The entire system.}
        \label{sec4:fig8_a}
    \end{minipage}
    \begin{minipage}{0.25\hsize}
        \centering
        \includegraphics[width=1\textwidth]{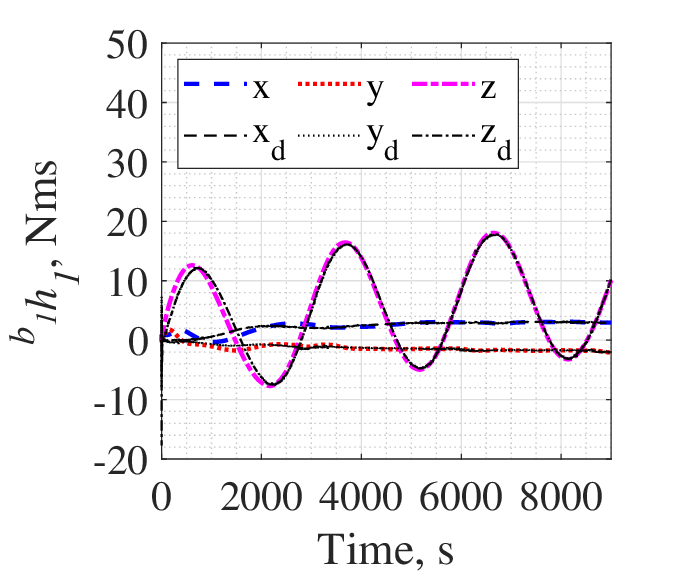}
        \subcaption{RWs of 1st satellite.}
        \label{sec4:fig8_b}
  \end{minipage}\\
%  \begin{comment}
  \begin{minipage}{0.25\hsize}
        \centering
        \includegraphics[width=1\textwidth]{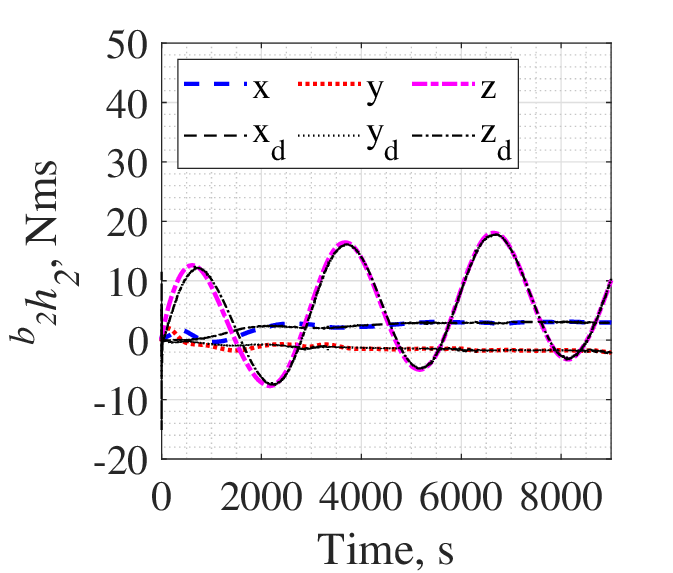}
        \subcaption{RWs of 2nd satellite.}
        \label{sec4:fig8_c}
    \end{minipage}
    \begin{minipage}{0.25\hsize}
        \centering
        \includegraphics[width=1\textwidth]{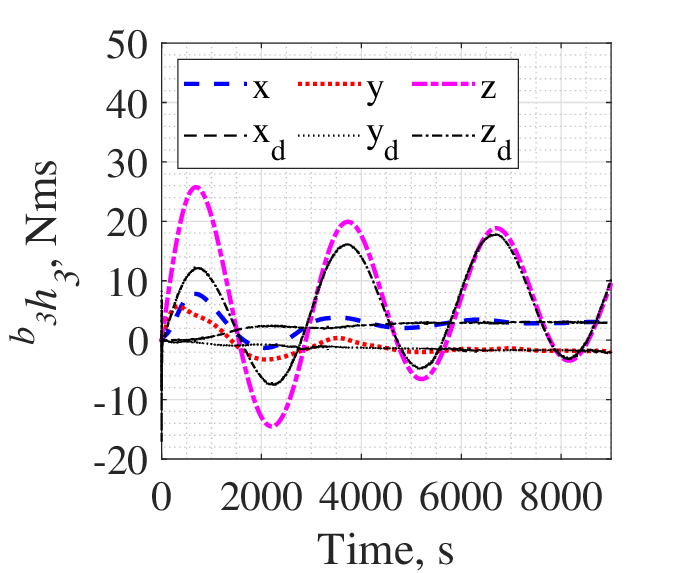}
        \subcaption{RWs of 3rd satellite.}
        \label{sec4:fig8_d}
  \end{minipage}\\
%\end{comment}
  \caption{Simulated results (Section~\ref{sec4-2}) with respect to angular momentum without the unloading control for the: (a) entire system $^{i} L$, (b) 1st satellite $^{b_1} h_1$, (c) 2nd satellite $^{b_2} h_2$, (d) 3rd satellite $^{b_3} h_3$.}
   \label{sec4:fig8}
\end{center}
\end{figure}

Fig.~\ref{sec4:fig8} shows the formation reconfiguration simulation results for $^iL$ and $^{b_j}h$. Although $^iL$ exhibits a bias owing to the initial random values of $^or$ and $\sigma$, $^{b_j}h$ converges to values that divide the biased periodic $^iL$ into three equal parts related to the number of active RWs. Therefore, a non-uniform distribution of the RW angular momentum does not occur. %Thus, it is shown that the proposed kinematics control law can control more than three satellites without all satellites being equipped with RWs.
It should be noted that the uncontrolled RW angular momentum of the 3rd satellite tends to overshoot, which necessitates the introduction of a saturation prevention function. In addition, compared to formation maintenance, formation reconstruction results in more significant fluctuations in the control input values due to the randomness of the initial values. Therefore, a smaller gain and a more significant AC frequency $\omega_f$ are needed to stabilize the dipole inversion process.
% \begin{figure}[bt!]
%\centering
%\includegraphics[width=.5\textwidth]{simu2-RW-proposed-paperdate.eps}
%\caption{Simulated results for five satellite formation reconfiguration to RW angular momentum:
%(a) entire system $^{i} L$, (b) 1st satellite $^{b_1} h_1$, (c) 2nd satellite $^{b_2} h_2$, and (d) 3rd satellite $^{b_3} h_3$.}
%\label{sec4:fig8}
%\end{figure}
%
\subsection{Formation Maintenance Using Five Satellites with Five Sets of RWs and Single MTQ}
\label{sec4-3}
The numerical calculation described in this section demonstrates that the proposed control law with MTQs mitigates the accumulation of angular momentum in the entire system. 
%As demonstrated in Sec.~\ref{sec4-2}, the proposed control law cannot eliminate the angular momentum accumulated in the system because the EMFF can only output internal force. Combined with only the chief satellite's actuators that can output force in any direction, such as thrusters, the proposed control law theoretically eliminates the accumulation of angular momentum in the entire system. 
%under conditions where the angular momentum of the entire system $\bm L$ accumulates over time.
As shown in Table~\ref{sec4:tableX-1}, the RWs of all satellites are activated and the chief satellite uses MTQs to interact with Earth's magnetic field for the unloading control in a similar manner as monolithic satellites. From the general unloading control law for the chief satellite, the dipole moment of the chief satellite $\bm\mu_{5}$ is expressed as
\begin{equation}
  \label{sec4:eq2-5}
  \begin{aligned}
        &\bm\mu_5(t)=\bm\mu_{(DC)5}+\bm\mu_{(\sin)5}\sin(\omega_{f}t)+\bm\mu_{(\cos)5}\cos(\omega_{f}t),\\
        &\bm\mu_{(DC)5}=\frac{k_{DC}}{\bm B_e \cdot \bm B_e}(\bm h_5 \times \bm B_e),\quad k_{DC}=0.02.\\
  \end{aligned}
\end{equation}
Here, the current to excite MTQs is equivalent to the DC component $\bm\mu_{(DC)5}$ of the electromagnetic coil.
%This accumulation can be theoretically eliminated by the simple unloading control only by the chief satellite using additional actuators. 
%As mentioned in Sec.~\ref{sec1}, disturbances will occur in each satellite if all satellites within the EMFF system generate DC for MTQs simultaneously. Therefore, it is desirable that only a single satellite unloads at a given time. In this case, it is assumed that all five satellites are equipped with RWs. Moreover, only the chief (5th) satellite is equipped with MTQs. The target relative values $^ir_d$ are listed in Table~\ref{sec4:table4}. Initially, $^ir$ and $\sigma$ are assigned the same values as $^ir_d$ and $\sigma_d$. 

\begin{figure}[bt!]
  \begin{center}
    \begin{minipage}{0.25\hsize}
        \centering
        \includegraphics[width=1\textwidth]{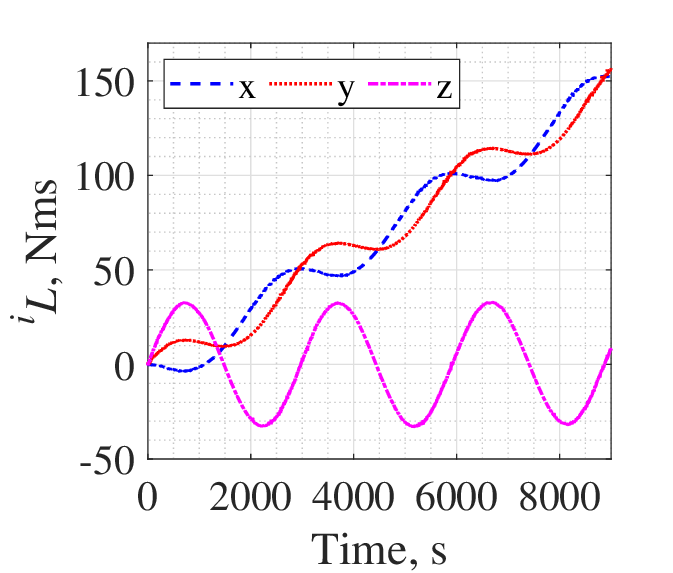}
        \subcaption{The entire system.}
        \label{sec4:fig9_a}
    \end{minipage}
    \begin{minipage}{0.25\hsize}
        \centering
        \includegraphics[width=1\textwidth]{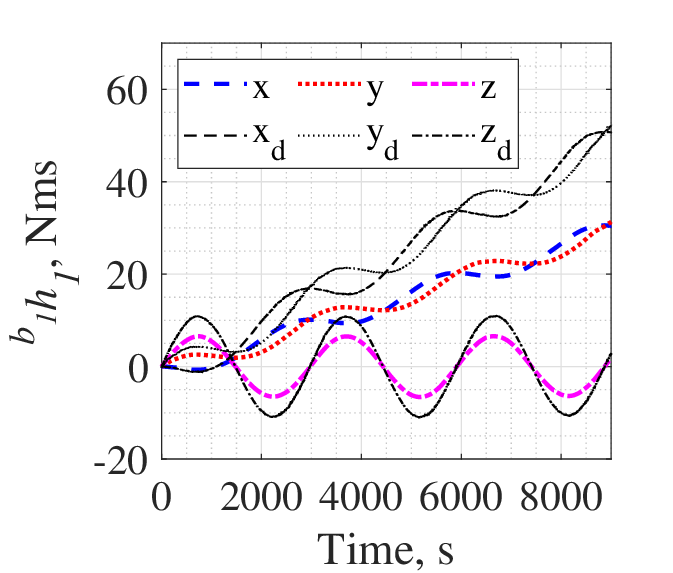}
        \subcaption{RWs of 1st satellite.}
        \label{sec4:fig9_b}
  \end{minipage}
%  \begin{comment}
  \begin{minipage}{0.25\hsize}
        \centering
        \includegraphics[width=1\textwidth]{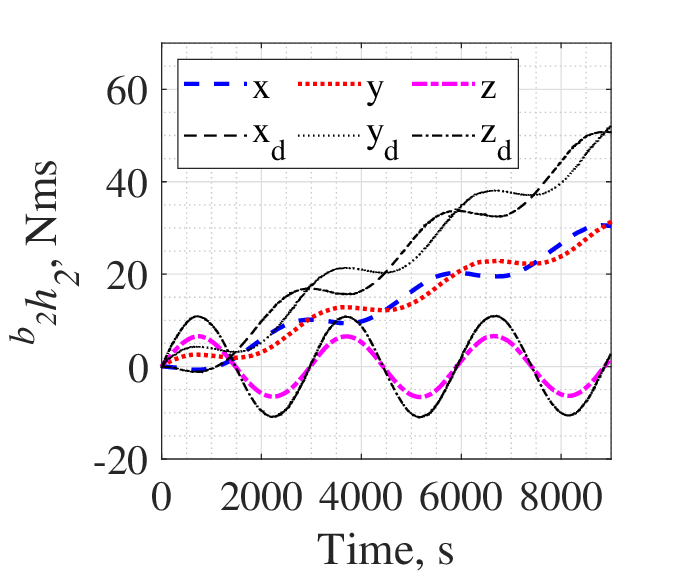}
        \subcaption{RWs of 2nd satellite.}
        \label{sec4:fig9_c}
    \end{minipage}\\
    \begin{minipage}{0.25\hsize}
        \centering
        \includegraphics[width=1\textwidth]{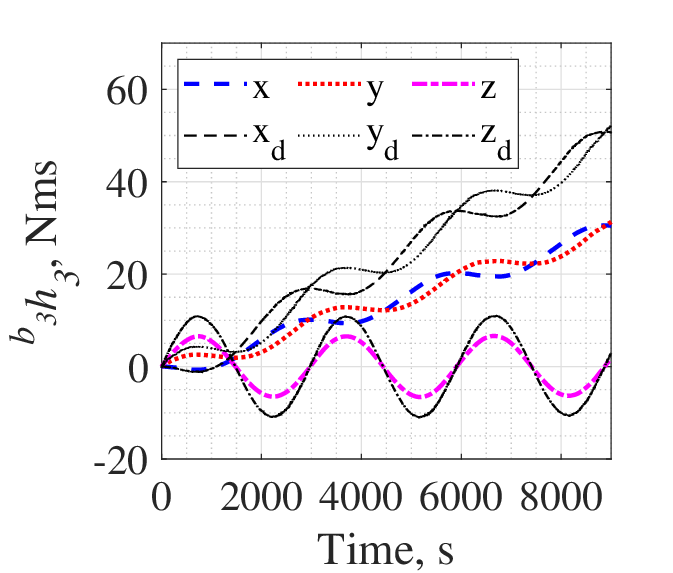}
        \subcaption{RWs of 3rd satellite.}
        \label{sec4:fig9_d}
  \end{minipage}
  \begin{minipage}{0.25\hsize}
        \centering
        \includegraphics[width=1\textwidth]{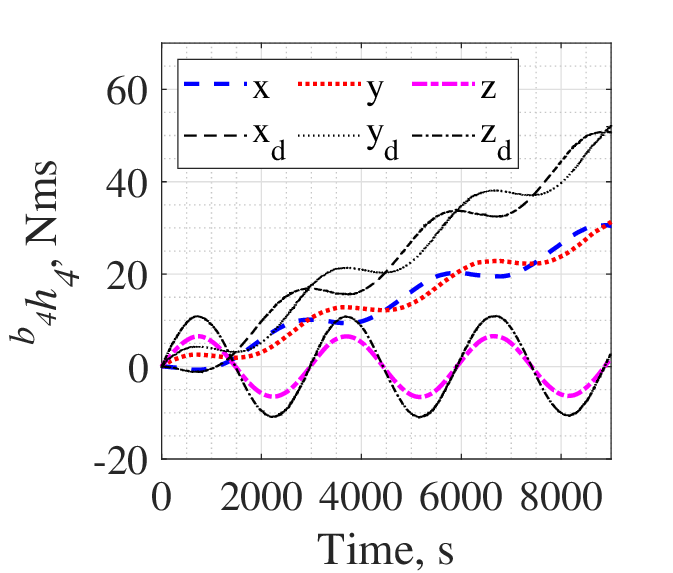}
        \subcaption{RWs of 4th satellite.}
        \label{sec4:fig9_e}
    \end{minipage}
    \begin{minipage}{0.25\hsize}
        \centering
        \includegraphics[width=1\textwidth]{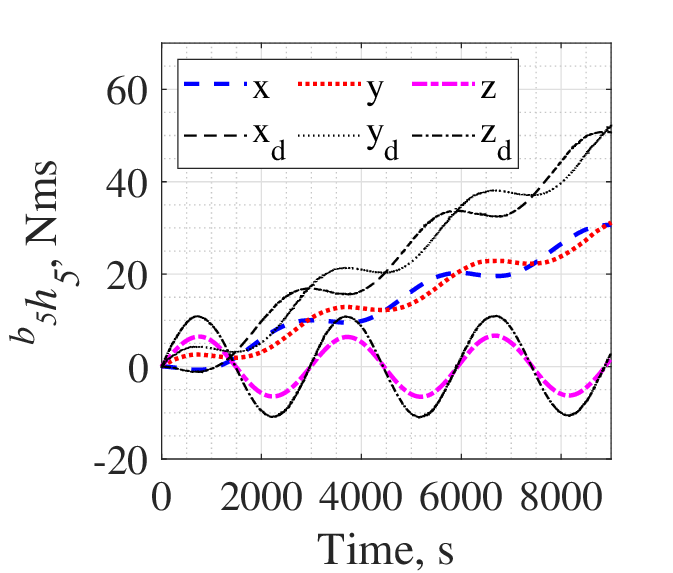}
        \subcaption{RWs of 5th satellite.}
        \label{sec4:fig9_f}
  \end{minipage}\\
%\end{comment}
  \caption{Simulated results (Section~\ref{sec4-3}) with respect to angular momentum with the unloading control for the: (a) entire system $^{i} L$, (b) 1st satellite $^{b_1} h_1$, (c) 2nd satellite $^{b_2} h_2$, (d) 3rd satellite $^{b_3} h_3$, (e) 4th satellite $^{b_4} h_4$, and (f) 5th satellite $^{b_5} h_5$.}
   \label{sec4:fig9}
\end{center}
\end{figure}

\begin{figure}[bt!]
  \begin{center}
    \begin{minipage}{0.25\hsize}
        \centering
        \includegraphics[width=1\textwidth]{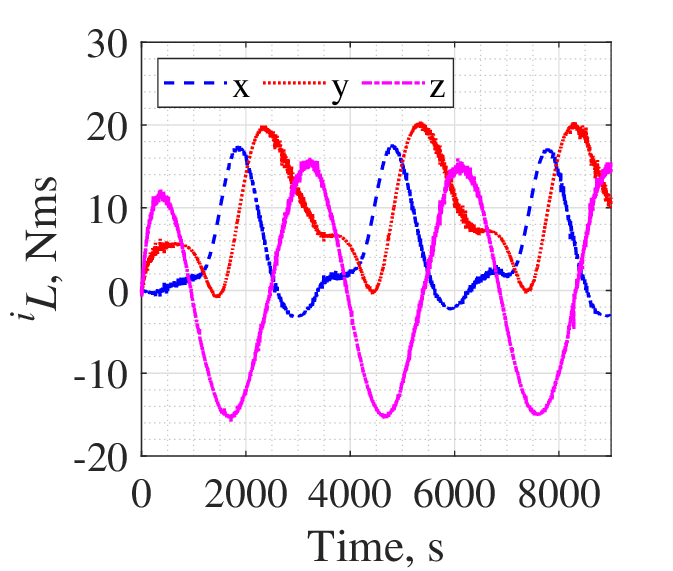}
        \subcaption{The entire system.}
        \label{sec4:fig12_a}
    \end{minipage}
    \begin{minipage}{0.25\hsize}
        \centering
        \includegraphics[width=1\textwidth]{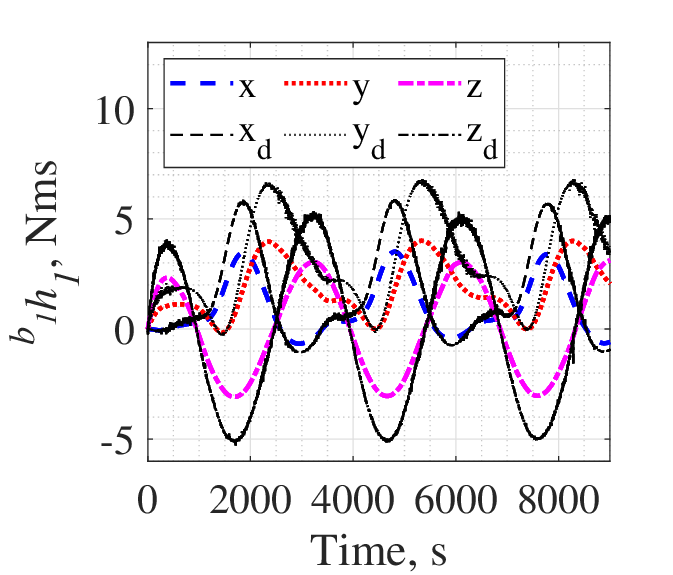}
        \subcaption{RWs of 1st satellite.}
        \label{sec4:fig12_b}
  \end{minipage}
%  \begin{comment}
  \begin{minipage}{0.25\hsize}
        \centering
        \includegraphics[width=1\textwidth]{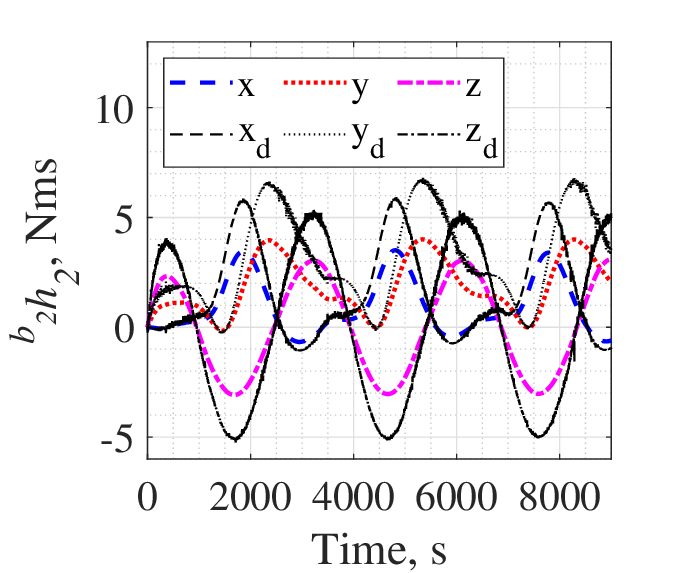}
        \subcaption{RWs of 2nd satellite.}
        \label{sec4:fig12_c}
    \end{minipage}\\
    \begin{minipage}{0.25\hsize}
        \centering
        \includegraphics[width=1\textwidth]{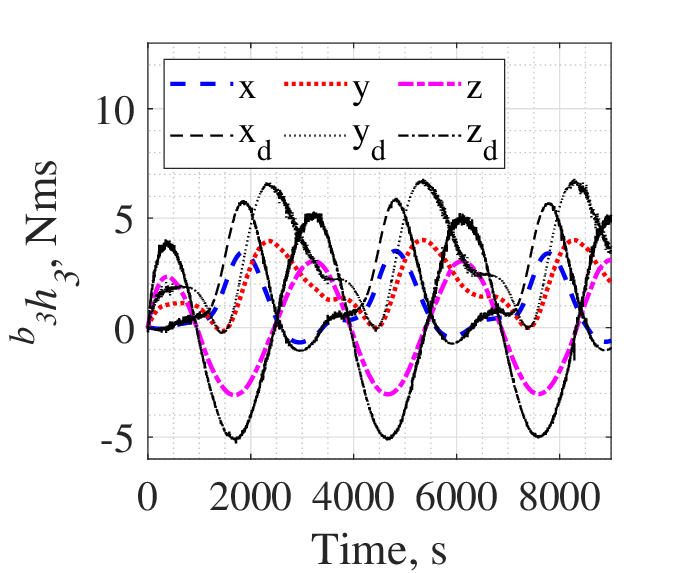}
        \subcaption{RWs of 3rd satellite.}
        \label{sec4:fig12_d}
  \end{minipage}
  \begin{minipage}{0.25\hsize}
        \centering
        \includegraphics[width=1\textwidth]{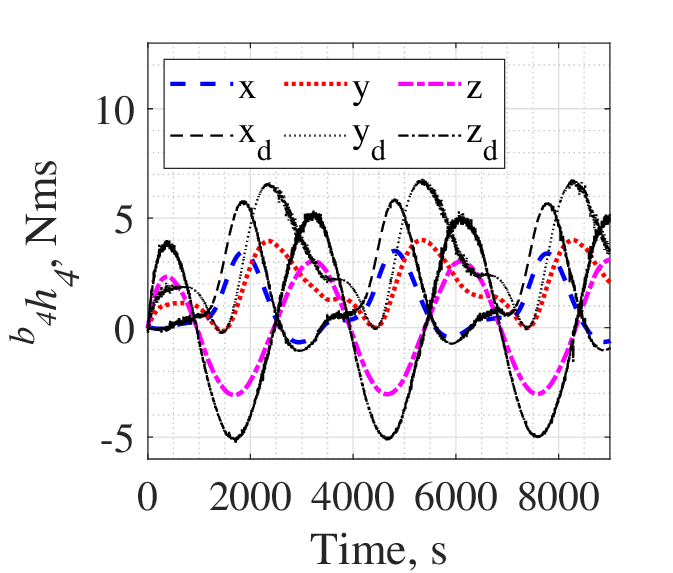}
        \subcaption{RWs of 4th satellite.}
        \label{sec4:fig12_e}
    \end{minipage}
    \begin{minipage}{0.25\hsize}
        \centering
        \includegraphics[width=1\textwidth]{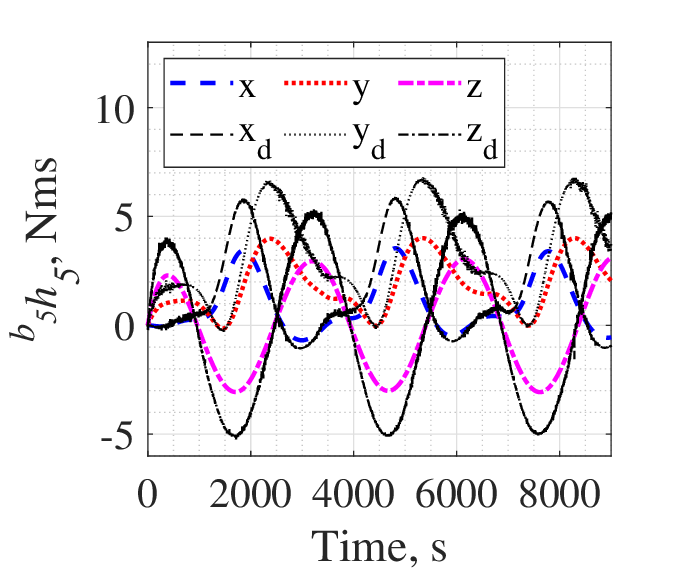}
        \subcaption{RWs of 5th satellite.}
        \label{sec4:fig12_f}
  \end{minipage}\\
%\end{comment}
  \caption{Simulated results (Section~\ref{sec4-3}) with respect to angular momentum for the: (a) entire system $^{i} L$, (b) 1st satellite $^{b_1} h_1$, (c) 2nd satellite $^{b_2} h_2$, (d) 3rd satellite $^{b_3} h_3$, (e) 4th satellite $^{b_4} h_4$, and (f) 5th satellite $^{b_5} h_5$.}
   \label{sec4:fig12}
\end{center}
\end{figure}

First, the results corresponding to the MTQ-based unloading control described in Eqs.~\eqref{sec4:eq2-5} are presented. As the z-axis target values of the 2nd and 3rd satellites are switched, as shown in Table~\ref{sec4:tableX-2}, the torques that make the $z$-axis of the formation symmetric with respect to the Earth are always applied to the system. This leads to the accumulation of $\bm L$ and $\bm h_j$ over time, as shown in Fig.~\ref{sec4:fig9}.
%\begin{table}[bt!]
%\caption{\label{sec4:table4} Simulation conditions.}
%\centering
%\begin{tabular}{ccc}
%\hline\hline
%Num&$^ir_d=[^ir_{dx},^ir_{dy},^ir_{dz}]$&Activated\\\hline%& Attitude$[\sigma_1,\sigma_2,\sigma_3]$\\\hline
%1st&$\left[10\cos(105^\circ),10\sin(105^\circ),-2\right]$& RWs\\%&[0,0,0]\\
%2nd&$\left[10\cos(165^\circ),10\sin(165^\circ),-2\right]$&RWs\\%&[0,0,0]\\
%3rd&$\left[10\cos(285^\circ),10\sin(285^\circ),2\right]$&RWs\\%&[0,0,0]\\
%4th&$\left[10\cos(-15^\circ),10\sin(-15^\circ),2\right]$&RWs\\%&[0,0,0]\\
%5th&$\left[0,0,0\right]$&RWs \& MTQs \\%&[0,0,0]\\
%\hline\hline
%\end{tabular}
%\end{table}
%\begin{figure}[bt!]
%\centering
%\includegraphics[width=.5\textwidth]{simu3-RW-proposed-paperdate.eps}
%\caption{Simulated results for five satellite formation maintenance to RW angular momentum without MTQs:
%(a) entire system $^{i} L$, (b) 1st satellite $^{b_1} h_1$, (c) 2nd satellite $^{b_2} h_2$, (d) 3rd satellite $^{b_3} h_3$, (e) 4th satellite $^{b_4} h_4$, and (f) 5th satellite $^{b_5} h_5$.}
%\label{sec4:fig9}
%\end{figure}
In contrast, Fig.~\ref{sec4:fig12} shows that the combination of the proposed controller with a simple unloading control suppresses the increases in $\bm L$ and $\bm h_j$, and $\bm h_j$ is uniform. It should be noted that MTQ-based unloading control can only mitigate the angular momentum of the system because the electromagnetic torque generated via the Earth's magnetic field cannot be output in a specific direction. For the same reason, the unloading control is not considered in the model; RW angular momentums do not converge asymptotically to the target values. By using the actuators that can output force in any direction, such as thrusters, these accumulations can be theoretically eliminated.
%\begin{figure}[bt!]
%\centering
%\includegraphics[width=.5\textwidth]{simu3-2-RW-proposed-paperdate.eps}
%\caption{Simulated results for five satellite formation maintenance to RW angular momentum with MTQs for the 5th satellite:
%(a) entire system $^{i} L$, (b) 1st satellite $^{b_1} h_1$, (c) 2nd satellite $^{b_2} h_2$, (d) 3rd satellite $^{b_3} h_3$, (e) 4th satellite $^{b_4} h_4$, and (f) 5th satellite $^{b_5} h_5$.}
%\label{sec4:fig12}
%\end{figure}

\begin{figure}[bt!]
  \begin{center}
    \begin{minipage}{0.25\hsize}
        \centering
        \includegraphics[width=.7\textwidth]{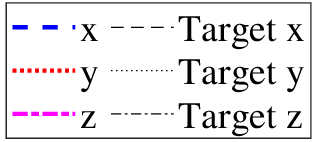}
        \subcaption{Legend.}
        \label{sec4:fig10_a}
    \end{minipage}
    \begin{minipage}{0.25\hsize}
        \centering
        \includegraphics[width=1\textwidth]{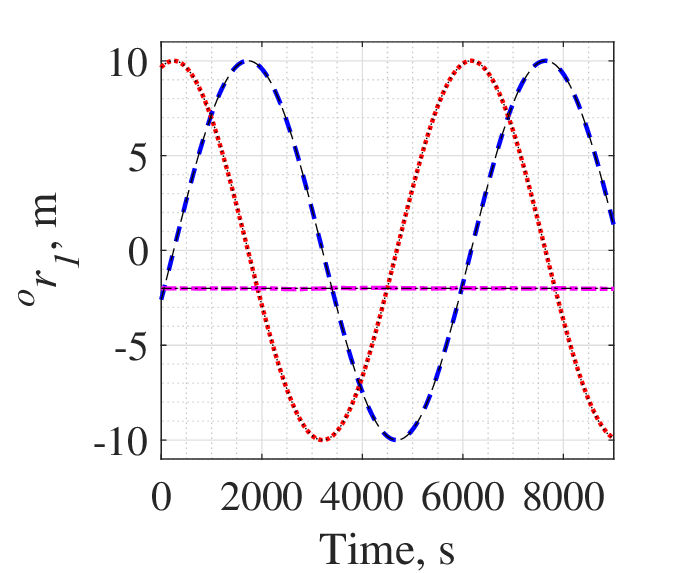}
        \subcaption{1st satellite.}
        \label{sec4:fig10_b}
  \end{minipage}
  \begin{minipage}{0.25\hsize}
        \centering
        \includegraphics[width=1\textwidth]{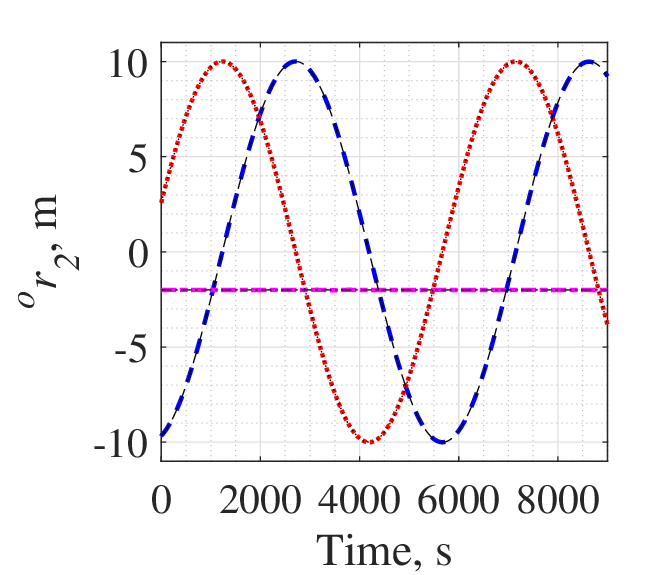}
        \subcaption{2nd satellite.}
        \label{sec4:fig10_c}
    \end{minipage}\\
    \begin{minipage}{0.25\hsize}
        \centering
        \includegraphics[width=1\textwidth]{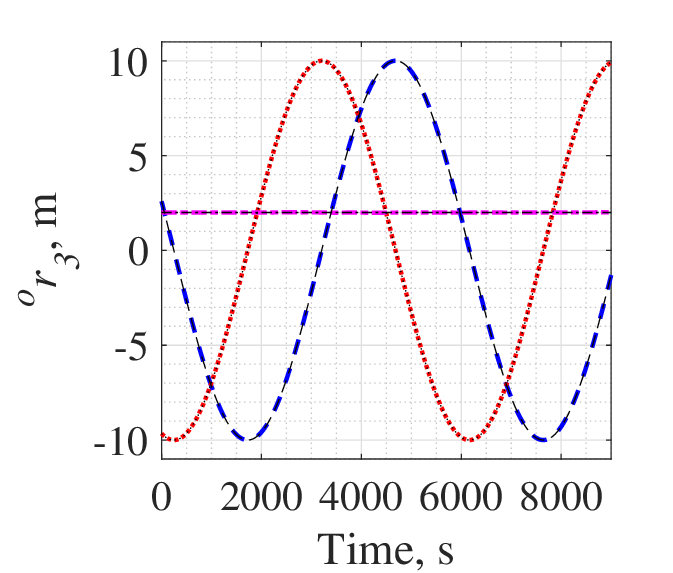}
        \subcaption{3rd satellite.}
        \label{sec4:fig10_d}
  \end{minipage}
  \begin{minipage}{0.25\hsize}
        \centering
        \includegraphics[width=1\textwidth]{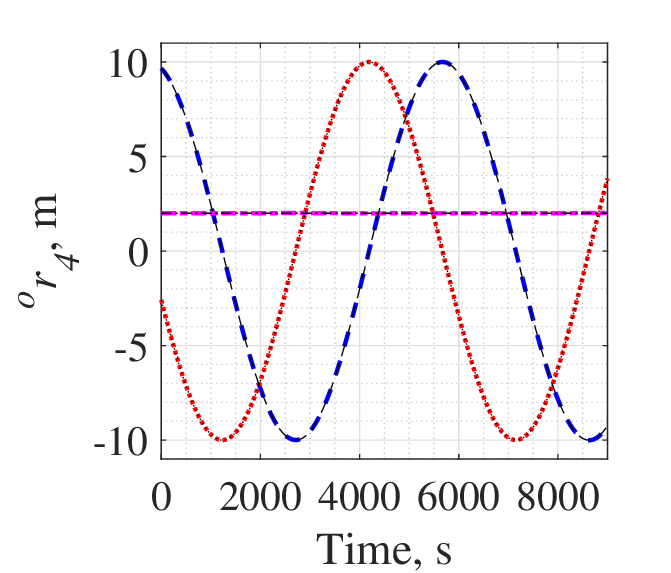}
        \subcaption{4th satellite.}
        \label{sec4:fig10_e}
    \end{minipage}
  \begin{minipage}{0.25\hsize}
        \centering
        \includegraphics[width=1\textwidth]{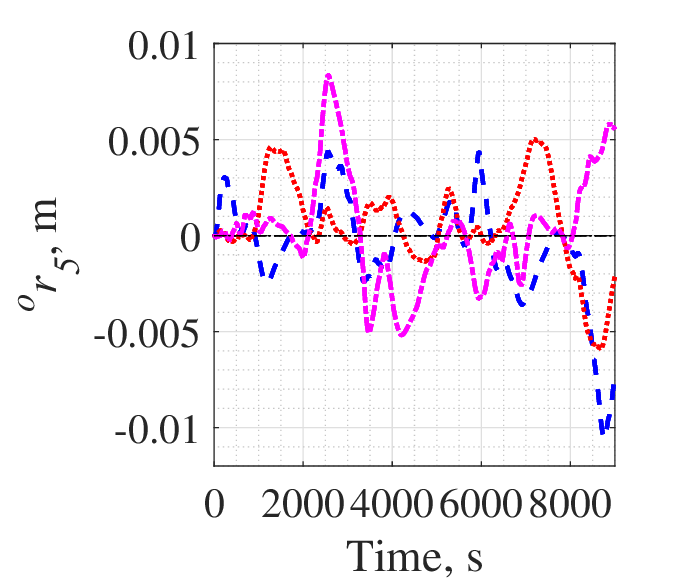}
        \subcaption{5th satellite.}
        \label{sec4:fig10_f}
    \end{minipage}\\
  \caption{Simulated results (Section~\ref{sec4-3}) with respect to the relative position for the: (a) 1st satellite $^or_1$, (b) 2nd satellite $^or_2$, (c) 3rd satellite $^or_3$, (d) 4th satellite $^or_4$, and (e) 5th satellite $^or_5$.}
   \label{sec4:fig10}
\end{center}
\end{figure}

\begin{figure}[bt!]
  \begin{center}
    \begin{minipage}{0.25\hsize}
        \centering
        \includegraphics[width=0.7\textwidth]{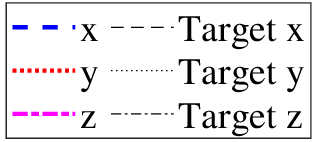}
        \subcaption{Legend.}
        \label{sec4:fig11_a}
    \end{minipage}
    \begin{minipage}{0.25\hsize}
        \centering
        \includegraphics[width=1\textwidth]{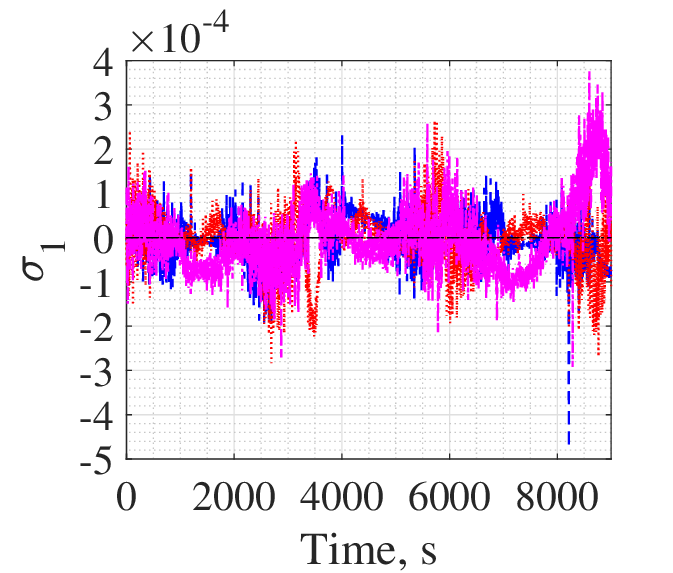}
        \subcaption{1st satellite.}
        \label{sec4:fig11_b}
  \end{minipage}
  \begin{minipage}{0.25\hsize}
        \centering
        \includegraphics[width=1\textwidth]{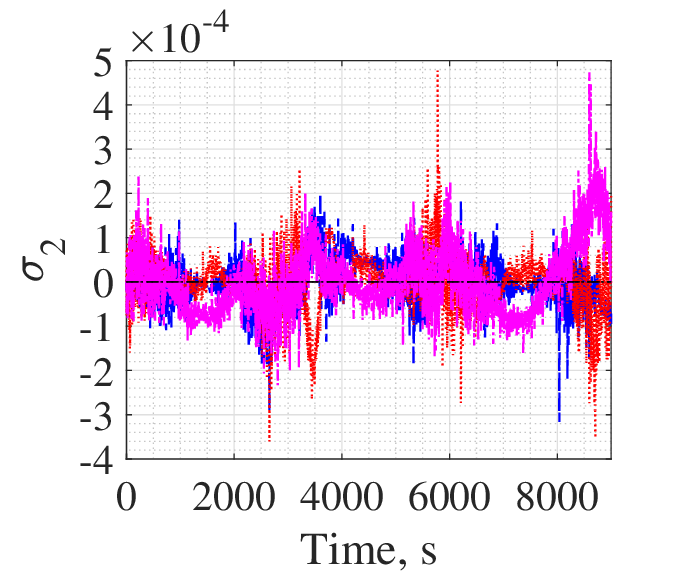}
        \subcaption{2nd satellite.}
        \label{sec4:fig11_c}
    \end{minipage}\\
    \begin{minipage}{0.25\hsize}
        \centering
        \includegraphics[width=1\textwidth]{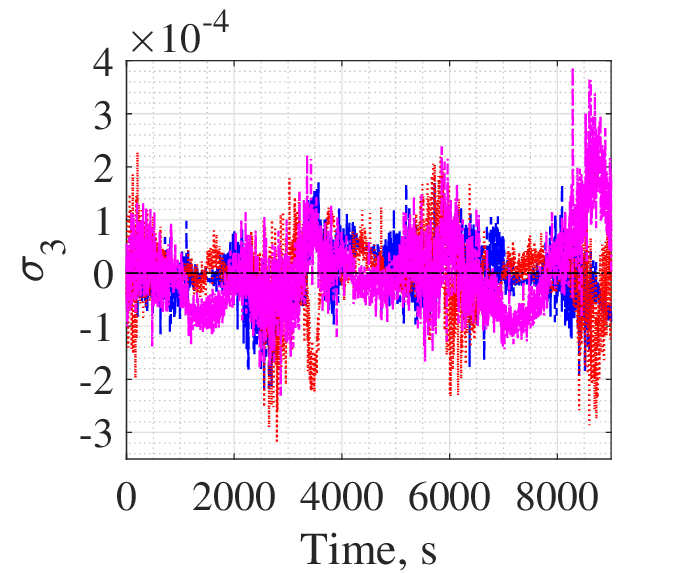}
        \subcaption{3rd satellite.}
        \label{sec4:fig11_d}
  \end{minipage}
  \begin{minipage}{0.25\hsize}
        \centering
        \includegraphics[width=1\textwidth]{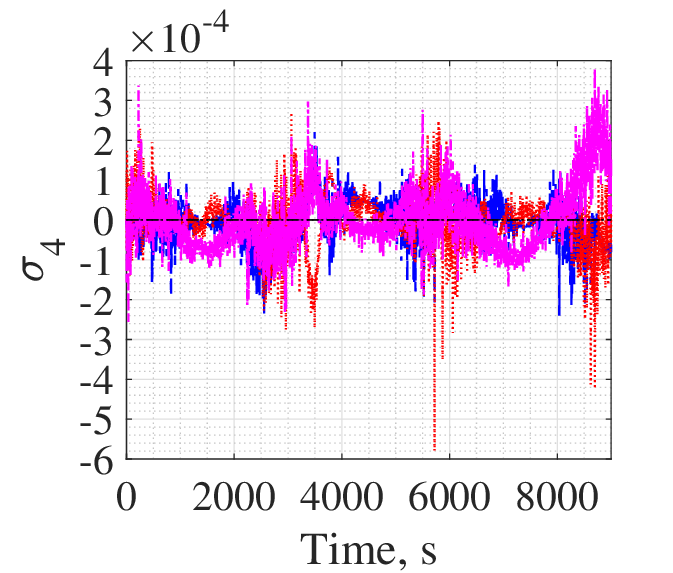}
        \subcaption{4th satellite.}
        \label{sec4:fig11_e}
    \end{minipage}
  \begin{minipage}{0.25\hsize}
        \centering
        \includegraphics[width=1\textwidth]{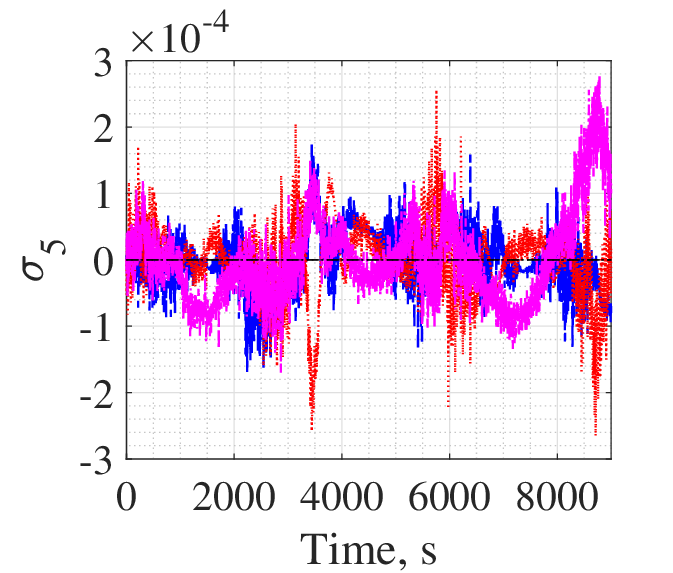}
        \subcaption{5th satellite.}
        \label{sec4:fig11_f}
    \end{minipage}\\
  \caption{Simulated results (Section~\ref{sec4-3}) with respect to the absolute attitude for the: (a) 1st satellite $\sigma_1$, (b) 2nd satellite $\sigma_2$, (c) 3rd satellite $\sigma_3$, (d) 4th satellite $\sigma_4$, and (e) 5th satellite $\sigma_5$.}
   \label{sec4:fig11}
\end{center}
\end{figure}

As illustrated in Figs.~\ref{sec4:fig10} and~\ref{sec4:fig11}, the satellite system maintains the shape and attitude of each satellite.
Although the coupling between AC-based control and the Earth's magnetic field is smaller than the control, the disturbance affects the control performance and is comparable to to the results presented in Section~\ref{sec4-1}. Furthermore, the oscillation term in Eq.~\eqref{sec2:eq14-2} of the AC EMFF control also contributes to the decrease in the control performance. Therefore, the AC frequency $\omega_f$ must also be considered in the control system's design to improve the control performance.
%This is attributed to the proposed control law approximating the electromagnetic torque as the control torque during the period $T$, as expressed by Eq.~\eqref{sec2:eq15}. %Therefore, the satellites do not cope well with the small disturbance torques caused by Earth's magnetic field compared with the attitude control, which was provided solely by RWs in previous research \cite{Ayyad}.

%\begin{figure}[bt!]
%\centering
%\includegraphics[width=.5\textwidth]{simu3-position-proposed-paperdate.eps}
%\caption{Simulated results for five satellite formation maintenance to relative position: 
%(a) 1st satellite $^or_1$, (b) 2nd satellite $^or_2$, (c) 3rd satellite $^or_3$, (d) 4th satellite $^or_4$, and (e) 5th satellite $^or_5$.}
%\label{sec4:fig10}
%\end{figure}
%\begin{figure}[bt!]
%\centering
%\includegraphics[width=.5\textwidth]{simu3-attitude-proposed-paperdate.eps}
%\caption{Simulated results for five satellite formation maintenance to attitude: 
%(a) 1st satellite $\sigma_1$, (b) 2nd satellite $\sigma_2$, (c) 3rd satellite $\sigma_3$, (d) 4th satellite $\sigma_4$, and (e) 5th satellite $\sigma_5$.}
%\label{sec4:fig11}
%\end{figure}
\section{Conclusion}
\label{sec5}
%This paper presents a new kinematics control law for more flexible and functional future space missions by the electromagnetic formation flight (EMFF) system. This control law theoretically realizes a completely uniform distribution of reaction wheel loaded angular momentum among the satellites. Combined with a simple unloading control governed by the chief satellite, the proposed control law theoretically eliminates the accumulation of angular momentum for the entire system. Moreover, this control law enables the relative position, absolute attitude, and reaction wheel loaded angular momentum to be controlled for $n$ satellites while using only one set of 3-axis reaction wheels among all satellites. The extensibility of the alternating current method for outputting an arbitrary electromagnetic force and torque is shown via the polynomial representation of the alternating current method. The reaction wheel conditions for smooth feedback stabilizability and the control target for avoiding the non-uniform distribution of angular momentum were summarized from the viewpoint of nonholonomic mechanical systems. Then, the kinematics of EMFF were derived from a nonholonomic system constraint. A kinematics controller capable of controlling the electromagnetic force and torque was designed using EMFF kinematics. Finally, three numerical calculations were performed for a system comprising five satellites, which demonstrate the validity of the proposed kinematics controller. In addition, remaining problems associated with the proposed controller are identified for future studies.
This study presented a new kinematics control law based on the electromagnetic formation flight (EMFF) system to increase the flexibility and functionality of future space missions. This control law theoretically realizes a completely uniform distribution of reaction wheel (RW) loaded angular momentum among the satellites. Combined with a simple unloading control governed by the chief satellite, the proposed control law theoretically eliminates the accumulation of angular momentum for the entire system. Moreover, it enables the control of the relative position, absolute attitude, and RW-loaded angular momentum of $n$ satellites while using only one set of 3-axis reaction wheels for all satellites. The extensibility of the alternating current method for outputting an arbitrary electromagnetic force and torque was shown via the polynomial representation of the alternating current method. The RW conditions required for smooth feedback stabilizability and the control target for avoiding the non-uniform distribution of RW angular momentum were summarized from the viewpoint of nonholonomic mechanical systems. Then, the kinematics of EMFF and averaged dynamics for ``Redundant EMFF'' were derived from a nonholonomic system constraint. A kinematics controller capable of controlling the electromagnetic force and torque was designed simultaneously using EMFF kinematics. Finally, three numerical calculations were performed for a system comprising five satellites, which demonstrates the validity of the proposed kinematics controller.
\section*{Appendix}
\subsection{\textit{Proof of Theorem 1}}
\textbf{Proof.} The matrix $R$ holds the relation of
 \begin{equation}
  \label{sec3:eq11-5}
  \begin{aligned}
   R[B]^{-1}[M]S=AS=0 \ (\because S \in Null\ Space(A)).
    \end{aligned}
\end{equation}
Based on Eqs.~\eqref{sec3:eq11-4-2} and \eqref{sec3:eq11-5}, $^i\dot{L}$ caused by the proposed control law is always 0, as shown in Eq.~\eqref{sec3:eq11-6}.
  \begin{equation}
  \label{sec3:eq11-6}
    {^{i}\dot{L}}=R(u_c+u_d)=Ru_d.
\end{equation}
Thus, Theorem 1 is proved.
\subsection{\textit{Proof of Theorem 2}}
%\subsection{Proof of Theorem 2 in Sec.~\ref{sec3-4}}
%\label{secA-2}
\textbf{Proof.} First, it is shown that EMFF system can output the values of $u_c$. However, it is necessary that ${\overline{B}_r^{-1}}=[B]^{-1}[M]S(S^{\mathrm{T}}[M]S)^{-1}$ is always defined in any states. This can be confirmed by showing that $S^{\mathrm{T}}[M]S$ has full rank. $S^{\mathrm{T}}[M]S$ can be decomposed as
\begin{equation}
  \label{sec3:eq11-X2}
%A_s^{\mathrm{T}}A_s=PDP^{-1},
S^{\mathrm{T}}[M]S=[\hat{M}]+{A}_s^{\mathrm{T}}{A}_s=[\hat{M}]+\hat{A}_s^{\mathrm{T}}\hat{A}_s,
\end{equation} 
where $[\hat{M}]$ is a $(6n+3m-6)\times (6n+3m-6)$ positive definite matrix and $\hat{A}_s$ is the square matrix $\hat{A}_s$ as follows:
\begin{equation}
  [\hat{M}]=
  \begin{bmatrix}
    [M_p]&&0\\
    &[M_a]&\\
    0&&E_{3m-3}
  \end{bmatrix},\qquad \hat{A}_s=
  \begin{bmatrix}
    A_s\\
   0_{((6n+3m-9)\times (6n+3m-6))}
  \end{bmatrix}.
\end{equation}
In this case, ${A}_s^{\mathrm{T}}{A}_s$ is a semi-positive definite matrix, because ${A}_s^{\mathrm{T}}{A}_s$ can be represented as $\hat{A}_s^{\mathrm{T}}\hat{A}_s$ using the square matrix. Therefore, $S^{\mathrm{T}}[M]S$ is a positive definite matrix and is always regular. Combining this fact with Theorem 1 shows that the EMFF system can output the values of the control law $u_c$.

Next, we consider the following Lyapunov function candidate $V(\delta v,\delta q_s)$:
\begin{equation}
  \label{sec3:eq12}
  \begin{aligned}
    V(\delta q_s,\delta v)&=V_{s1}+V_{s2}=\left(\frac{1}{2}\delta v^{\mathrm{T}}\ \overline{M}(q_s)\delta v
    \right)%+\frac{1}{2}(q_s-q_{sd})^{\mathrm{T}}K_1(q_s-q_{sd}),
    +\left(\frac{1}{2}\delta{^i{r}}^{\mathrm{T}}K_1\delta{^i{r}}+2K_a\sum_{j=1}^n{\mathrm{In}(1+\delta\sigma_j^{\mathrm{T}}\delta\sigma_j)}\right),
  \end{aligned}
\end{equation}
where $V(\delta v,\delta q_s)$ is a scalar function 
with continuous first partial derivatives 
and $V(\delta v,\delta q_s) \rightarrow \infty$ as:
%$\left|\left|\begin{bmatrix}
%   v^{\mathrm{T}},\ q_s^{\mathrm{T}}
%    \end{bmatrix}^{\mathrm{T}}\right|\right| \rightarrow \infty$. 
$||\delta v||,||\delta q_s|| \rightarrow \infty$. 
Moreover, $V \geq 0$ and $V=0$ if and only if $\delta v=0$ and $\delta q_s=0$. Note that the state $q_s$ contain all of the relative positions $^ir$ and absolute attitudes $\sigma$ of the n satellites when $m$ satellites (from a total of $n$) are equipped with RWs $(1 \leq m \leq n)$. Taking the derivative of $\ V_{s1}$, and using Eqs.~\eqref{sec3:eq10-X} and \eqref{sec3:eq11}, we obtain
\begin{equation}
  \label{sec3:eq13-X1}
  \begin{aligned}
    \dot{V}_{s1}(\delta q_s,\delta v)=&\frac{1}{2}\delta v^{\mathrm{T}}\ \dot{\overline{M}}\delta v+{\delta v^{\mathrm{T}}}\ \overline{M}\delta\dot{v}
           %=&(v-v_d)^{\mathrm{T}}\left(-\overline{C}v+\overline{B}(q_s)u_c+\overline{u}_d-\overline{M}\dot{v}_d\right)\\
           %&+\frac{1}{2}(v-v_d)^{\mathrm{T}}\ \dot{\overline{M}}(v-v_d)\ (\because Eq.~\eqref{sec3:eq10}) \\
           %=&\frac{1}{2}(v-v_d)^{\mathrm{T}}\left(\dot{\overline{M}}-2\overline{C}\right)(v-v_d)-(q_s-q_{sd})^{\mathrm{T}}K_1(\dot{q}_s-\dot{q}_{sd})-(v-v_d)^{\mathrm{T}}K_2(v-v_d)\\%\ (\because Eqs.~\eqref{sec3:eq10-X}\ \mathrm{and}\ \eqref{sec3:eq11})\\
           =-\delta q_s^{\mathrm{T}}K_1\delta\dot{q}_s-\delta v^{\mathrm{T}}K_2\delta v.\\
           %=&-(v-v_d)^{\mathrm{T}}K_2(v-v_d)+\frac{1}{2}(v-v_d)^{\mathrm{T}}(\dot{\overline{M}}-2\overline{C})(v-v_d)\\
          % =&-(v-v_d)^{\mathrm{T}}K_2(v-v_d).
          \end{aligned}
\end{equation}
Furthermore, the derivative of $\ V_{s2}$ \cite{Tsiotras} is
\begin{equation}
  \label{sec3:eq13-X12}
  \begin{aligned}
    \dot{V}_{s2}(\delta q_s,\delta v)=\delta{^i{r}}^{\mathrm{T}}K_1\delta{^i\dot{r}}+\delta\sigma^{\mathrm{T}}K_a{{^{b}\delta\omega}}.\\
          \end{aligned}
\end{equation}
From Eqs.~\eqref{sec3:eq13-X1} and \eqref{sec3:eq13-X12}, the time derivative of $V(\delta q_s,\delta v)$ is
\begin{equation}
  \label{sec3:eq13}
  \begin{aligned}
    \dot{V}(\delta q_s,\delta v)=&\dot{V}_{s1}(\delta q_s,\delta v)+\dot{V}_{s2}(\delta q_s,\delta v)
           =-\delta v^{\mathrm{T}}K_2\delta v.
          \end{aligned}
\end{equation}
Clearly, $\dot{V}(\delta q_s,\delta v)<0$ as long as $\delta v\neq 0$, so that manifold
\begin{equation}
   \label{sec3:eq13-X2}
\bm{R}=\left\{(\delta {q}_{s},\delta {v},{^{b_m}\delta \xi_m}) \mid \left(\overline{M}(q_s)\delta\dot{v}+\overline{C}(q_s,v)\delta v=-K_1\delta q_s-K_2\delta v\right),\ \delta {v}=0\right\}
\end{equation}
satisfying $V(\delta q_s,\delta v)=0$ is a globally asymptotically stable equilibrium manifold by combining Eqs.~\eqref{sec3:eq10} and \eqref{sec3:eq11}. Then, the global invariant set theorem \cite{Slotine} concludes that all solutions globally asymptotically converge to a three-dimensional equilibrium manifold
\begin{equation}
   \label{sec3:eq15-X3}
\bm{M}=\left\{(\delta {q}_{s},\delta {v},{^{b_m}\delta \xi_m}) \mid \delta q_s=0,\ \delta v=0\right\}
\end{equation}
that is the largest invariant set in $\bm R$. 
Subsequently, since EMFF holds the conservation of angular momentum in the entire system, i.e., the kinematics of the EMFF system in Eq.~\eqref{sec3:eq7} is established.
\begin{equation}
   \lim_{t \to \infty} {^{b_m}\xi_m}=-C^{B_m/I}A_s(q_{sd})v_d \Leftrightarrow \lim_{t \to \infty}{^{b_m}\xi_m}={^{b_m}\xi_{md}}.
%\bm{M}_e=\{({q}_{s},{v},{^{b_m}\xi_m}) \mid  {q}_{s}={q}_{sd},{v}={v}_d,^{b_m}\xi_m=^{b_m}\xi_{md})\}
\end{equation} 
In conclusion, the states converge to the target values $q_{sd}$ and $\zeta_d$ asymptotically as $t \rightarrow \infty$. Thus, Theorem 2 is proved.
%\input{M1_1_appendix.tex}
%\section*{Acknowledgments}
%An Acknowledgments section, if used, \textbf{immediately precedes} the References. Individuals other than the authors who contributed to the underlying research may be acknowledged in this section. The use of special facilities and other resources also may be acknowledged. 
%\bibliography{sample}
%\begin{comment}
%\bibliography{./library}

\begin{thebibliography}{99}
  %\bibitem{Scharf}Scharf, D. P., Hadaegh, F. Y., and Ploen, S. R., ``A Survey of Spacecraft Formation Flying Guidance and Control (Part 1): Guidance,'' \textit{Proceedings of the 2003 American Control Conference}, Denver, CO, USA, 2003, pp. 1733-1739. https://doi.org/10.1109/ACC.2003.1239845.
  %\bibitem{Scharf2}Scharf, D. P., Hadaegh, F. Y., and Ploen, S. R., ``A Survey of Spacecraft Formation Flying Guidance and Control. Part II: Control,'' \textit{Proceedings of the 2004 American Control Conference}, Vol.4, IEEE, Boston, MA, USA, 2004, pp. 2976-2985.  https://doi.org/10.23919/ACC.2004.1384365.
  %\bibitem{Alfriend}Alfriend, K., Vadali, S. R., Gurfil, P., How, J., and Breger, L., \textit{Spacecraft Formation Flying: Dynamics, Control and Navigation}, Vol. 2, Butterworth–Heinemann, Amsterdam, 2009.
  \bibitem{Skinner}Skinner, G. K., Dennis, B. R., Krizmanic, J. F., and Kontar, E. P., ``Science Enabled by High Precision Inertial Formation Flying,'' \textit{International Journal of Space Science and Engineering}, Vol. 1, No. 4, 2013, pp. 331-348. https://doi.org/10.1504/IJSPACESE.2013.059271.
  %\bibitem{Aung}Aung, M., Ahmed, A., Wette, M., Scharf, D., Tien, J., Purcell, G., and Regehr, M., ``An Overview of Formation Flying Technology Development for the Terrestrial Planet Finder Mission,'' \textit{Proceedings of the IEEE Aerospace Conference}, IEEE Publ., Piscataway, NJ, 2004, pp. 2667-2679. https://doi.org/10.1109/AERO.2004.1368062.
  \bibitem{Quadrelli1}Quadrelli, B.M. ``Modeling and Dynamics Analysis of Tethered Formations for Space Interferometry,'' 11th AAS/AIAA Space Flight Mechanics Meeting, Santa Barbara, California, USA, February 11-14, 2001. http://hdl.handle.net/2014/16379. 
  %\bibitem{Quadrelli2}Quadrelli, B.M. ``Dynamics and Control of Novel Orbiting Formations with Internal Dynamics,'' in The Journal of the Astronautical Sciences, vol.51, no. 3, July-September 2003, pp. 319-337. https://doi.org/10.1007/BF03546301.
  \bibitem{Lawson}Lawson, P. and Dooley, J., ``Technology Plan for the Terrestrial Planet Finder Interferometer,'' Tech. Rep. 05-5, NASA Jet Propulsion Lab, June 2005. http://hdl.handle.net/2014/37428.
  \bibitem{Arya}Arya, M., Lee, N., and Pellegrino, S., ``Ultralight Structures for Space Solar Power Satellites,'' \textit{3rd AIAA Spacecraft Structures Conference}, AIAA Paper, 2016. https://doi.org/10.2514/6.2016-1950.
  \bibitem{Quadrelli}Quadrelli, M. B., Hodges, R., Vilnrotter, V., Bandyopadhyay, S., Tassi, F., and Bevilacqua, S., ``Distributed Swarm Antenna Arrays for Deep Space Applications,'' \textit{2019 IEEE Aerospace Conference}, Big Sky, MT, USA, 2019, pp. 1-15. https://doi.org/10.1109/AERO.2019.8742019.
  %\bibitem{Kapila}Kapila, V., Sparks, A. G, Buffington, J. M., and Yan, Q., ``Spacecraft Formation Flying: Dynamics and Control,'' \textit{Journal of Guidance, Control, and Dynamics}, Vol. 23, No. 3, 2000, pp. 561-564. https://doi.org/10.2514/2.4567.
  %\bibitem{Foust}Foust, R. C., Nakka, Y. K., Saxena, A., Chung, S., and Hadaegh, F., ``Automated Rendezvous and Docking Using Tethered Formation Flight,'' \textit{9th International Workshop on Satellite Constellations and Formation Flying}, Boulder, Colorado, USA, 2017. https://resolver.caltech.edu/CaltechAUTHORS:20170630-100154807.
  %\bibitem{Xu}Xu, G. and Wang, D., ``Nonlinear Dynamic Equations of Satellite Relative Motion Around an Oblate Earth,'' \textit{Journal of Guidance, Control, and Dynamics}, Vol. 31, No. 5, 2008, pp. 1521-1524. https://doi.org/10.2514/1.33616.
  \bibitem{Morgan2}Morgan, D., Chung, S.-J., Blackmore, L., Acikmese, B., Bayard, D., and Hadaegh, F. Y., ``Swarm-Keeping Strategies for Spacecraft under J2 and Atmospheric Drag Perturbations,'' \textit{Journal of Guidance, Control, and Dynamics}, Vol. 35, No. 5, 2012, pp. 1492-1506. https://doi.org/10.2514/1.55705.
  %\bibitem{Kristiansen}Kristiansen. R., and Nicklasson,P. J., ``Spacecraft Formation Flying: A Review and New Results on State Feedback Control,'' \textit{Acta Astronautica}, Vol. 65, 2009, pp. 1537-1552. https://doi.org/10.1016/j.actaastro.2009.04.014.
  \bibitem{Kwon}Kwon, D. W., Sedwick, R. J., Lee, S.I., and Riberos, J. R., ``Electromagnetic Formation Flight Testbed Using Superconducting Coils,'' \textit{Journal of Spacecraft and Rockets}, Vol. 48, No. 1, 2011, pp. 124-134. https://doi.org/10.2514/1.45017.
  \bibitem{Sakaguchi}Sakaguchi, A., “Micro-Electromagnetic Formation Flight of Satellite Systems,'' Ph.D. Thesis, Massachusetts Institute of Technology, Cambridge, MA, 2007, https://dspace.mit.edu/handle/1721.1/39708.
  %\bibitem{Ahsun2}Ahsun, U., Miller, D. W., and Ramirez, J. L., ``Control of Electromagnetic Satellite Formations in Near-Earth Orbits,'' \textit{Journal of Guidance, Control, and Dynamics}, Vol. 33, No. 6, 2010, pp. 1883-1891. https://doi.org/10.2514/1.47637.
  \bibitem{Ahsun}Ahsun, U., ``Dynamics and Control of Electromagnetic Satellite Formations,'' Ph.D. Thesis, Massachusetts Institute of Technology, 2007, https://dspace.mit.edu/handle/1721.1/40894.
  %\bibitem{Huang}Huang, H., Zhu, Y. W., Yang, L. P., and Zhang, Y. W., ``Stability and Shape Analysis of Relative Equilibrium for Three-Spacecraft Electromagnetic Formation,'' \textit{Acta Astronautica}, Vol. 94, No. 1, 2014, pp. 116-131. https://doi.org/10.1016/j.actaastro.2013.08.011.
  \bibitem{Elias}Elias, L., Kwon, D., Sedwick, R., and Miller, D., ``Electromagnetic Formation Flight Dynamics Including Reaction Wheel Gyroscopic Stiffening Effects,'' \textit{Journal of Guidance, Control, and Dynamics}, Vol. 30, No. 2, 2007, pp. 499-511. https://doi.org/10.2514/1.18679.
  \bibitem{Fabacher}Fabacher, E., Lizy-Destrez, S., Alazard, D., Ankersen, F., and Profizi, A., ``Guidance of Magnetic Space Tug,'' \textit{Advances in Space Research}, Vol. 60, 2017, pp. 14-27. https://doi.org/10.1016/j.asr.2017.03.042.
  \bibitem{Schweighart}Schweighart, S. A., ``Electromagnetic Formation Flight Dipole Solution Planning,'' Ph.D. Thesis, Massachusetts Institute of Technology, Cambridge, MA, 2005, https://dspace.mit.edu/handle/1721.1/32464.
  \bibitem{Fan}Fan, L., Hu, M., and Yang, M., ``Rotational and Relative Translational Control for Satellite Electromagnetic Formation Flying in Low Earth Orbit,'' \textit{Aircraft Engineering and Aerospace Technology}, Vol. 89, No. 6, 2017, pp. 815-825. https://doi.org/10.1108/AEAT-01-2016-0007.
  %\bibitem{Buck}Buck. A., ``Path Planning and Position Control and of an Underactuated Electromagnetic Formation Flight Satellite System in the Near Field,'' Master's Thesis, Massachusetts Institute of Technology, 2013, https://dspace.mit.edu/handle/1721.1/82502.
  %\bibitem{Zhendong}Zhendong, H., Jinxiu, Z., and Xibin, C., ``Distributed Electromagnetic Control for Satellite Swarms with Local Sensing and Communicating,'' Conference Proceedings, 2013.
  \bibitem{Ramirez-Riberos}Ramirez-Riberos, J. L., ``New Decentralized Algorithms for Spacecraft Formation Control Based on a Cyclic Approach,'' Ph.D. Thesis, Massachusetts Institute of Technology, Cambridge, MA, 2010, https://dspace.mit.edu/handle/1721.1/59666.
  \bibitem{Huang2}Huang, X., Zhang, C., and Ban, X., ``Dipole Solution and Angular-Momentum Minimization for Two-Satellite Electromagnetic Formation Flight,'' \textit{Acta Astronautica}, Vol. 119, 2016, pp. 79-86. https://doi.org/10.1016/j.actaastro.2015.11.009.
  \bibitem{Ayyad}Ayyad, A., ``Optimal Guidance and Control for Electromagnetic Formation Flying,'' Master’s Thesis, The University of Tokyo, 2019. http://hdl.handle.net/2261/00078563.
  \bibitem{Kaneda}Kaneda, R., Yazaki, F., Sakai, S., Hashimoto, T., and Saito, H., ``The Relative Position Control in Formation Flying Satellites Using Super-Conducting Magnets,'' \textit{International Symposium on Formation Flying Missions and Technologies}, Washington DC, 2004.
  \bibitem{Sakai}Sakai, S., Kaneda, R., Maeda, K., Saitoh, T., Saito, H., and Hashimoto, T., ``Electromagnetic Formation Flight for LEO Satellites,'' \textit{International Symposium on Formation Flying, Missions and Technologies}, 2008.
  \bibitem{Zhang}Zhang, C., and Huang, X., ``Angular-Momentum Management of Electromagnetic Formation Flight Using Alternating Magnetic Fields,'' \textit{Journal of Guidance Control Dynamics}, Vol. 39, No. 6, 2016, pp. 1292-1302. https://doi.org/10.2514/1.G001529.
  \bibitem{Porter}Porter, A., Alinger, D., Sedwick, R., Merk, J., Opperman, R., Buck, A., Eslinger, G., Fisher, P., Miller, D., and Bou, E., ``Demonstration of Electromagnetic Formation Flight and Wireless Power Transfer,'' \textit{Journal of Spacecraft and Rockets}, Vol. 51, No. 6, 2014, pp. 1914-1923. https://doi.org/10.2514/1.A32940.
  \bibitem{Youngquist}Youngquist, R. C., Nurge, M. A., and Starr, S. O., ``Alternating Magnetic Field Forces for Satellite Formation Flying,'' \textit{Acta Astronautica}, Vol. 84, March–April 2013, pp. 197-205. https://doi.org/10.1016/j.actaastro.2012.11.012.
  %\bibitem{Abbasi}Abbasi, Z., Hoagg, J., Seigler, T. M., ``Decentralized Position and Attitude Control for Electromagnetic Formation Flight,'' AIAA Conf., 2019.
  \bibitem{Nurge}Nurge, M. A., Youngquist, R. C., and Starr, S. O., ``A Satellite Formation Flying Approach Providing both Positioning and Tracking,'' \textit{Acta Astronautica}, Vol. 122, 2016, pp. 1-9. https://doi.org/10.1016/j.actaastro.2016.01.010.
  \bibitem{Sunny}Sunny, A., ``Single-Degree-of-Freedom Experiments Demonstrating Electromagnetic Formation Flying for Small Satellite Swarms using Piecewise Sinusoidal Controls,'' M.S. Thesis, University of Kentucky, 2019. https://doi.org/10.13023/etd.2019.466.
  \bibitem{Abbasi3}Abbasi, Z., Sunny, A., Hoagg, J., and Seigler, T. M. ``Relative-Position Formation Control of Satellites Using Electromagnetic Actuation with Piecewise-Sinusoidal Controls,'' \textit{2020 American Control Conference (ACC)}, Denver, CO, USA, 2020, pp. 4951-4956. https://doi.org/10.23919/ACC45564.2020.9147724.
  %\bibitem{Abbasi}Abbasi, Z., Hoagg, J., Seigler, T. M., ``Decentralized Position and Attitude Control for Electromagnetic Formation Flight,'' \textit{Proceedings of AIAA Guidance, Navigation, and Control Conference}, 2019-0908. https://doi.org/10.2514/6.2019-0908.
  %\bibitem{Abbasi2}Abbasi, Z., Hoagg, J. B., Seigler, T. M., ``Decentralized Position and Attitude Based Formation Control for Satellite Systems with Electromagnetic Actuation,'' AIAA Conf., 2020.
  %\bibitem{Abbasi2}Abbasi, Z., Hoagg, J. B., Seigler, T. M., ``Decentralized Position and Attitude Based Formation Control for Satellite Systems with Electromagnetic Actuation,'' \textit{Proceedings of AIAA Guidance, Navigation, and Control Conference}, 2020-0617. https://doi.org/10.2514/6.2020-0617.
  \bibitem{Schaub}Schaub, H., and Junkins, J., \textit{Analytical Mechanics of Space Systems}, Reston, Virginia, USA, AIAA, 2012, pp. 96, 109, 110, 147, 258, 484. https://doi.org/10.2514/4.867231.
  \bibitem{Wen}Wen, J. T.-Y., and Kreutz-Delgado, K., “The Attitude Control Problem,” \emph{IEEE Transactions on Automatic Control}, Vol. 36, No. 10, 1991, pp. 1148–1162. https://doi.org/10.1109/9.90228.
  \bibitem{Brockett}Brockett, R. W., ``Asymptotic Stability and Feedback Stabilization,'' \textit{Differential Geometric Control Theory}, 1983, pp. 181-191.
  %\bibitem{Coron}Coron, J.-M., ``Global Asymptotic Stabilization for Controllable Systems without Drift,'' \textit{Mathematics of Control, Signals, and Systems}, Vol. 5, 1992, pp. 295-312. https://doi.org/10.1007/BF01211563.
  \bibitem{Nakamura}Nakamura, Y., and Mukherjee, R., ``Nonholonomic Path Planning of Space Robots via a Bidirectional Approach,'' \textit{IEEE Transactions on Robotics and Automation}, Vol. 7, No. 4, 1991, pp. 500-514. https://doi.org/10.1109/70.86080.
  \bibitem{Bloch}Bloch, A., and McClamroch, N.H., ``Control of Mechanical Systems with Classical Nonholonomic Constraints,'' \textit{Conference on Decision and Control}, 1989, 201-205. https://doi.org/10.1109/CDC.1989.70103.
  %\bibitem{Oriolo}Oriolo, G., and Nakamura, Y., ``Control of Mechanical Systems with Second Order Nonholonomic Constraints: Underactuated Manipulators,'' \textit{Proceedings of the 30th IEEE Conference on Decision and Control}, Brighton, England, 1991, pp 306-308. https://doi.org/10.1109/CDC.1991.261620.
  \bibitem{Fierro}Fierro, R., and Lewis, F. L., ``Control of a Nonholomic Mobile Robot: Backstepping Kinematics into Dynamics,'' \textit{Journal of Robotic Systems}, Vol. 14, No. 3, 1997, pp. 149-163. https://doi.org/10.1109/CDC.1995.479190.
  %\bibitem{Coront}Coront, J. M., ``Global asymptotic stabilization for controllable systems without drift,'' Mathematics of Control, Signals, and Systems, Vol. 5, 1992, pp. 295-312.
  %\bibitem{Pomet}Pomet, J. B., ``Explicit design of time-varying stabilizing control laws for a class of controllable systems without drift,'' Systems $\&$ Control Letters, Vol. 18, 1992, pp. 147-158.
  %\bibitem{Morin}Morin, P., Pomet, J.B., and Samson, C., ``Design of homogeneous time-varying stabilizing control laws for driftless controllable systems via oscillatory approximation of Lie brackets in closed loop,'' SIAM Journal on Control and Optimization, Vol.38, No. 1, 1999, pp. 22–49.
  %\bibitem{M’Closkey}M’Closkey, R. T., and Murray, R. M., ``Nonholonomic systems and exponential convergence: Some analysis tools,'' in Proc. 32nd IEEE Conf. Decision Contr., 1993, pp. 943–948.
  %\bibitem{Morin2}Morin, P. and Samson, C., ``Time-Varying Exponential Stabilization of a Rigid Spacecraft with Two Control Torques,'' IEEE Transactions on Automatic Control, Vol. 42, No. 4, 1997, pp. 528-534.
  %\bibitem{Pettersen}Pettersen, K. Y., and Egeland, O., ``Position and attitude control of an underactuated autonomous underwater vehicle,'' Proc. of the 35th IEEE Conf. on Decision and Control (Kobe, Japan), 1996, pp. 987–991.
  \bibitem{Morgan}Morgan, A., \emph{Solving Polynomial Systems Using Continuation for Engineering and Scientific Problems}, Prentice-Hall, Inc., Englewood Cliffs, New Jersey, 1987, Chap. 3-3. https://doi.org/10.1137/1.9780898719031.
  %\bibitem{Kane}Kane, T.R., Likins, P.W., and Levinson, D.A., \emph{Spacecraft Dynamics}, McGraw-Hill Book Co., New York, 1983.
  \bibitem{Poedts}Poedts, S. and Goedbloed, H., \emph{Principles of Magnetohydrodynamics: With Applications to Laboratory and Astrophysical Plasmas}, Cambridge University Press, 2004, Chap. 8-3. https://doi.org/10.1017/CBO9780511616945.
  \bibitem{Sanders}Sanders, J. A., Verhulst, F., and Murdock, J. A., \emph{Averaging Methods in Nonlinear Dynamical Systems, vol. 2}, Springer, 2007, Chap. 2. https://doi.org/10.1007/978-0-387-48918-6.
\bibitem{Tsiotras}Tsiotras, P., “Stabilization and Optimality Results for the Attitude Control Problem,” \emph{Journal of Guidance, Control, and Dynamics}, Vol. 19, No. 4, 1996, pp. 772–779. https://doi.org/10.2514/3.21698.
  \bibitem{Slotine}Slotine, J.-J.E. and Li, W., \emph{Applied Nonlinear Control}, Prentice Hall,. New Jersey, 1991, Chaps. 3, 4-5-2.%3-4,
  %\bibitem{Olfati-Saber}Olfati-Saber, R., and Murray, R. M., ``Consensus problems in networks of agents with switching topology and time-delays,'' in IEEE Transactions on Automatic Control, Vol. 49, No. 9,  2004, pp. 1520-1533.
  %\bibitem{Mesbahi}Mesbahi, M., and Egerstedt, M., ``Graph theoretic methods in multiagent networks'', Princeton University Press, 2010.
  %\bibitem{Ren}Ren, W., and Beard, R. W., ``Distributed Consensus in Multi-Vehicle Cooperative Control,'' ser. Communications and Control Engineering. London, U.K.: Springer-Verlag, 2008.
  %\bibitem{Chung}Chung, S. J., Ahsun, U., Slotine, J. J. E., ``Application of Synchronization to Formation Flying Spacecraft: Lagrangian Approach,'' Journal of Guidance, Control, and Dynamics, Vol. 32, 2009, pp. 512-526.
  \end{thebibliography}
  %\bibliographystyle{./new-aiaa}
%\begin{comment}  

%\end{comment}
%\end{multicols}
%\tableofcontents
\end{document}